\documentclass[twocolumn,usenatbib]{aastex62}
\usepackage{aasmacros}
\usepackage{graphics}
\usepackage{graphicx}
\usepackage{amssymb}
\usepackage{amsmath}
\usepackage{amstext}  
\usepackage{amsfonts}    
\usepackage{floatrow}
\usepackage{relsize}	
\usepackage{booktabs}
\PassOptionsToPackage{hyphens}{url}
\usepackage{url}
\usepackage{hyperref}
\usepackage{mathtools}
\usepackage{bm}
\usepackage{esvect}
\usepackage{natbib}
\usepackage{tabu}
\usepackage{subfigure}
\usepackage{float}  		
\usepackage{dcolumn}
\usepackage{mathtools}
\usepackage{chngcntr}

\let\cite=\citen

\def\baas{AAS}
\def\apj{ApJ}
\def\aj{AJ}
\def\mnras{MNRAS}

\bibpunct{(}{)}{;}{a}{}{,}

\newcolumntype{d}{D{.}{.}{-1}} 
\newcolumntype{e}{D{,}{\pm}{-1}} 

\makeatletter
\newcommand*{\rom}[1]{\expandafter\@slowromancap\romannumeral #1@}
\makeatother

\shortauthors{\small{Vernstrom et al.}} 
\shorttitle{\small{Pairs vs Doubles}}

\begin{document}

\label{firstpage}
\title{Differences in Faraday Rotation Between Adjacent Extragalactic Radio Sources as a Probe of Cosmic Magnetic Fields}

\correspondingauthor{Tessa Vernstrom}
\email{tessa.vernstrom@csiro.au}
\author{T. Vernstrom}
\affiliation{CSIRO Astronomy $\&$ Space Science, Kensington, Perth 6151, Australia}
\affiliation{Dunlap Institute for Astronomy and Astrophysics University of Toronto, Toronto, ON M5S 3H4, Canada}

\author{B.M. Gaensler}
\affiliation{Dunlap Institute for Astronomy and Astrophysics University of Toronto, Toronto, ON M5S 3H4, Canada}
\affiliation{Department of Astronomy and Astrophysics, University of Toronto, M5S 3H4, Ontario, Canada}

\author{L. Rudnick}
\affiliation{Minnesota Institute for Astrophysics, School of Physics and Astronomy, University of Minnesota, 116 Church Street SE, Minneapolis, MN 55455, USA}

\author{H. Andernach}
\affiliation{Depto. de Astronom\'{i}a, DCNE, Universidad de Guanajuato, Apdo. Postal 144, Guanajuato, CP 36000, Gto., Mexico}

\begin{abstract}
 Faraday rotation measures (RMs) of extragalactic radio sources provide information on line-of-sight magnetic fields, including contributions from our Galaxy, source environments, and the intergalactic medium (IGM). Looking at differences in RMs, $\Delta$RM, between adjacent sources on the sky can help isolate these different components. In this work, we classify adjacent polarized sources in the NVSS as random or physical pairs. We recompute and correct the uncertainties in the NVSS RM catalog, since these were significantly overestimated. Our sample contains 317 physical and 5111 random pairs, all with Galactic latitudes $|b|\ge20\degr$, polarization fractions $\ge2\%$, and angular separations between $1.5\arcmin$ and $20\arcmin$. We find an rms $\Delta$RM of $14.9\pm0.4\,$rad m$^{-2}$ and $4.6\pm1.1\,$rad m$^{-2}$ for random and physical pairs, respectively. This means polarized extragalactic sources that are close on the sky, but at different redshifts, have larger differences in RM than two components of one source. This difference of $\sim10\,$rad m$^{-2}$ is significant at $5\sigma$, and persists in different data subsamples. While there have been other statistical studies of $\Delta$RM between adjacent polarized sources, this is the first unambiguous demonstration that some of this RM difference must be extragalactic, thereby providing a firm upper limit on the RM contribution of the IGM. If the $\Delta$RMs originate local to the sources, then the local magnetic field difference between random sources is a factor of two larger than between components of one source. Alternatively, attributing the difference in $\Delta$RMs to the intervening IGM yields an upper limit on the IGM magnetic field strength of $40\,$nG.
\end{abstract}


\section{Introduction}
\label{sec:intro}

Magnetic fields are found on almost all scales: from stars in our Galaxy, to the interstellar medium, to powerful active galactic nuclei (AGN), the intracluster medium (ICM), and the intergalactic medium (IGM). What is not yet well known are the detailed properties of these magnetic fields, such as the field strengths, or how they vary with spatial scale, distance or cosmic time.

While we cannot measure cosmic magnetic fields directly, they do affect light in ways we can observe. One of the most powerful techniques to investigate magnetic fields is the Faraday rotation effect \citep[e.g.][]{Gardner66,Carilli02, Govoni04}. The linear polarization angle $\psi_0$ of the signal emitted by a radio source along a given line of sight through a foreground magneto-ionic medium is rotated at a wavelength $\lambda$ such that:
\begin{equation}
\psi_{\rm obs}= \psi_{0} + \rm{RM} \lambda^2,
\label{eq:rm0}
\end{equation}
where $\psi_{\rm{obs}}$ is the observed polarized angle and RM is the rotation measure, and $\lambda$ is in the observer's frame. 
The RM is related to the properties of the Faraday rotating plasma by the equation
\begin{equation}
\rm{RM} \, = \, 0.812 \int_{z_s}^{0} \frac{n_e(z)\,\vv{\bm{B}}(z)}{(1+z)^2}\, \cdot\frac{d\vv{\bm{l}}}{dz} \, dz \,\,\, {\rm rad} \,\,\, {\rm m}^{-2},
\label{eq:RM1}
\end{equation}
where $z_s$ is the redshift of the source, $n_e$ (cm$^{-3}$) is the thermal electron density, $\vv{\bm{B}}$ ($\mu$G) is the magnetic field vector, and $\vv{\bm{l}}$ (pc) is the path element along the line of sight (LOS). When observing a distant polarized source, the rotation of its linear polarization angle is due to the contribution of different magneto-ionic regions along the line of sight. The measured RM is a sum of contributions from multiple sources, or regions, such that
\begin{equation}
\rm{RM} \, = \,\rm{RM}_{\rm{source}}+\rm{RM}_{\rm{IGM}}+\rm{RM}_{\rm{Gal}}+\rm{RM}_{\rm{noise}}.
\label{eq:dsum}
\end{equation}
Here RM$_{\rm{source}}$ is the RM from inside, or around, the source, RM$_{\rm{IGM}}$ is the contribution originating in the foreground intergalactic medium, RM$_{\rm{Gal}}$ is the RM contribution from our Galaxy,  and RM$_{\rm{noise}}$ is measurement uncertainty. 

Magnetic fields in structures such as galaxies, clusters and perhaps filaments were amplified from their origins at nG levels \citep[estimated for the IGM from models,][]{Dolag99,Ryu08,Vazza14a,Vazza17} to the $\mu$G levels we observe today. While large-scale fields have been detected in galaxies and clusters, signatures of the fields' origins have been erased due to strong modification \citep{Vazza15}. Knowledge of the overall distribution of magnetic fields in the IGM, and their dependence on redshift, is a prerequisite to uncovering the history of magnetic field evolution. This is a challenging task because there are several contributors to the observed rotation measures (eq.~\ref{eq:dsum}). The different RM contributions need to be separated if we want to understand what RM$_{\rm{source}}$ and RM$_{\rm{IGM}}$ are, and how they evolve with redshift. 

The large-scale structure of the Galactic magnetic field has been studied using Faraday rotation measurements from pulsars and distant AGN \citep[e.g.][]{Simard-Normandin80,Han06,Brown03,Men08}. Efforts have been made to map the Galactic RM contributions using RM grids and Bayesian analysis \citep{Oppermann12, Oppermann15}.  

By observing the RMs of pairs of extragalactic polarized sources that are located adjacent to each other on the sky, we can use the difference between their RMs to subtract out the Galactic contribution and isolate the contribution of other Faraday components in eq.(~\ref{eq:dsum}). By observing {\it many} pairs with different angular separations, $\Delta r$, and redshifts, one can begin to break apart the different contributions and isolate the redshift dependence. It is known from looking at the RM structure functions (the difference in RM of the components of a pair, $\Delta$RM, as a function of $\Delta r$), that there is a dependence on the RM difference with separation \citep{Simonetti84,Minter96, Stil11}. At higher Galactic latitudes the RM variance remains relatively flat as a function of angular separation \citep{Simonetti84, Leahy87,Sun04}, while at lower Galactic latitudes the difference in RMs appears to increase with increasing separation \citep{Leahy87,Simonetti92,Sun04} with a break in the change in the increase at separations of $\ga1{\degr}$, a separation which can thus be interpreted as an upper bound on the scale in the turbulence in the Galactic ISM.

Pairs with a small angular separation are thus likely to have very similar Galactic contributions. For those with larger separations, the trend of increasing RM difference with increasing separation will have a contribution from variations in the Galactic foreground. 

The dependence of the $\Delta$RM on separation has not previously been explored separately for physical pairs (components of the same source) and random pairs (two physically unrelated sources with a given angular separation on the sky). Since random pairs are typically made up of physically unrelated sources at differing redshifts, the more distant source of the pair has contributions from more of the intergalactic medium, as well as possible contributions from the local environment of the closer source. This extra contribution to the RM from the intervening medium between the background and foreground source should result in larger $\Delta \rm{RM}$s for random pairs than for physical pairs. Thus by separating the pairs into physically related and unrelated ones, trends in $\Delta$RM as a function of angular separation, redshift, and the $\Delta z$ of the pairs can provide information about the magnetic field of the Galactic foreground on different angular scales, as well as of the IGM at different redshifts. However, to isolate this change in RM due to the IGM, differences from other factors such as source type and local environment must also be accounted for (with the aid of other source properties such as spectral indices, polarization fractions, source size, and more).

In this work we examine whether physically connected pairs of sources have different $\Delta \rm{RM}$s than random associations of source pairs coincidentally located close to each other on the sky, and what that can tell us about the source environments and intervening medium. Section~\ref{sec:data} details the data used for this analysis. In Section~\ref{sec:method} we describe the methods used for classifying source pairs, how the samples are defined, and which parameters are measured. Section~\ref{sec:results} presents the results from comparing the two different classes of source pairs, and in Sec.~\ref{sec:discussion} we discuss the possible physical interpretations of the results. Throughout the paper, we assume a standard $\Lambda$CDM concordance cosmology with $H_0=70 \,$km s$^{-1}$ Mpc$^{-1}$, $\Omega_M= 0.3$, and $\Omega_{\Lambda}=0.7$ and define the spectral index, $\alpha$ such that the observed flux density $I$ at frequency $\nu$ follows the relation $I_{\nu}\propto \nu^{+\alpha}$.

\section{Data}
\label{sec:data}

We make use of the \citet{Taylor09} catalog, which provides $1.4\,$GHz polarized intensities, polarized fractions, and rotation measures for $37,543$ sources and source components from the NRAO VLA Sky Survey \citep[NVSS,][]{Condon98} catalog, covering the sky north of Declination $-40{\degr}$. Both the NVSS and \citet{Taylor09} catalogs include entries for individually detected peaks, even if multiple peaks, or entries, are part of one larger source. These are therefore catalogs of components rather than of complete sources. In what follows we refer to the catalog entries in \citet{Taylor09} as ``RM sources". Based on NVSS data, these authors obtained rotation measures from the difference of the polarization angle at two frequencies around $1.4\,$GHz separated by $70\,$MHz (eq.~\ref{eq:rm0}). The \citet{Taylor09} catalog has an RM source density of $\sim 1\,$ per deg$^{2}$.

In the original NVSS release, off-axis leakage corrections were applied to the data \citep[see ][for more details]{Condon98}. However, for the \citet{Taylor09} catalog the data were split into two frequency bands and reimaged without the off-axis leakage corrections. For sources far away from a pointing center ($\ga 10{\arcmin}- 15{\arcmin}$) the off-axis leakage fraction can be significant for weakly polarized sources (up to a few percent of the total flux). Therefore for any components of the Taylor catalog used in this analysis, we compute the leakage fractions following the method used in \citet{Ma18}, which uses the distance of the components to the nearest NVSS pointing center and a parabolic fit to leakage measurements of calibrator sources as a function of angular offset.\footnote{The document giving the parabolic fit to the leakage measurements of calibrator sources can be found at \url{https://web.archive.org/web/20170705091828/ftp://ftp.aoc.nrao.edu/pub/software/aips/TEXT/PUBL/AIPSMEMO86.PS}} While this is a crude estimate of the leakage fraction it still yields information on which sources may have more inaccurate polarization properties.

The accuracy of the uncertainties in the rotation measures reported by \citet{Taylor09} has previously been questioned \citep[e.g.][]{Stil11}. As the RM uncertainties are relevant for the analysis in this work, we investigated this issue. While \citet{Stil11} found the uncertainties to be underestimated, our findings show that for, at least a subset of the catalog entries, the uncertainties are actually overestimated. The process used by \citet{Taylor09} for obtaining the RM uncertainties is not clear, and thus we cannot be sure of the exact reason for the overestimation. However, the most likely explanation would be that real polarized signal was, inadvertently, included in the image noise measurement when there were nearby sources or additional source components. The full set of images at the two frequencies is not publicly available. However, using a subset of images, made available to us by J. Stil, we are able to determine a relation for estimating the RM uncertainties that yields more accurate values. For full details on this on this issue refer to Appendix~\ref{sec:rmuncert}.

In order to aid in classification of ambigous pairs as physical or random associations we make use of a growing compilation of optically identified extended radio galaxies (ERGs) maintained by one of us (H.A.). This compilation was originally aimed at compiling giant radio galaxies \citep[GRGs, i.e. larger than $1\,$Mpc in projection on the sky, e.g.][]{Andernach12}, but has evolved into a collection of radio galaxies potentially larger than $\sim 200\,$kpc, requiring an angular extent of $\ga 30{\arcsec}$. It is limited to sources with a clear, or highly likely, optical identification and includes, apart from a careful measurement of the largest angular size (LAS), spectroscopic redshifts when available. In the absence of spectroscopic redshifts, photometric redshifts are collected and averaged from various sources like \citet{Beck16,Brescia14,Bilicki16,Bilicki14a} and for QSO candidates from \citet{DiPompeo15}. While most of the largest sources were found from visual inspections by H.A. and his students, a significant fraction of sources were drawn from the volunteer work of the Radio Galaxy Zoo project \citep{Banfield15}. The compilation has made use of object lists of over 500 references, including several dozen dedicated to GRGs \citep[like those of][]{Lara01,Proctor16}. For the present paper we used the June 2018 version of the compilation, hereafter referred to as the ERG catalog, comprising about 5780 radio sources larger than $1.5{\arcmin}$, with an additional $\sim$ 1100 candidates (Andernach et al., in prep).

For redshifts of the sources we utilize the ERG catalog as well as the catalogs of \citet{Hammond12} and \citet{Kimball08,Kimball14} (for details see Sec.~\ref{sec:redz}). Additionally, we include source spectral index properties from the spectral index catalog of \citet{deGasperin18}. These spectral indices were derived from crossmatching the NVSS survey with the GMRT 150 MHz all-sky radio survey first alternative data release \citep[TGSS ADR1,][]{Intema17}.

\section{Method}
\label{sec:method}

\subsection{Classification of Two Types of Pairs}
\label{sec:classify}

\begin{figure*}
\includegraphics[scale=0.37]{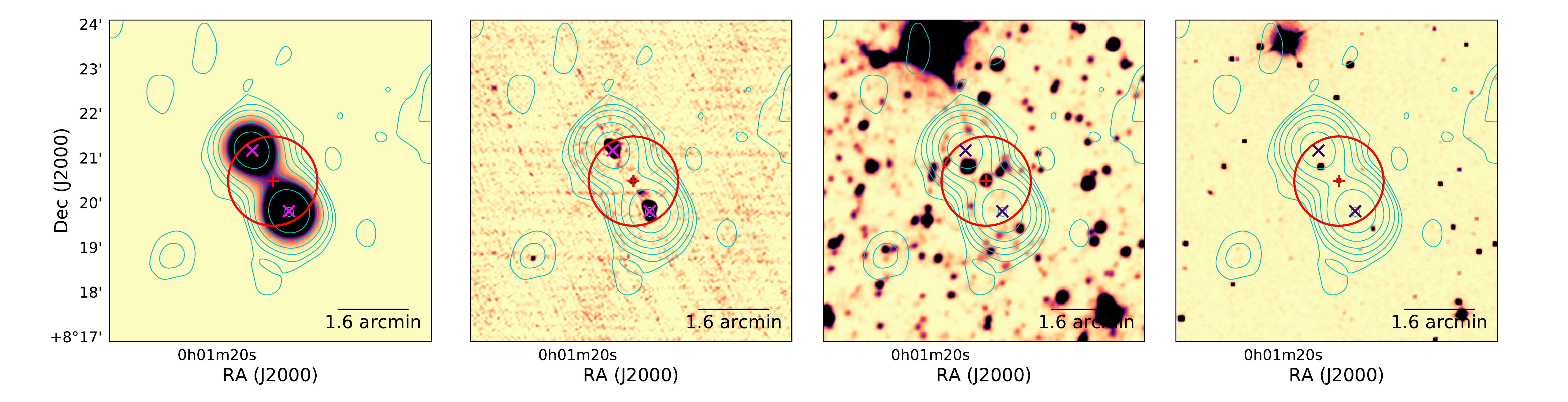}
\includegraphics[scale=0.37]{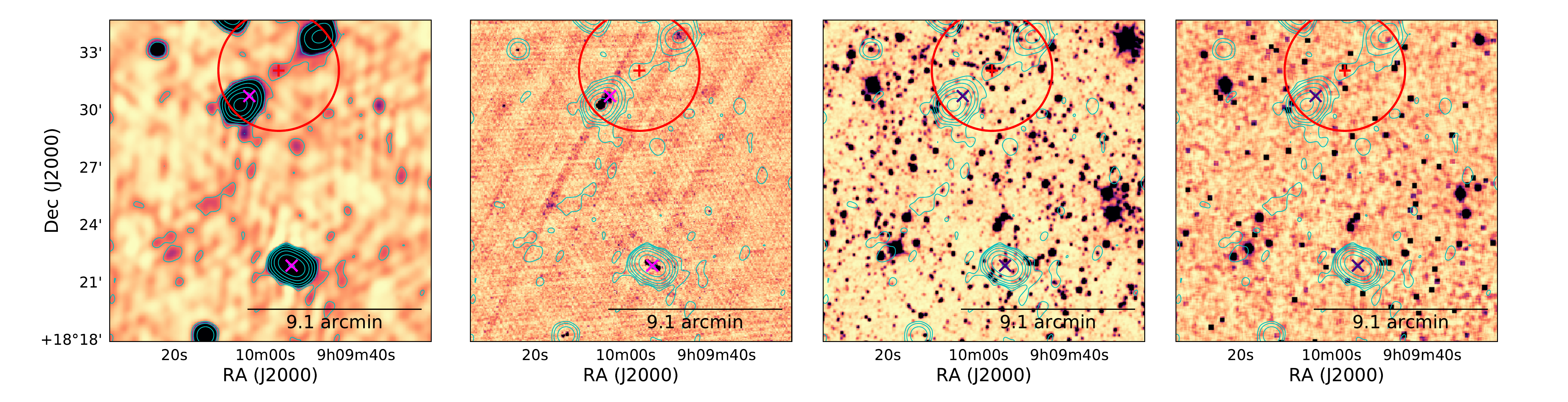}
\includegraphics[scale=0.37]{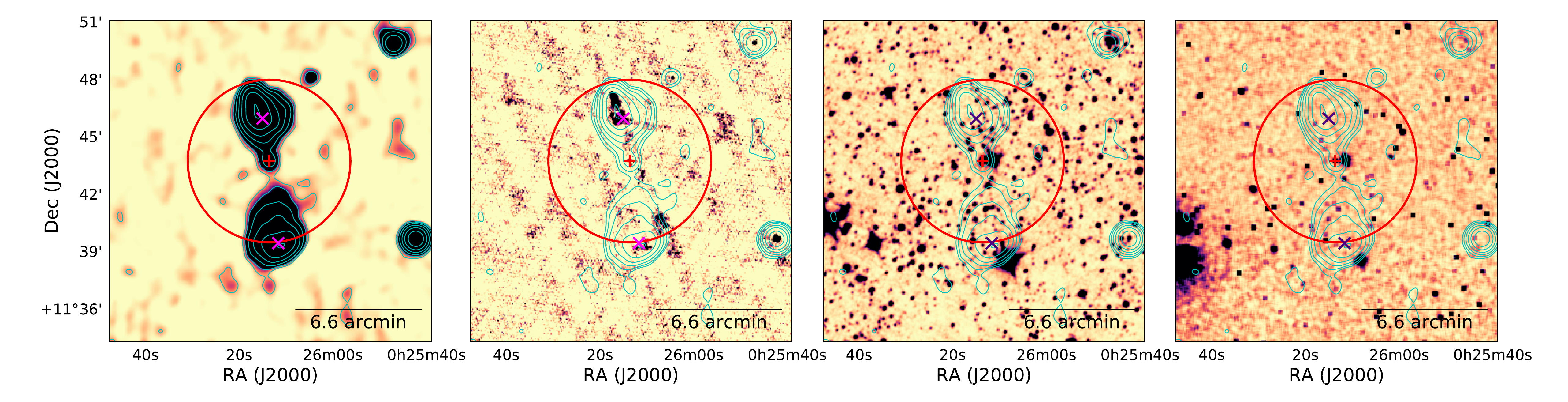}
\caption{Example images used in the process of pair classification. Top row is pair J000115+082029 (components J000113+081948 and J000116+082110), classified as a physical pair; middle row is for pair J090958+182617 (components J090954+182151 and J091003+183042) classified as a random pair; and bottom row is for pair J002613+114241 (components J002611+113926 and J002615+114556), classified as a physical pair (with the names here corresponding to the mean RA and Dec of the pair). The light blue contours are from NVSS. The green crosses are the positions of the RM sources from \citet{Taylor09}. The red $+$ marks the location of any matched source in the ERG compilation by Andernach et al., with the circle corresponding to the largest angular size as listed in the ERG compilation. The separation of the RM components is marked in the bottom right corner. From left to right, the color images are $1.4\,$GHz NVSS, $1.4\,$GHz FIRST, {\it WISE} $3.4 \, \mu$m and SDSS $g$-band images. \\ }
\label{fig:classify1}
\end{figure*}

We start by cross-matching the \citet{Taylor09} catalog against itself, finding all pairs with separations $\Delta r \leq 30{\arcmin}$. Excluding self-matches, this results in $19,510\,$ pairs. We exclude pairs where the polarized fraction of either source is less than $2\%$, which ensures the source's polarized fractions are larger than their reported uncertainties in polarized fraction. We also exclude pairs with $\Delta r < 1.5{\arcmin}$, i.e. less than two times the synthesized beam size of the NVSS survey of $45{\arcsec}$. This is to prevent, or minimize, the effect of source blending either on the detection of sources or measurement of the RMs. We include the additional constraint that the pairs are located away from the Galactic plane, $|b|\geq 20{\degr}$. This threshold is comparable to that chosen by other studies \citep{Welter84, Oren95, Hammond12} and avoids the increased spatial fluctuations of the Galactic RMs \citep{Schnitzeler10}.  

Using the angular size distribution of physical sources from the ERG catalog, we exclude any pairs with angular sizes greater than $20{\arcmin}$, as the number of physical pairs at larger separations would be very small, especially considering the number of random pairs is expected to increase greatly with larger separations. This leaves a sample of 5428 pairs to classify. 

Each pair could be a random association or a physical pair, the latter being either separate components of a multi-component radio galaxy (e.g. AGN lobes, cores, hotspots, or jets) or multiple RMs within one of the components (e.g. two RM measurements within one AGN jet or lobe). To classify a pair as physical or random requires visual inspection. For each pair, we obtain the NVSS image (centered on the mean RA and Dec of the pair), an infrared image from the Wide-field Infrared Survey Explorer \citep[{\it WISE}, ][]{Wright10} at $3.4 \, \mu$m, the optical $g$-band image (where available) from the Sloan Digital Sky Survey (SDSS) data release 9 \citep{Ahn12}, and the $1.4\,$GHz image from the Faint Images of the Radio Sky at Twenty-cm \citep[FIRST,][]{Becker95} survey. If the pair is outside of the FIRST coverage area but has been observed in the first epoch of observing for the new VLA Sky Survey (VLASS, Lacy et al., in prep)\footnote{\url{https://science.nrao.edu/science/surveys/vlass}}, then the VLASS ``Quicklook" image is obtained instead. Plots are made using the available images for each pair, with NVSS contours overlaid and the positions of the RM sources highlighted, in order to visually inspect the sample of pairs and determine classification. 

All of the pairs are cross-matched with the ERG catalog using the larger of either the pair separation or the largest angular size (LAS) of the ERGs as a search radius. If a possible ERG match is found, the listed RA and Dec position of the ERG and the LAS are added to plots used for classification. Some examples of these images are shown in Figure~\ref{fig:classify1}.

A number of different criteria were used to determine whether a pair should be classified as physical. These include: 
\begin{itemize}
\setlength\itemsep{0em}
\item A visible and/or cataloged optical and/or infrared counterpart near the mean RA and Dec of the pair or near the position of what appears to be a radio core between AGN jets or lobes;
\item An ERG source with center near the mean RA and Dec of the pair and the pair separation close to the LAS listed in the ERG compilation;
\item No visually discernible or cataloged (bright) optical and/or infrared counterparts corresponding to the positions of either of the two RM sources;
\item Source blending in NVSS, or overlapping or connecting NVSS $3\sigma$ contours;
\item Complex (resolved) radio emission seen in NVSS and/or FIRST/VLASS associated with the positions of the RM source(s); 
\item Clear presence of multiple resolved source components in higher resolution radio images (e.g. jets, lobes, hot spots, and/or a discernible core).
\end{itemize}
The majority of pairs examined either clearly met a majority of the physical criteria or they clearly did not. However, due to the fact that optical and/or high resolution radio and/or ERG matches were not available for all the pairs, the exact criteria required to be considered physical varied depending on the pair being considered. Nevertheless, at least two of these criteria needed to be met for a pair to be classified as physical. 

All of the pairs were visually inspected by the lead author. In cases where the classification was not immediately clear, additional verification was provided by at least one of the co-authors. The ERG catalog was heavily used to identify physical pairs, meaning for any pairs with a match in the ERG catalog were examined by T. Vernstrom for this work and H. Andernach during the compilation of the ERG catalog. We note that while machine learning algorithms for the purpose of finding or classifying extended radio galaxies are becoming more prevalent \citep[e.g.][]{Proctor16}, at this stage however, there is usually some amount of manual inspection or verification required with the results from these algorithms. Given that the goal of this work was not to create a full catalog of ERGs and only a relatively small sample was being examined, a machine learning algorithm was not developed for this case and all of the classification was done manually. 

\begin{figure}
\includegraphics[scale=0.38]{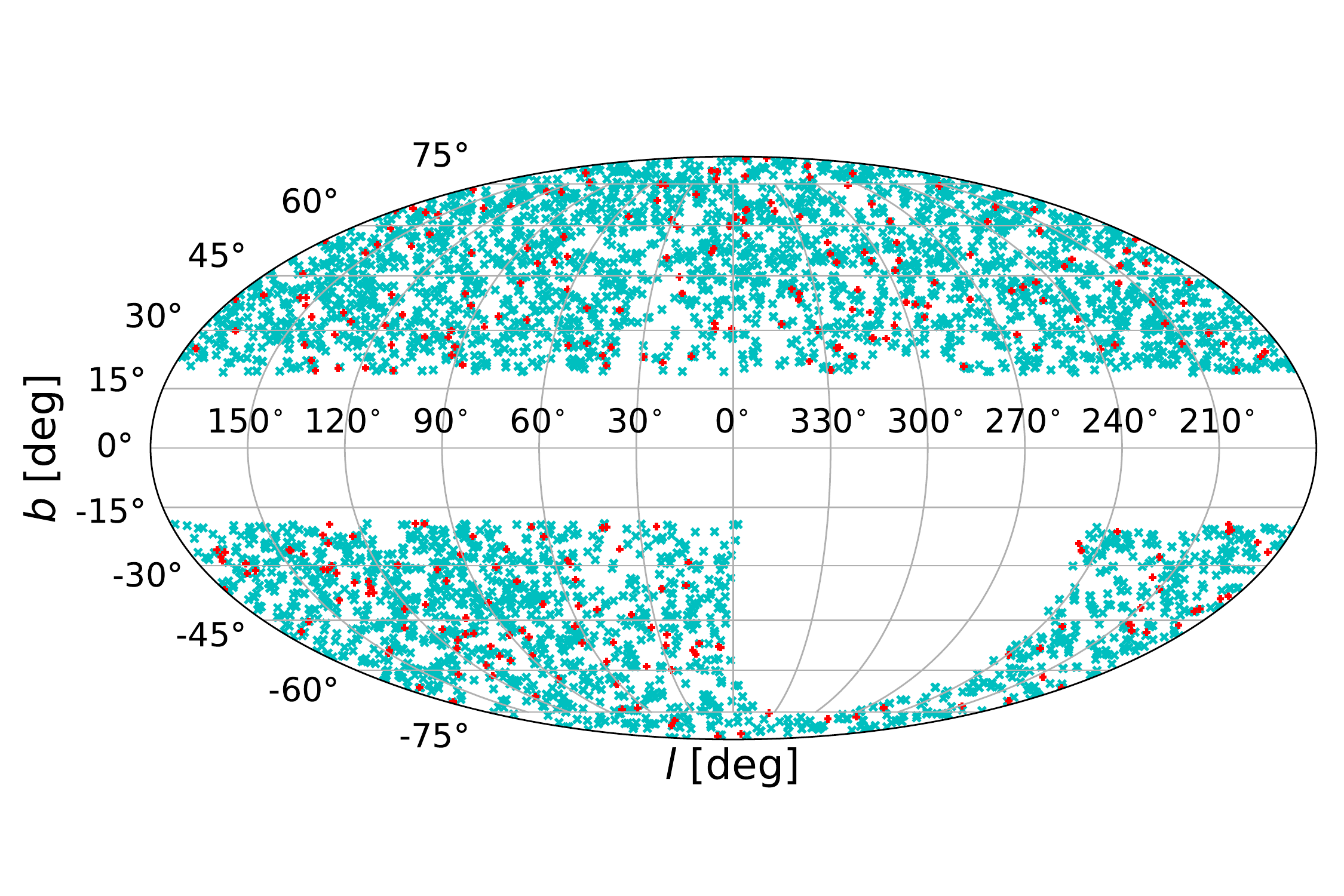}
\caption{Positions of classified RM pairs in Galactic coordinates. The light blue points are the random pairs, while the red points are the physical pairs. }
\label{fig:galcoords}
\end{figure}

This resulted in 317 physical pairs and 5111 random pairs all with $|b|\geq \, 20{\degr}$, $\pi_{\rm min}\geq 2 \%$, and $1.5{\arcmin}\leq \Delta r \leq 20{\arcmin}$. The spatial distribution of pairs is shown in Fig.~\ref{fig:galcoords}, which shows a fairly uniform sampling of positions within the NVSS sky coverage. The physical pairs, being separated by more than the NVSS beam size, are either completely or partially resolved. The sources in random pairs tend to be more compact, with a mean fitted major axis from NVSS of $54\arcsec$ with a $45\arcsec$ beam size (the mean fitted component size of the physical pair sources being $65\arcsec$) and over $30\%$ of the NVSS sizes for the random pair sources are marked as upper limits. Thus compared to the physical sources the random pair sources are on average smaller, or more unresolved.

It should be noted that while there are no duplicate pairs, individual components from \citet{Taylor09} may appear in multiple pairs. One catalog entry can be matched with any number of other sources or components. A component of a physical pair may also be matched with other non-associated (random) sources to make random pairs. Likewise, physical sources with more than two polarized components will comprise multiple pairs. The highest number of pairs that any one catalog entry is included in is five, with the average number being 1.3 (and a median of 1).

\subsection{The Difference in RMs}
\label{sec:rmdifs}
For each pair we compute the difference in the RMs, or $\Delta {\rm RM} = RM_1 -RM_2$. The RMs can have negative or positive values, which should result in a mean $\Delta$RM over many sources of approximately zero. In order to explore the spread of the distribution of the $\Delta$RMs we compute the root mean square (rms) for the $\Delta$RMs as 

\begin{equation}
\sqrt{ \langle (\Delta {\rm RM})^2 \rangle } = \sqrt{ \frac{1}{N}\sum_i{(RM_1 -RM_2)^2_i}},
\label{eq:rmsrm}
\end{equation}
where the sum is over all the pairs in a group (physical or random pairs) and $N$ is the total number of pairs in the group. We can compute this rms for an entire group (e.g. all random pairs) or as a function of some parameter such as angular separation or polarization fractions. 

There are several possible contributions to the variance, or rms, such that
\begin{equation}
(\Delta {\rm RM})^2_{\rm obs} = (\Delta {\rm RM})^2_{\rm signal} +(\Delta {\rm RM})^2_{\rm noise}.
\label{eq:delrmsum}
\end{equation}
Here $(\Delta {\rm RM})^2_{\rm obs}$ is the value computed from the raw RMs, with no corrections, and $(\Delta {\rm RM})^2_{\rm signal}$ is made up of contributions from different astrophysical sources such as the Galaxy, the local source environments, and the IGM (more discussion on these is given in Sec.~\ref{sec:discussion}). 

The noise term is the variance resulting from measurement errors and instrumental noise. The noise variance term must be subtracted from $(\Delta {\rm RM})^2$ to obtain the contributions from the physical signals \citep[e.g.][]{Haverkorn04,Stil11}. This term, in theory, can be calculated from the errors on the pair RMs, where the noise term for the $i$th pair is
\begin{equation}
(\Delta {\rm RM}_{\rm noise})^2_i = ( \sigma_{\rm RM1}^2 + \sigma_{\rm RM2}^2)_i.
\label{eq:rmnosie}
\end{equation}
Here $\sigma_{\rm RM1}$ and $\sigma_{\rm RM2}$ are the reported RM uncertainties for the two sources in the $i$th pair. The mean from all of the noise terms can then be subtracted off from $\langle (\Delta {\rm RM})^2 \rangle$ \citep[see][appendix A for a full derivation]{Haverkorn04}. This noise correction yields
\begin{equation}
\sqrt{ \langle (\Delta {\rm RM})^2_{\rm signal} \rangle } = \sqrt{   \langle (\Delta {\rm RM})^2_{\rm obs} \rangle -  \langle (\Delta {\rm RM})^2_{\rm noise} \rangle}.
\label{eq:ncor}
\end{equation}

However, this procedure assumes that the RM uncertainties in the RM catalog provide a realistic estimate of the measurement errors. As discussed in Sec.~\ref{sec:data}, we found the RM uncertainties reported in the \citet{Taylor09} catalog to be overestimated. Therefore, to get new RM uncertainties we use the approximation 
\begin{equation}
\sigma_{\rm RM} = 150 \, [{\rm rad \, \, m^{2}}] \, \frac{\sigma_{QU}}{P},
\label{eq:newrm}
\end{equation}
where $\sigma_{\rm RM}$ is the uncertainty in the RM, $\sigma_{QU}$ is the average instrumental noise in the Stokes $Q$ and $U$ images near the source, and $P$ is the peak polarized intensity of the source (for full details and discussion on this see Appendix~\ref{sec:rmuncert}). Throughout the rest of this work, all results presented will have been corrected for the measurement noise unless otherwise specified ($\Delta {\rm RM}_{\rm signal}$, the ``signal" subscript specifies this is after noise correction). 

\subsection{Redshifts}
\label{sec:redz}

After classifying the pairs based on visual inspection of the images, the RM source positions from \citet{Taylor09} are cross-matched with the \citet{Hammond12} and \citet{Kimball08} catalogs, which include redshifts \footnote{The updated version of \citet{Kimball08}, \citet{Kimball14}, was used in this work and can be accesed at \url{http://www.aoc.nrao.edu/~akimball/radiocat_2.0.shtml}}. \citet{Hammond12} includes \citet{Taylor09} components cross-matched with spectroscopic redshifts from the SDSS data release 8 \citep{Aihara11}, the Six-degree Field Galaxy Survey \citep[6dFGS,][]{Jones04,Jones09}, the Two-degree Field Galaxy Redshift Survey \citep[2dFGRS,][]{Colless01,Colless03}, and the 2Df QSO Redshift Survey (2QZ) and 6dF QSO Redshift Survey (6Qz) \citep{Croom04}. \citet{Kimball14} cross matched NVSS components with spectroscopic redshifts from SDSS DR9 \citep{Ahn12}.

The \citet{Kimball14} catalog matches NVSS sources with redshifts. However, if the NVSS entry is indeed a component of a (large) radio galaxy, and not the core component, it is likely not matched with a redshift (even if there exists an optical counterpart with a redshift for the radio galaxy). In the \citet{Hammond12} catalog this was taken into account. In these cases the ERG catalog is useful in that it not only lists the ERG host position, but also provides redshifts (where available). For physical pairs without a redshift match based on the RM source positions, a redshift search was performed using either the position of the core (as determined from the NVSS images) or, if no core is present, the mean RA and Dec of the RM source positions.  

\begin{figure}
\includegraphics[scale=0.37]{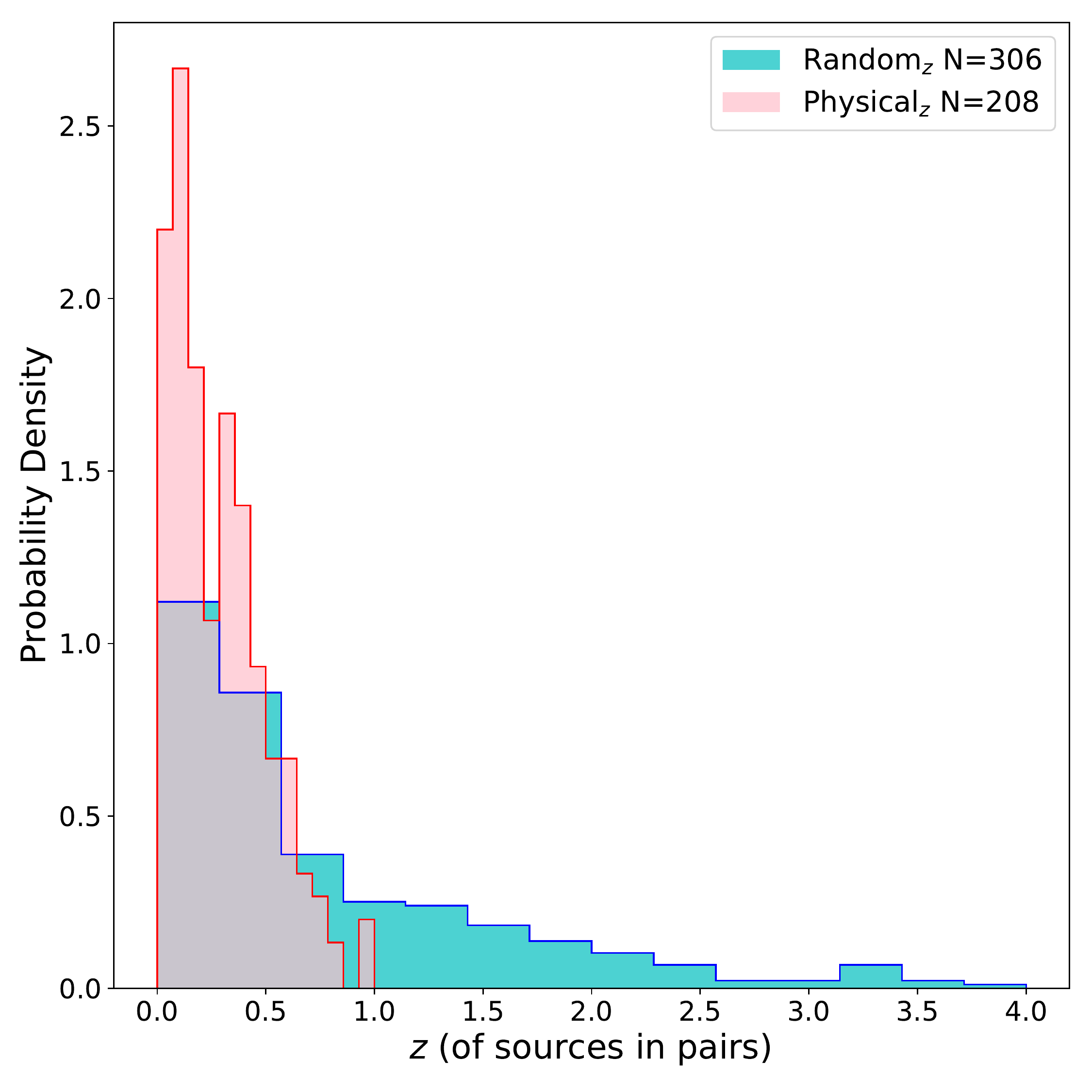}
\caption{The redshift histograms for physical (red) and random (blue) pairs (for pairs where both sources in the pair have a redshift). The redshift for each source in a random pair is included thus the number of redshifts in the random distribution is twice the number of random pairs.}
\label{fig:zhists}
\end{figure}

In the end our redshift subsample consists of 208 physical pairs with redshifts and 153 random pairs where both sources in the pair have a redshift, either spectroscopic or else photometric redshifts. There are 1411 random pairs where at least one source in the pair has a redshift. The median redshift of the physical pairs is $0.3^{+0.2}_{-0.2}$, and the median redshift of the random pairs is $0.5^{+0.9}_{-0.3}$, with the super and subscripts indicating the inner $68\%$ of the distributions. The distributions of redshifts can be seen in Fig.~\ref{fig:zhists}. 

Both the \citet{Hammond12} and \citet{Kimball14} catalogs only provided spectroscopic redshifts. However, the ERG catalog includes both spectroscopic and photometric redshifts. Thus of the 1411  sources with redshifts for components of random pairs, only 6 have photometric redshifts. Of the 208 physical pairs, the redshifts of 144 are spectroscopic and 64 those are photometric.  

For a particular source's RM, from eq~(\ref{eq:RM1}), we can see that the observed RM contribution from the source's intrinsic RM is reduced by a factor of $(1+z)^2$. The RM is an additive quantity and all intervening Faraday screens contribute to the RM of a background source. For continuous regions spread across redshift space, the observed RM is the integral, as seen in eq~(\ref{eq:RM1}). For discrete RM contributions, the observed RM is given by ${\rm RM}_{\rm obs}=\sum_F [{\rm RM}_i^F/(1+z_F)^2]$, where ${\rm RM}_i^F$ is the intrinsic RM of the Faraday screen at redshift $z_F$ and the sum is over all screens along the line of sight. Thus to study the intrinsic RM a correction for the redshift is necessary.

We do not know the nature or number of the intervening Faraday screens and their redshifts. However, we can adopt various simple models for the redshift distribution of the screens and examine their effects on the inferred intrinsic and line of sight RMs. The different models we assume are:
\begin{itemize}
\setlength\itemsep{0em}
\item Model A: Assume all or the majority of the pair RMs come from the redshift of the physical pair or the minimum/foreground redshift of a random pair, such that $\Delta {\rm RM_{\rm corrected}} =\Delta {\rm RM_{\rm observed}} \times (1+z_{\rm min})^2$
\item Model B: Assume all or the majority of each source's RM in a random pair comes from its local environment (or its redshift) and correct each source's RM by the respective redshift factor, $\Delta {\rm RM_{\rm corrected}} =((1+z_1)^2 \times {\rm RM}_1)-((1+z_2)^2 \times \rm{RM}_2)$
\item Model C: Assume all or the majority of $\Delta$RM for a random pair is due to an evenly distributed IGM in the intervening space from $z_1$ to $z_2$, $\Delta {\rm RM}_{\rm corrected}= \Delta {\rm RM}_{\rm uncorrected} / \int_{z_{\rm{max}}}^{z_{\rm{min}}}{(1+z)^{-2} \, dz}$ 

\end{itemize}

In the results presented below, for the random pairs where both components have a redshift, all three possibilities are computed in addition to the observed or uncorrected $\Delta$RM, and for the physical pairs with a redshift the uncorrected $\Delta$RM and redshift corrected $(1+z)^2\times \Delta$RM (model A) are considered.

\section{Results}
\label{sec:results}

\begin{table*}
{\scriptsize
 \setlength{\tabcolsep}{3.9pt}
\caption{Details of \citet{Taylor09} polarized pairs used in this analysis. The ID column is just an internal reference number and ends with an ``a" or ``b", indicating the two components of a pair. Under class, ``R" is for random and ``P" is for physical. In the RM column the uncertainty listed is not from the Taylor catalog, but the uncertainty derived in this work following the procedure described in Appendix~\ref{sec:rmuncert}. The $P$ column gives the frequency-averaged peak polarized intensity as reported by \citet{Taylor09}. The $\sigma_{QU}$ values given are the average of the $Q$ and $U$ image noise measured in this work from the single frequency NVSS postage stamps multiplied by a factor of $\sqrt{2}$ (for details see Appendix~\ref{sec:rmuncert}). The Name column gives the NVSS source name for sources in random pairs and gives the source name from published catalogs (usually the optical or infrared core counterpart) if it is a physical pair. This table is an excerpt, with the full table available online.}
\label{tab:pairs1}
\begin{tabular}{lcccdeecdddccl}
\hline 
\multicolumn{1}{c}{ID} & \multicolumn{1}{c}{Class} & \multicolumn{1}{c}{RA} & \multicolumn{1}{c}{Dec} &  \multicolumn{1}{c}{$\Delta r$} & \multicolumn{1}{c}{RM}& \multicolumn{1}{c}{$\Delta$RM$_{\rm obs}$} & \multicolumn{1}{c}{$P$}& \multicolumn{1}{c}{$\sigma_{QU}$} & \multicolumn{1}{c}{leakage}& \multicolumn{1}{c}{$\pi$} & \multicolumn{1}{c}{$z$} & \multicolumn{1}{c}{$\alpha$} &   \multicolumn{1}{c}{Name}\\

&   & \multicolumn{1}{c}{{\tiny (J2000)}} & \multicolumn{1}{c}{{\tiny (J2000)}} &  \multicolumn{1}{c}{{\tiny [arcmin]}}& \multicolumn{1}{c}{{\tiny [rad m$^{-2}$]}} & \multicolumn{1}{c}{{\tiny [rad m$^{-2}$]}} & \multicolumn{1}{c}{{\tiny [${\rm mJy}\over{\rm beam}$]}}& \multicolumn{1}{c}{{\tiny [${\rm mJy}\over{\rm beam}$]}}& \multicolumn{1}{c}{[$\%$]} & \multicolumn{1}{c}{[$\%$]} & & & \\[1.25ex]

\hline
1a&R&00:00:10.10&+30:55:59.5&16.7&-38,10.4&38,14&5.55&0.39&0.88&6.5&$1.8$&$-0.55$&NVSS J000010+305559\\
1b&R&23:59:58.49&+30:39:26.4&16.7&-76,9.0&38,14&6.39&0.39&0.35&6.5&$-$&$-0.67$&NVSS J235958+303926\\
2a&R&00:00:27.38&+20:35:53.1&10.4&-20,9.2&10,10&6.34&0.39&0.42&8.6&$-$&$-0.77$&NVSS J000027+203553\\
2b&R&23:59:42.74&+20:36:02.2&10.4&-30,5.0&10,10&11.67&0.39&0.23&8.6&$-$&$-0.79$&NVSS J235942+203602\\
3a&R&00:00:27.38&+20:35:53.1&9.5&-20,9.2&11,9&6.34&0.39&0.42&8.6&$-$&$-0.77$&NVSS J000027+203553\\
3b&R&23:59:46.95&+20:36:14.9&9.5&-31,1.8&11,9&33.14&0.39&0.17&8.6&$-$&$-0.79$&NVSS J235946+203614\\
4a&R&00:00:17.06&-34:10:28.9&14.5&-5,9.2&-28,12&6.53&0.40&0.20&7.9&$-$&$-$&NVSS J000017-341028\\
4b&R&00:00:40.47&-34:24:10.1&14.5&23,7.9&-28,12&7.59&0.40&0.83&7.9&$-$&$-0.91$&NVSS J000040-342410\\
5a&R&00:00:23.96&+12:29:50.7&17.7&-24,13.5&-24,14&4.35&0.39&0.94&6.4&$-$&$-0.79$&NVSS J000023+122950\\
5b&R&00:00:37.01&+12:12:26.6&17.7&0,4.4&-24,14&13.23&0.39&0.67&6.4&$0.202$&$-0.47$&NVSS J000037+121226\\
6a&R&00:00:17.06&$-$34:10:28.9&14.5&-5,9.2&-1,12&6.53&0.40&0.20&7.9&$-$&$-$&NVSS J000017-341028\\
6b&R&00:00:44.45&$-$34:23:51.5&14.5&-4,7.4&-1,12&8.11&0.40&0.72&7.9&$-$&$-0.88$&NVSS J000044-342351\\
7a&R&00:01:05.48&$-$16:59:24.9&16.6&-26,4.3&-13,13&12.95&0.37&0.83&2.2&$-$&$-0.85$&NVSS J000105-165924\\
7b&R&23:59:58.86&$-$17:04:00.7&16.6&-13,12.6&-13,13&4.39&0.37&0.03&2.2&$-$&$-0.7$&NVSS J235958-170400\\
8a&R&00:00:23.96&+12:29:50.7&12.9&-24,14.3&-24,15&4.35&0.41&0.94&6.4&$-$&$-0.79$&NVSS J000023+122950\\
8b&R&00:00:49.41&+12:18:32.1&12.9&0,4.7&-24,15&13.07&0.41&1.16&6.4&$0.202$&$-$&NVSS J000049+121832\\
9a&P&00:00:37.01&+12:12:26.6&6.8&0,4.3&0,6&13.23&0.38&0.67&21.6&$0.202$&$-0.47$&SDSS J000043.82+121608.3\\
9b&P&00:00:49.41&+12:18:32.1&6.8&0,4.4&0,6&13.07&0.38&1.16&21.6&$0.202$&$-$&SDSS J000043.82+121608.3\\
10a&R&00:01:01.13&+24:08:42.1&15.4&-63,15.9&10,20&3.65&0.39&0.76&7.8&$-$&$-0.78$&NVSS J000101+240842\\
10b&R&00:01:07.29&+23:53:23.1&15.4&-73,12.8&10,20&4.53&0.39&0.74&7.8&$0.073$&$-0.86$&NVSS J000107+235323\\

\hline
\end{tabular}
}
\end{table*}

\begin{figure}
\includegraphics[scale=0.37]{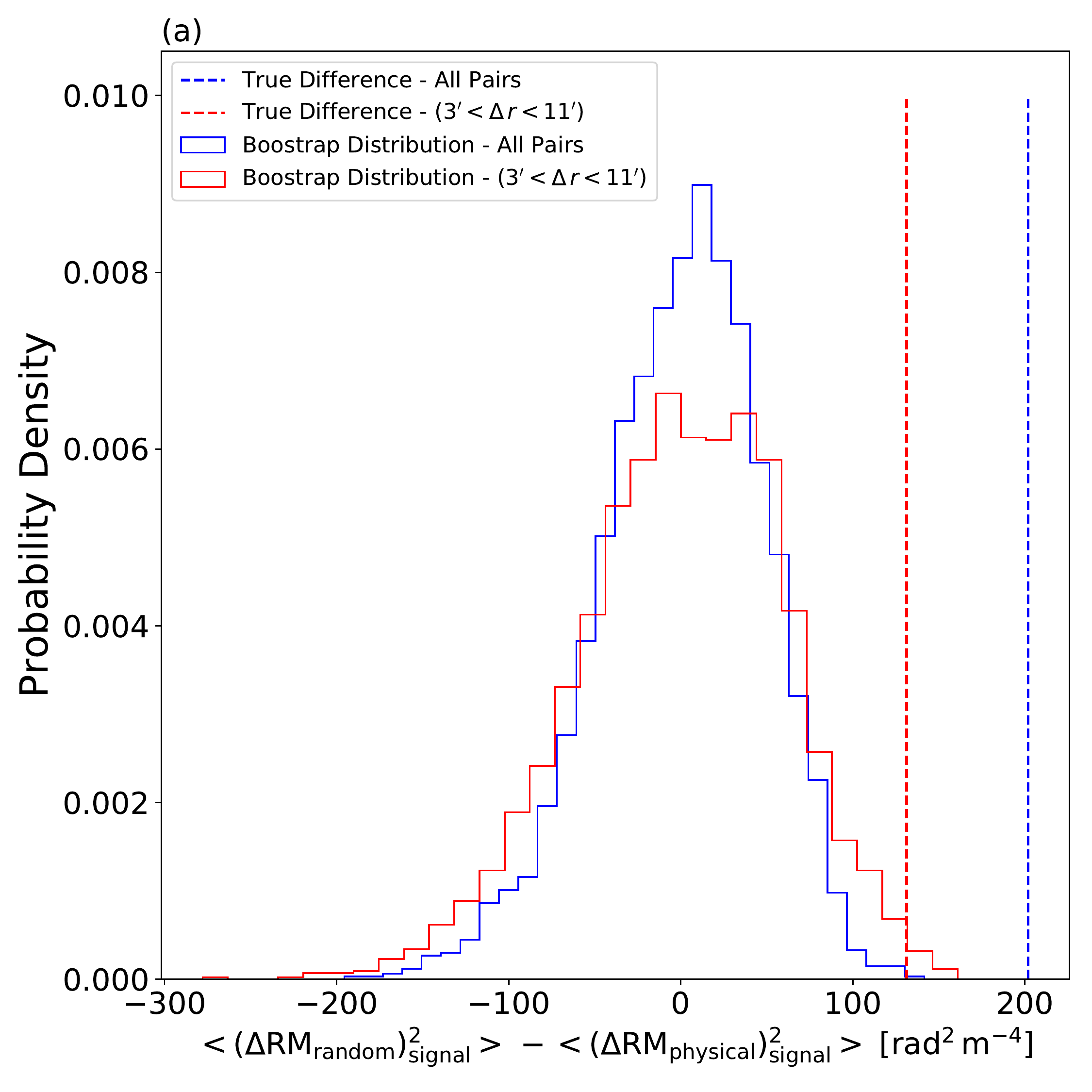}
\includegraphics[scale=0.37]{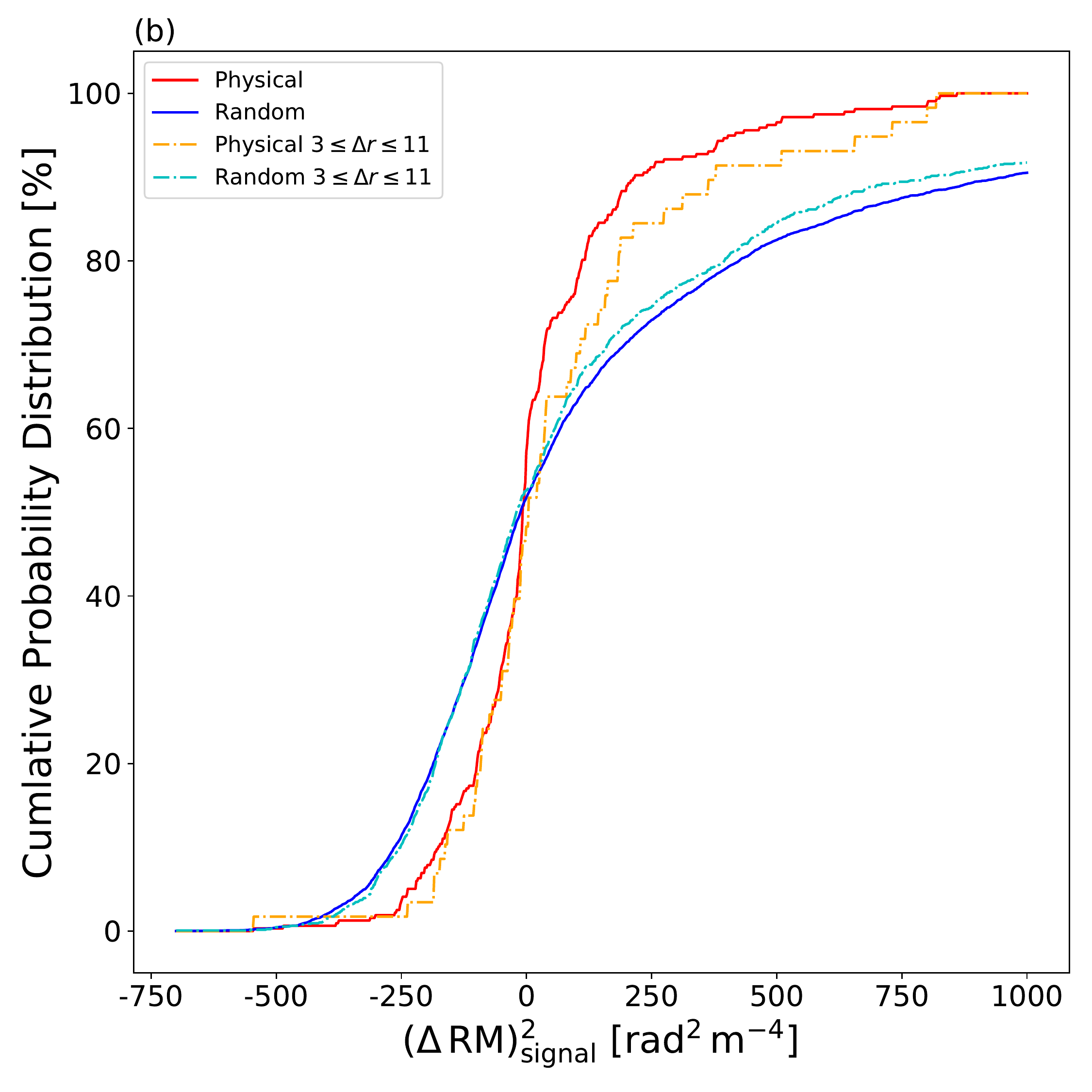}
\caption{Panel (a) shows the distributions from bootstrap realizations of the difference in the mean rms value ($\sqrt{\langle (\Delta \rm{RM})^2_{\rm signal} \rangle}$) of physical and random pairs for all pairs (blue lines) and in the region of greater separation overlap,  $3{\arcmin}\leq \Delta r \leq 11{\arcmin}$ (red lines). The vertical lines mark the actual difference in mean rms values between the random pairs sample and the physical pairs sample. The distribution comes from randomly resampling from the combined samples and recomputing the difference in the mean rms values 5000 times. Panel (b) shows the cumulative probability distributions for the physical and random pairs. The solid lines are the samples with separations $1.5{\arcmin}\leq \Delta r \, \leq 20{\arcmin}$ (red for physical and blue for random pairs), while the dot-dashed lines are the samples with separations $3\arcmin \leq \Delta r \, \leq 11{\arcmin}$ (orange for physical and cyan for random pairs). In both cases the noise-corrected $(\Delta \rm{RM})^2$ values are used. }
\label{fig:boots1}
\end{figure}

\begin{table*}
\caption{Total numbers and average parameters values for the different pair classifications. The uncertainties listed are the standard errors on the means (or $1\sigma/\sqrt{N}$). Here $N$ and $N_{\alpha}$ are the numbers of pairs, not numbers of sources, and $N_{\alpha}$ is the number of pairs where both pair components have a matched spectral index from the NVSS TGSS Catalog of \citet{deGasperin18}. \\}
\label{tab:means}
\begin{tabular}{lreeeere}
\hline
Classification & \multicolumn{1}{c}{$N$ }& \multicolumn{1}{c}{$\sqrt{ \langle (\Delta {\rm RM})^2_{\rm signal} \rangle }$ }& \multicolumn{1}{c}{$\langle \Delta r \rangle$} & \multicolumn{1}{c}{$\langle \pi \rangle$} & \multicolumn{1}{c}{$\langle z \rangle$} & \multicolumn{1}{c}{$N_{\alpha}$} & \multicolumn{1}{c}{$\langle \alpha \rangle$}  \\
 & & \multicolumn{1}{c}{[rad m$^{-2}$] }&  \multicolumn{1}{c}{[arcmin]} & \multicolumn{1}{c}{[ $\%$]} & & & \\
\hline
Random - all & 5111 &14.9,0.4 & 14.0,0.1&7.4,0.1 &-- &4097 &-0.70,0.01\\
Physical - all & 317 &4.6,1.1 &2.9,0.2 &11.1,0.4 &-- & 147&-0.78,0.01\\
Random - both with $z$ &153 & 16,3&13.7,0.3 &7.5,0.4 &0.84,0.06 &118 &-0.58,0.03\\
Physical - both with $z$ & 208 &5,1.5 &3.5,0.2 &11.8,0.5 &0.28,0.02 &77 &-0.77,0.02\\
\hline
\end{tabular}
\end{table*}

Table~\ref{tab:pairs1} lists our sample of 5428 pairs found in \citet{Taylor09} that meet the criteria defined in Sec.~\ref{sec:classify}, along with the pair component properties (full version of the table is available online). Table~\ref{tab:means} lists the number of pairs for each class (randoms, physicals, randoms with redshifts, etc), along with average values of the rms of the RM difference, the polarization fractions, redshifts, and spectral indices. 

The average values of $\Delta$RM are $-0.25\,$rad m$^{-2}$ and $0.9\,$rad m$^{-2}$ for random and physical pairs respectively, or approximately zero, as expected. The value of the rms, $\sqrt{ \langle (\Delta {\rm RM})^2_{\rm signal} \rangle}$, averaging over all separations and other parameters, is $4.6\pm1.1 \, \rm{rad} \, \rm{m}^{-2}$ for physical pairs and $14.9\pm0.4 \, \rm{rad} \, \rm{m}^{-2}$ for random pairs (with the uncertainties given as the standard errors on the mean). This yields a difference between physicals and randoms of $10.3 \pm 1.2 \,$rad m$^{-2}$. 

To determine the significance of the difference, we perform a bootstrap test where we construct a joint sample of $(\Delta \rm{RM})^2_{\rm signal}$ values of the random and physical pairs. We randomly draw, with replacement, the number in each category ($N_{\rm physical}$ and $N_{\rm random}$) and compute ${ \langle (\Delta \rm{RM})^2_{\rm signal} \rangle}$, and compute the difference between the two samples. We repeat this a few thousand times and compare the distribution of rms differences to the values obtained from the true (non-randomized) samples of physical and random pairs. The true mean difference of the rms between physical and random pairs is $10.3 \, \rm{rad} \, \rm{m}^{-2}$, which occurs at a $< 0.0001$ significance level (greater than $5\sigma$ difference). Figure~\ref{fig:boots1}a shows the distribution of bootstrap means compared with the true difference.

We also performed a Kolmogorov–Smirnov (KS) test and the two-sample Anderson-Darling (AD) test on the samples. The KS test returns a $p$-value of $\ll 0.0001$, similarly the AD test yields a $p$ value of $4\times10^{-5}$, in both cases greater than $5\sigma$ significance for the random and physical samples being drawn from different populations. Figure~\ref{fig:boots1}b shows the cumulative probability distributions for the different cases.

\subsection{Off-axis Leakage Polarization}
\label{sec:results_leak}

\begin{figure}
\includegraphics[scale=0.37]{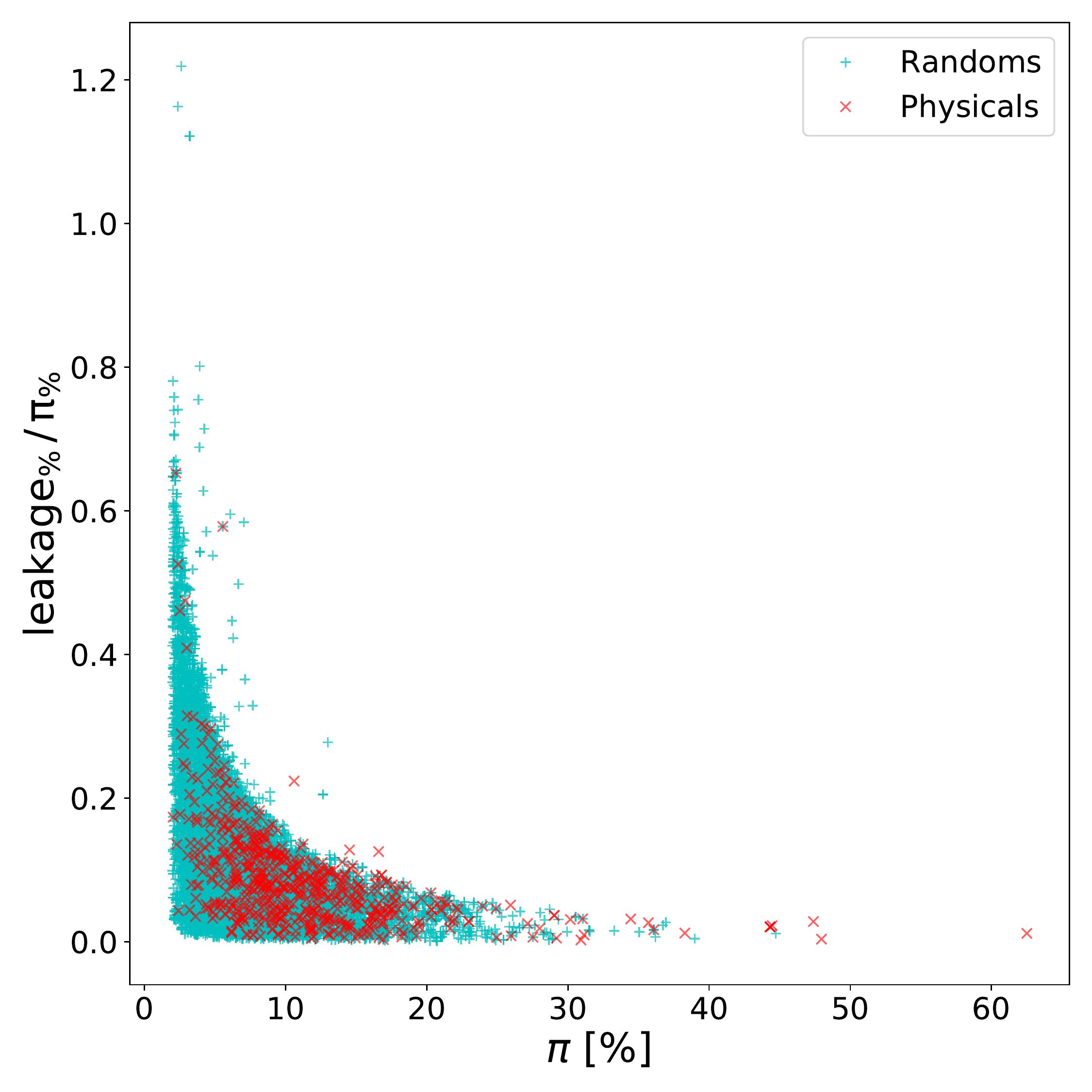}
\caption{Fractional amount of leakage per polarization fraction as a function of observed polarization fraction. The light blue plus signs are for the random pair sources and the red crosses are the physical pair components.\\}
\label{fig:leaks}
\end{figure}

In Sec~\ref{sec:data} we mentioned that the \citet{Taylor09} catalog does not include off-axis leakage corrections. We calculated the leakage values for each pair component based on their positions relative to the nearest NVSS pointing centres as per \citet{Ma18}. Figure~\ref{fig:leaks} shows the ratio of leakage fraction to observed polarization fraction as a function of the observed polarized fraction. This shows that the leakage is a dominant factor in the polarization fraction predominantly for those sources with $\pi \la 5\, \%$ when the source is significantly off axis. It is possible that this large amount of uncorrected leakage could lead to spurious or inaccurate RM and or $\Delta$RM values for those source or pairs.  

To test the significance of the leakage for our findings, we can use the leakage fraction as a criterion and select only pairs in which the leakage fraction to polarization fraction ratio is less than some amount. The ratio of leakage fraction to polarization fraction tells us how much of the measured polarized fraction signal is due to leakage rather than from true signal. If we require the minimum leakage fraction ratio of the pair to be less than $0.2$ (i.e. at least one component in the pair has a ratio less than this value) then $\sqrt{\langle (\Delta {\rm RM})^2_{\rm signal} \rangle}=14.5\,$rad m$^{-2}$ for randoms pairs and $4.4\,$rad m$^{-2}$ for physical pairs. For pairs with the minimum leakage fraction ratio of the pair less than $0.1$, we find $\sqrt{\langle (\Delta {\rm RM})^2_{\rm signal} \rangle}=13.9\,$rad m$^{-2}$ and $5.5\,$rad m$^{-2}$ for randoms and physicals respectively. If we instead require both components of the pair to have leakage fraction ratios less than $0.2$ (or the maximum of the pair $<0.2$) then $\sqrt{\langle (\Delta {\rm RM})^2_{\rm signal} \rangle}=13.7\,$rad m$^{-2}$ and $4.4\,$rad m$^{-2}$ for randoms and physical pairs. 

The changes from applying cuts based on the leakage fractions do not significantly alter the results of the current analysis. Therefore for the subsequent analysis we choose not to cut any additional pairs from the sample. Having established a difference in $\sqrt{(\Delta {\rm RM})^2_{\rm signal}}$, we now look at a number of variables that may be contributing to that difference.

\subsection{Angular Separation}
\label{sec:results_sep}

\begin{figure}
\includegraphics[scale=0.37]{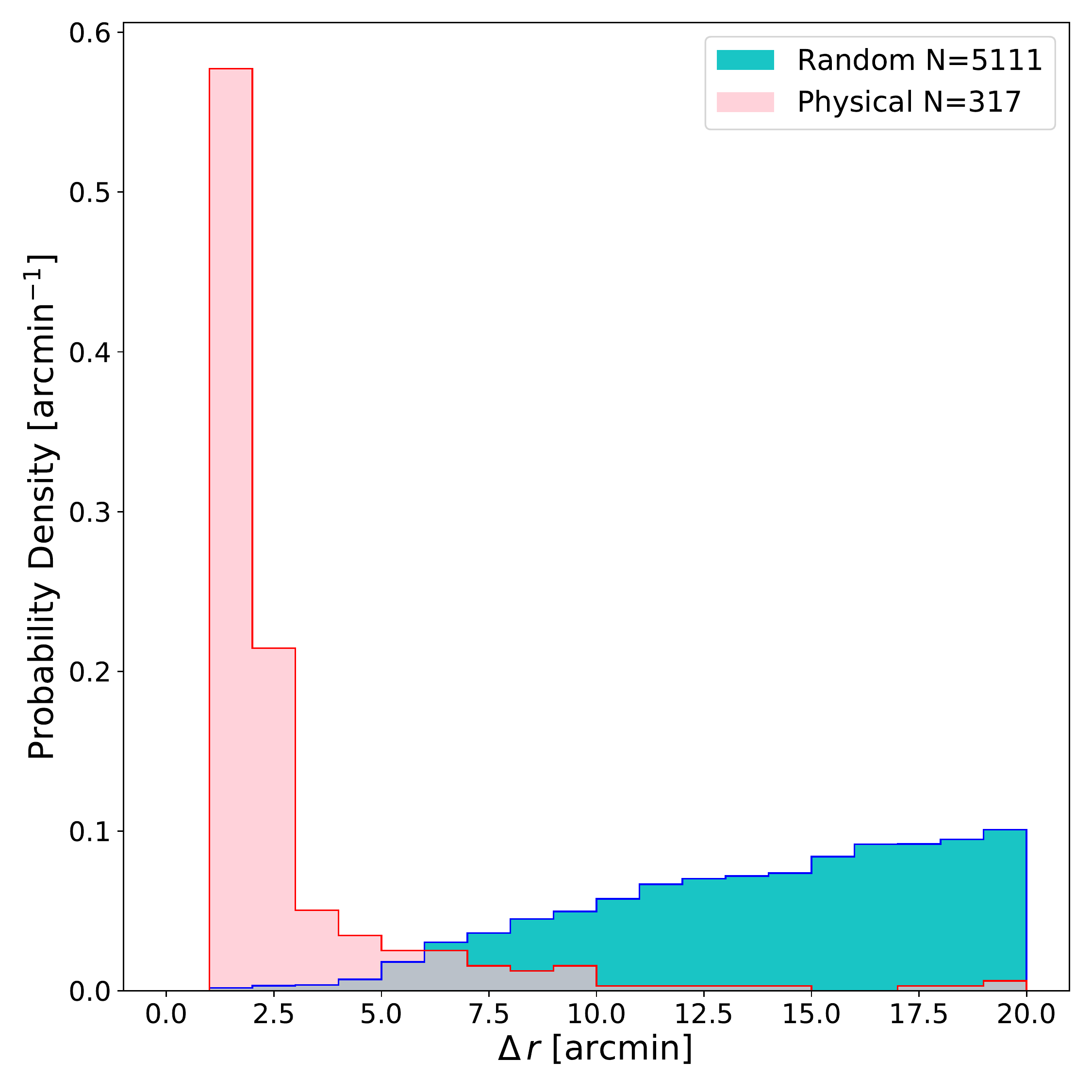}

\caption{Pair separation probability density functions for all physical (red) and random (blue) pairs.}
\label{fig:sephists}
\end{figure}

\begin{figure}
\includegraphics[scale=0.38]{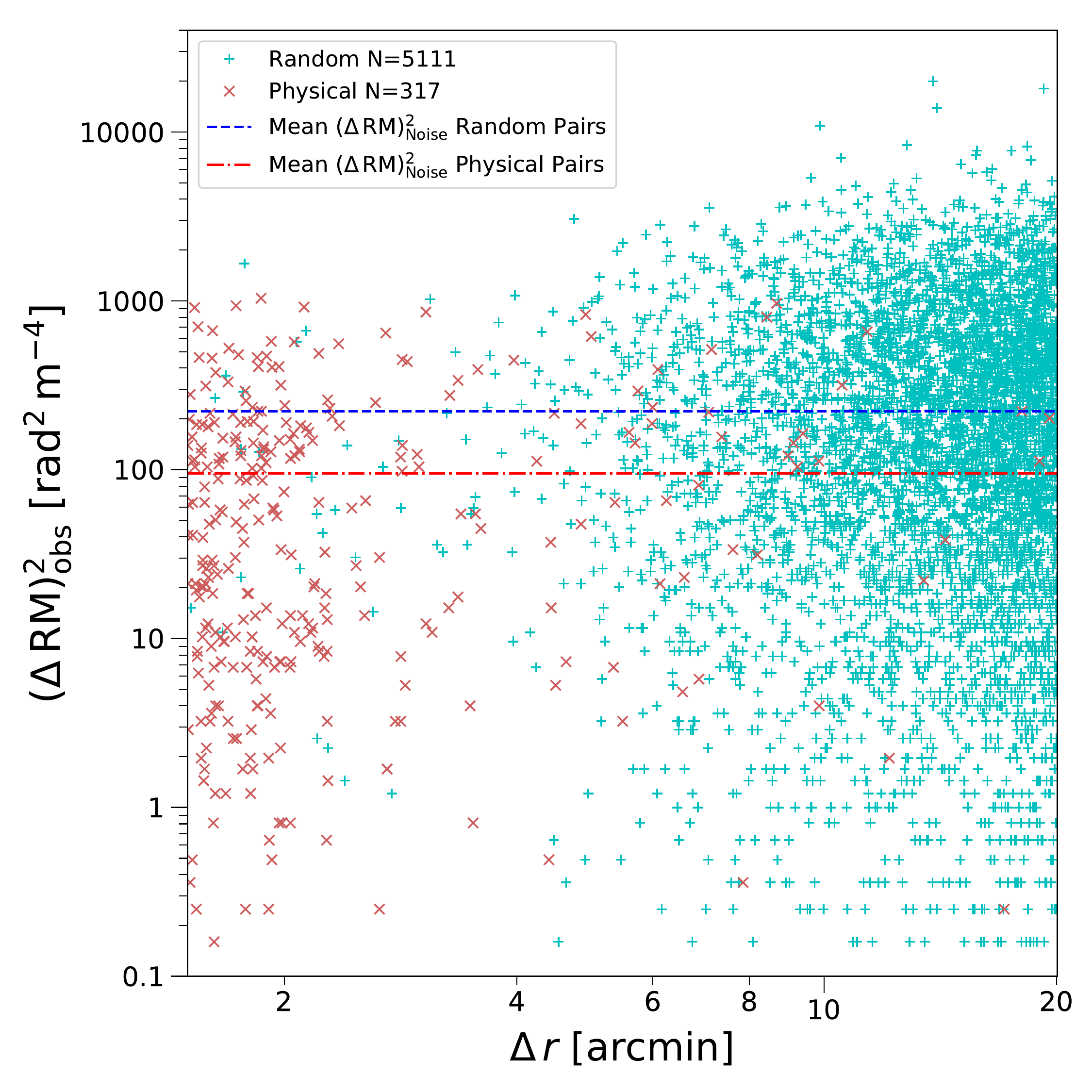}
\includegraphics[scale=0.38]{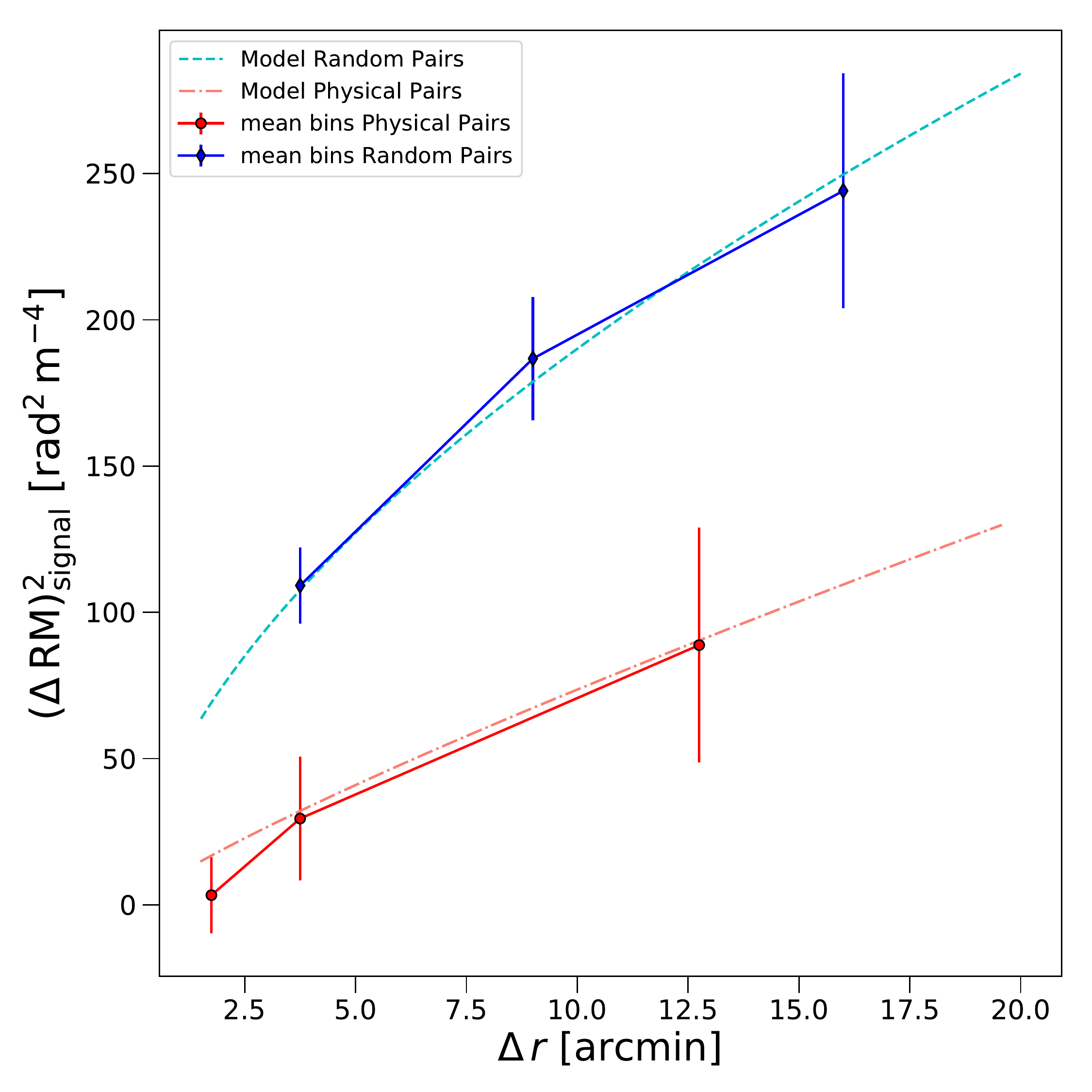}
\caption{$(\Delta \rm{RM})^2$ vs separation for physical and random pairs. The top panel shows $(\Delta \rm{RM})^2_{\rm obs}$ vs $\Delta r$ for the physical (pink crosses) and random (light blue plus signs) pairs {\bf before} correcting for any measurement noise. The dashed lines show the average value of the noise variance as a function of separation (red for physical pairs and blue for random pairs). The bottom panel shows binned average $(\Delta \rm{RM})^2_{\rm signal}$, or structure functions (blue for random pairs and red for physical pairs), with $1\sigma$ uncertainties. The light-blue dashed line is the power-law fit to the random pairs and the pink dot-dashed line is the fit to the physical pairs. This is after the correction, or subtraction, of the mean noise variance for each bin.\\ }
\label{fig:strfncs1}
\end{figure}

\begin{figure*}
\includegraphics[scale=0.26]{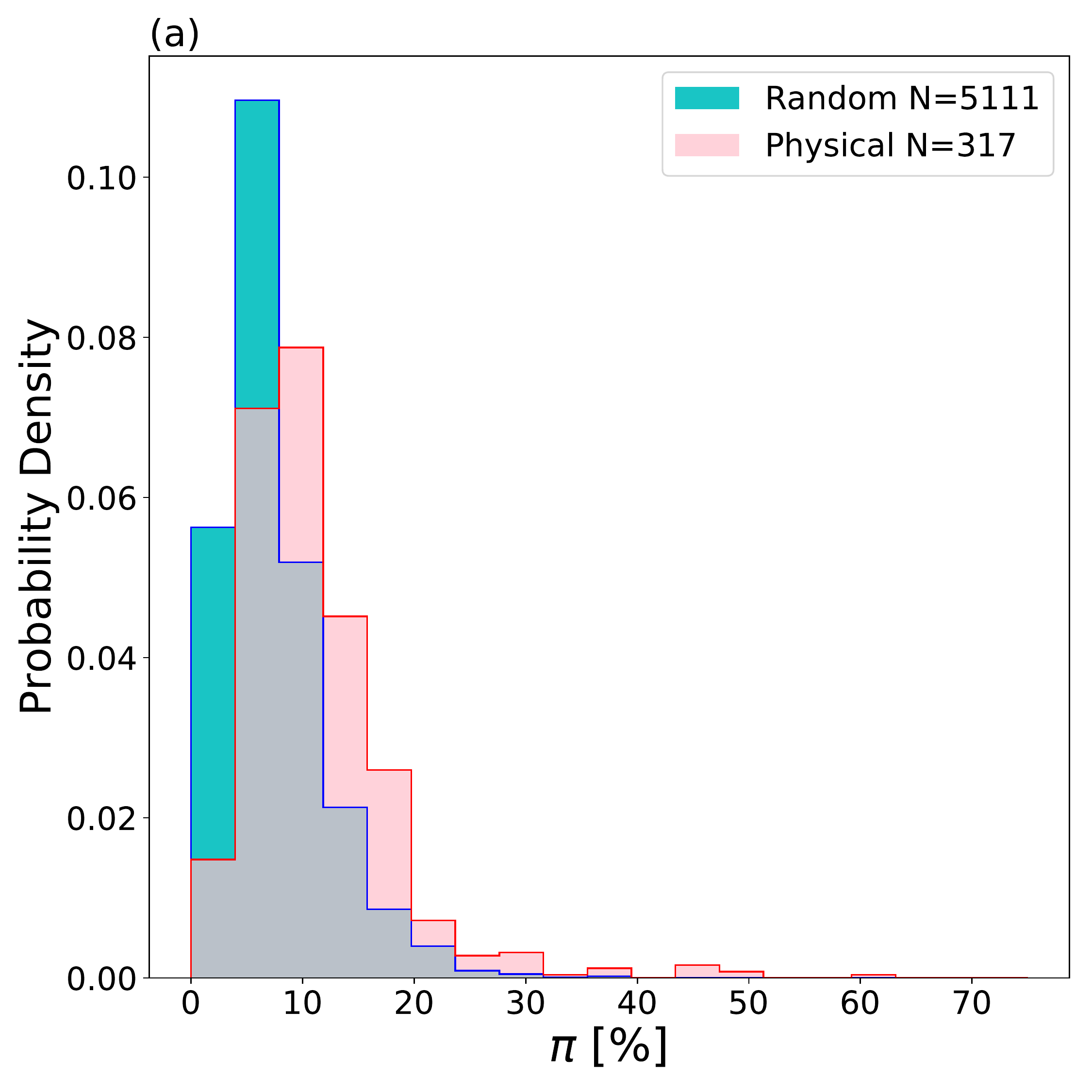}\includegraphics[scale=0.26]{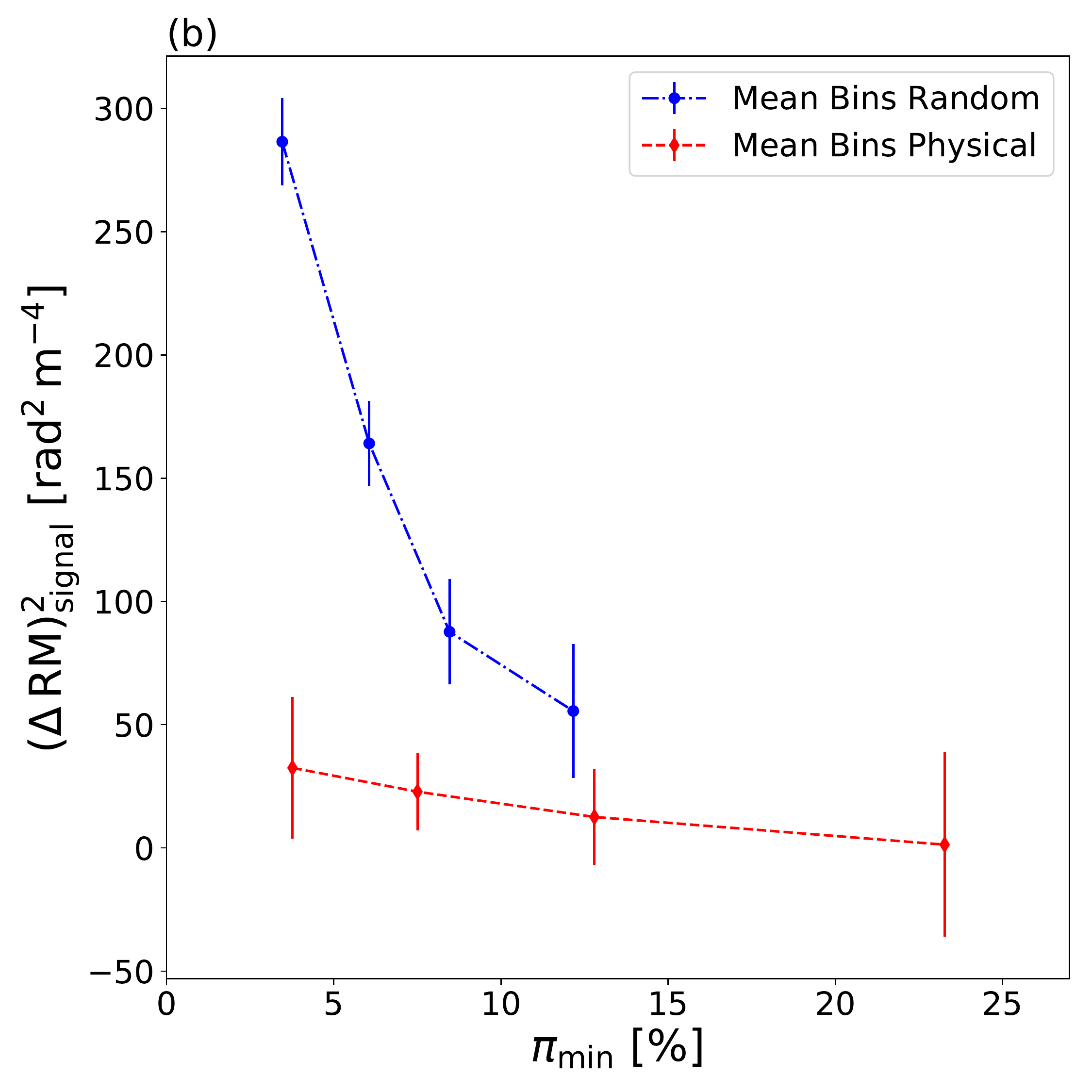}\includegraphics[scale=0.26]{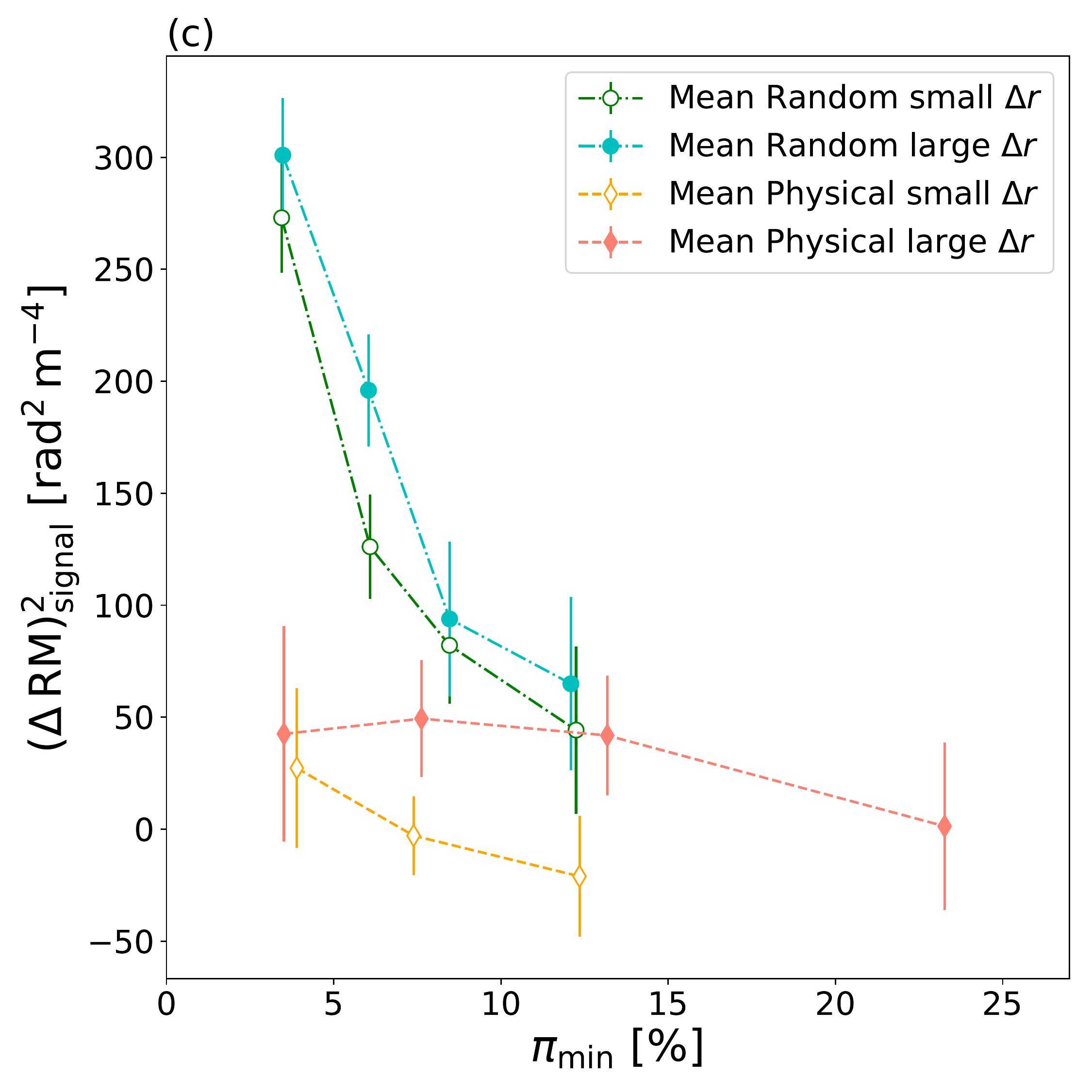}
\includegraphics[scale=0.26]{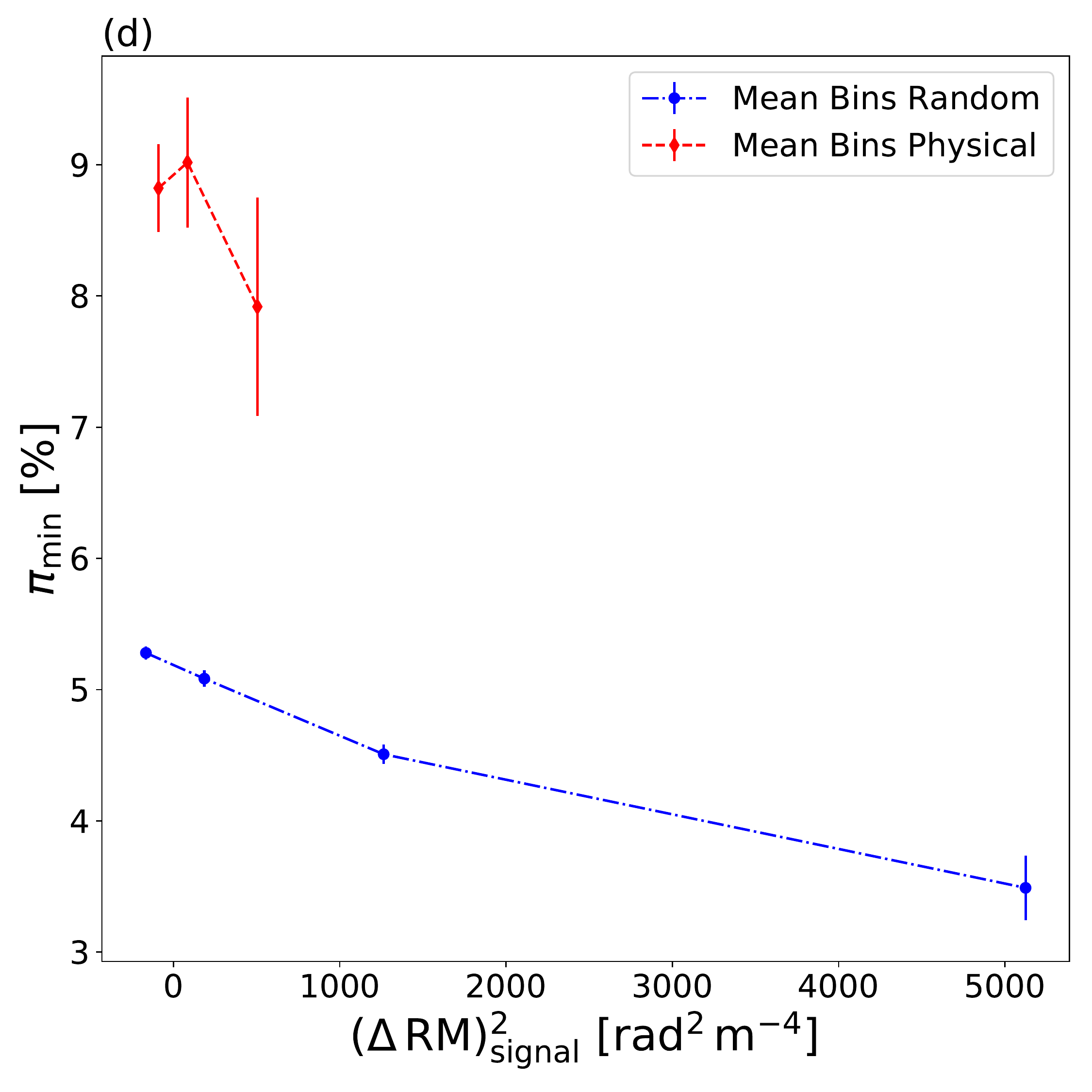}\includegraphics[scale=0.26]{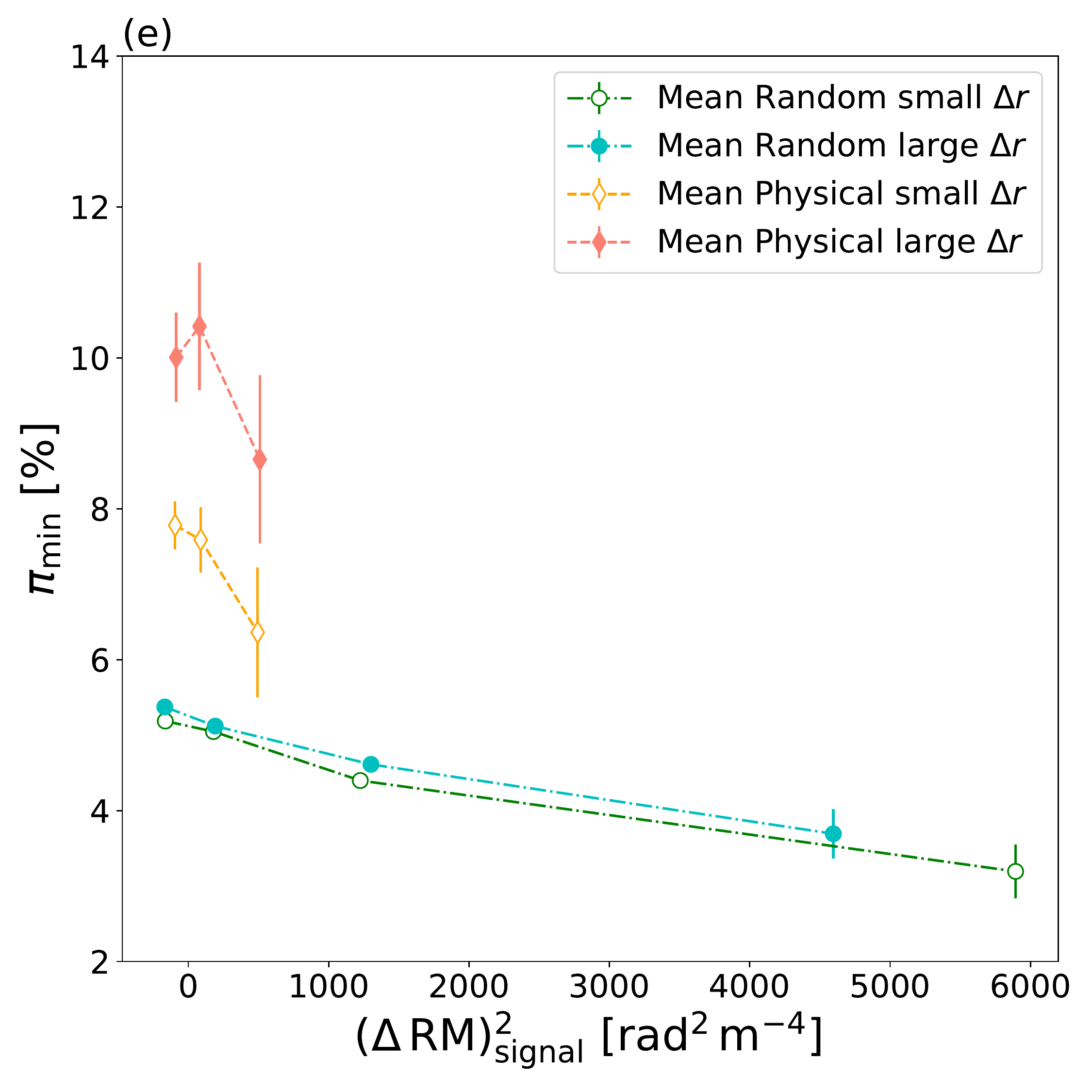}
\caption{Pair polarization fraction properties. Panel (a) shows the distribution of polarization fractions for random pair sources (blue) and physical pair sources (red). Panels (b) through (e) compare the minimum polarization fraction of the pair, $\pi_{\rm min}$ with $\langle(\Delta {\rm RM})^2_{\rm signal}\rangle$. Panels (b) and (c) show the mean $\langle(\Delta {\rm RM})^2_{\rm signal}\rangle$ vs $\pi_{\rm min}$, while panels (d) and (e) show the mean $\pi_{\rm min}$ as a function of $\langle(\Delta {\rm RM})^2_{\rm signal}\rangle$. In panels (c) and (e) each classification is divided into two groups based on the median separation for each class, with open-faced markers showing those with $\Delta r$ less than the medians (orange for physical pairs and green for random pairs) and the color-faced markers showing the averages for pairs with $\Delta r$ larger than the medians (pink for physical pairs and light blue for random pairs). All values of $(\Delta {\rm RM})^2_{\rm signal}$ have been corrected for the mean measurement noise variance and all error bars show the $1\sigma$ uncertainties. \\ }
\label{fig:pihists}
\end{figure*}

\begin{figure*}
\includegraphics[scale=0.26]{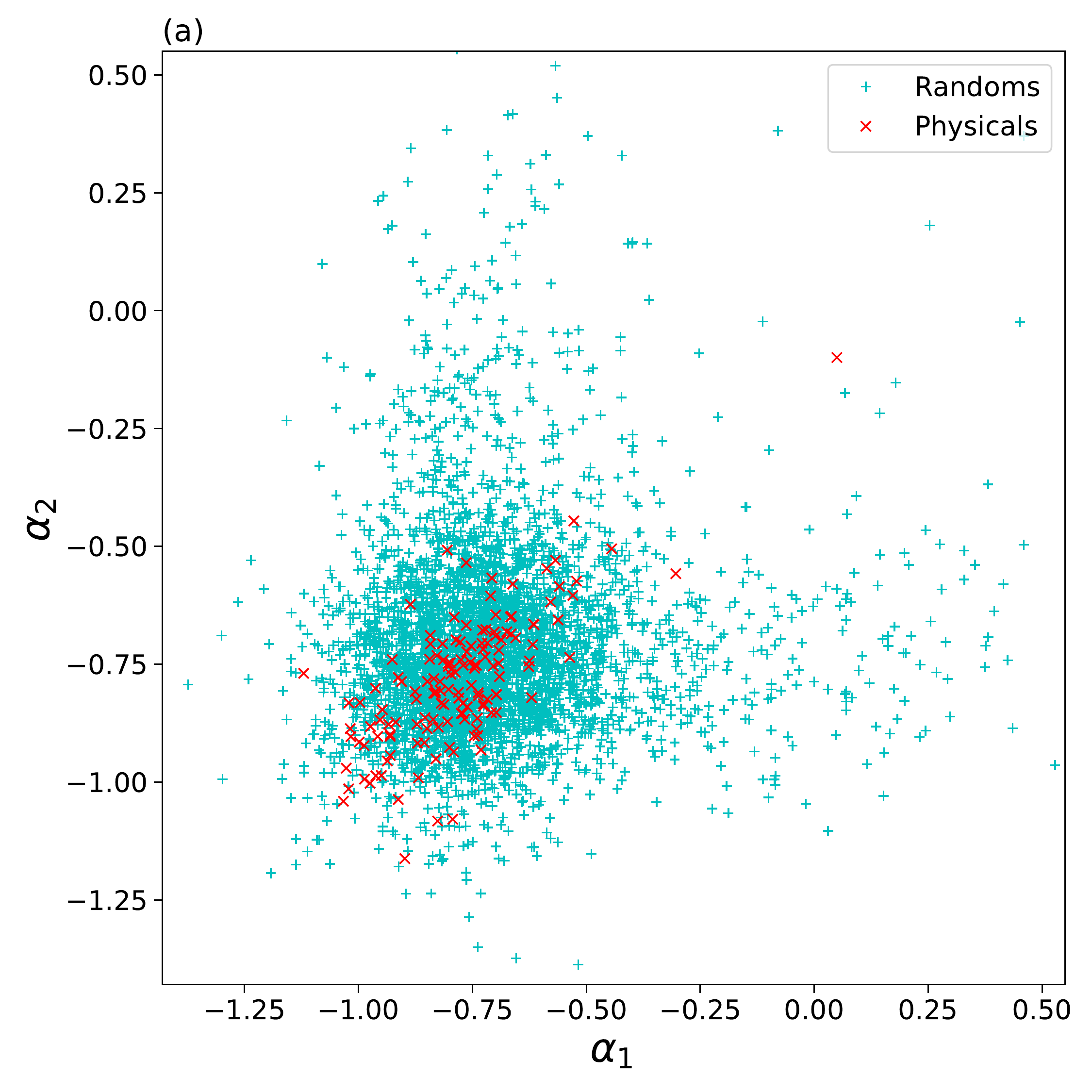}\includegraphics[scale=0.26]{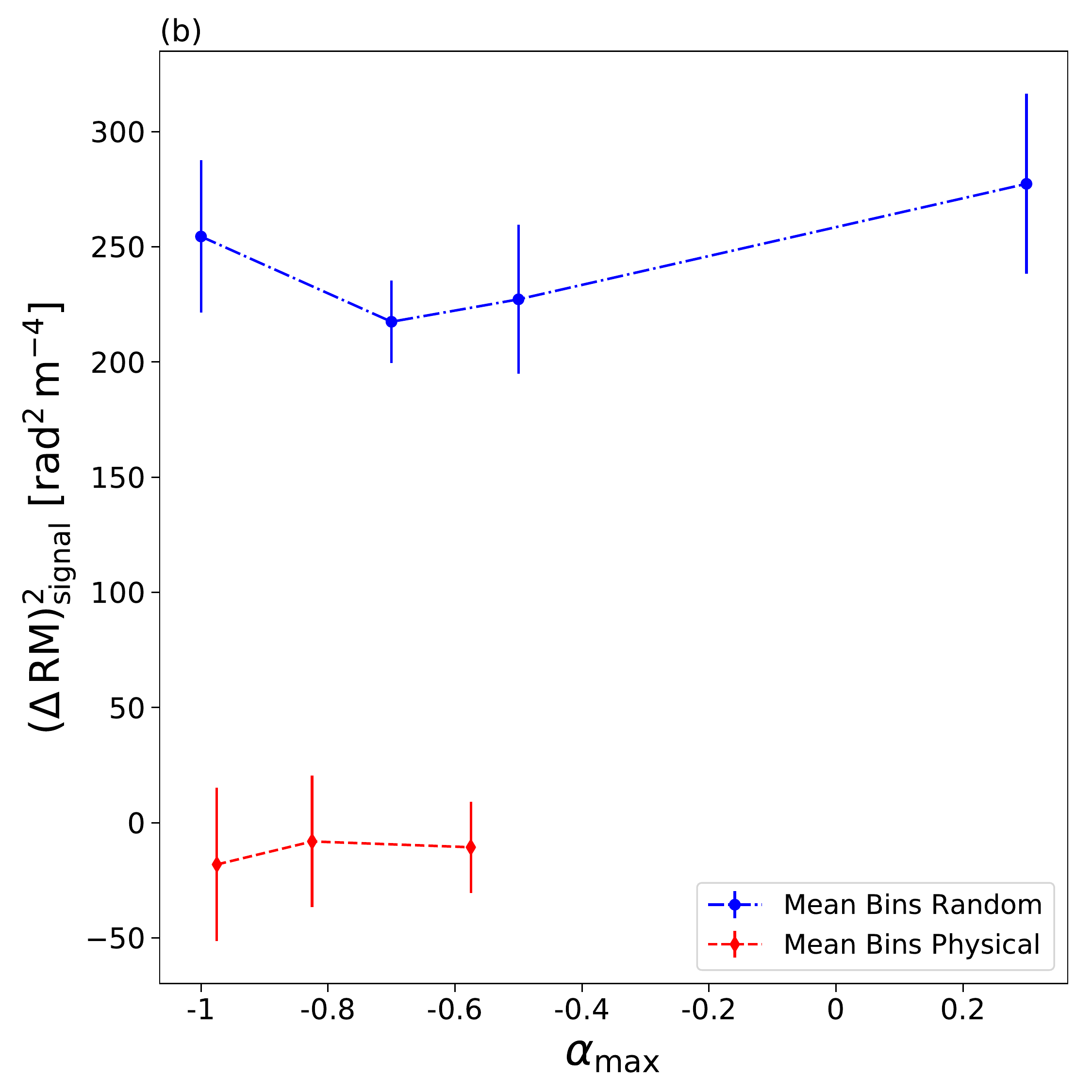}\includegraphics[scale=0.26]{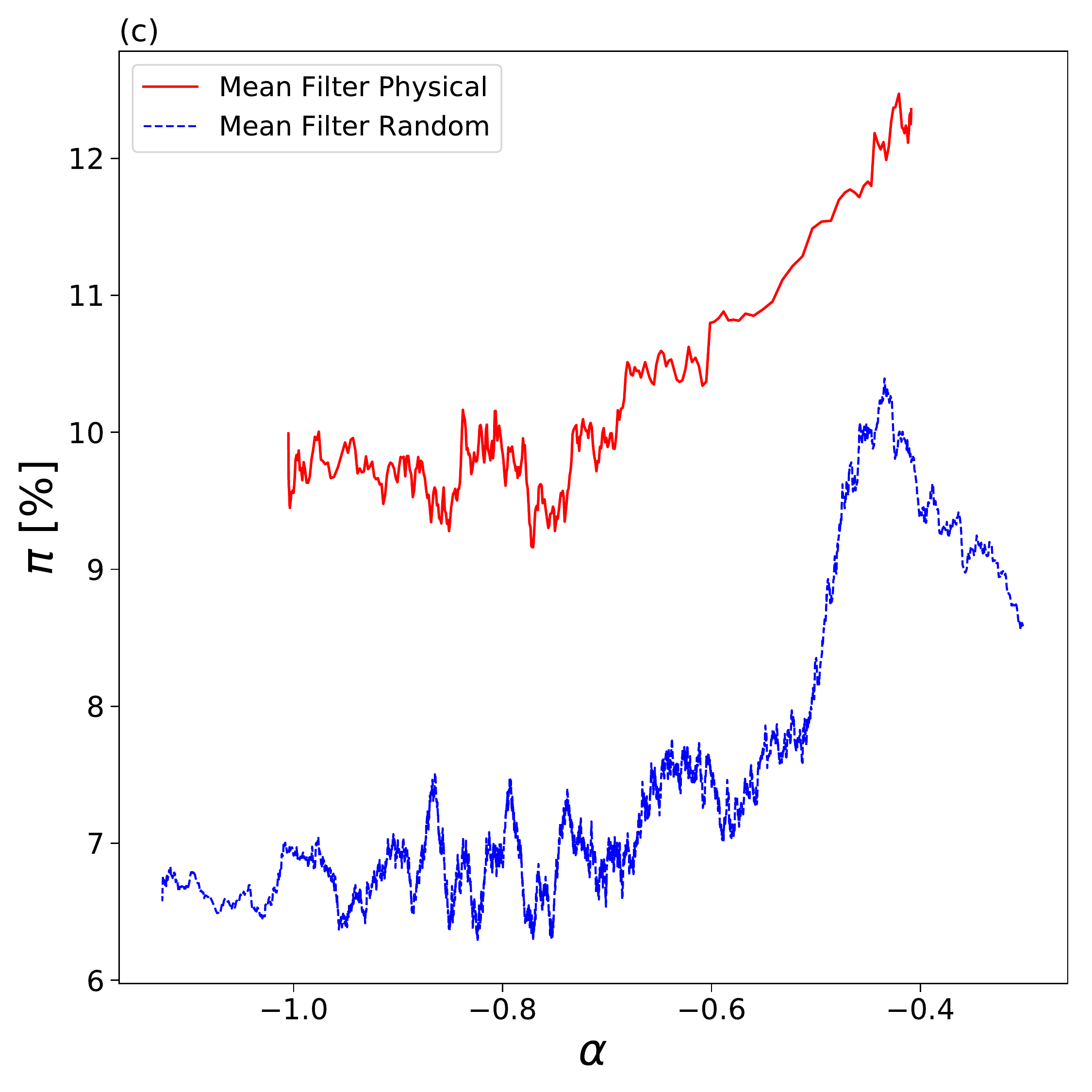}
\caption{Panel (a) shows the spectral indices of one pair component vs another. The light blue plus signs are for the random pairs and the red crosses are for the physical pairs. Panel (b) shows $ \langle (\Delta {\rm RM})^2_{\rm signal} \rangle $ as a function of the average maximum spectral index of the pair for random pairs (blue dot-dashed line) and physical pairs (red dashed line). In both panel(a) and (b) pairs are only included if both components of the pair have a matched spectral index from cross matching with the NVSS TGSS spectral index catalog \citep{deGasperin18}. The $(\Delta {\rm RM})^2_{\rm signal}$ values have been corrected for the mean squared measurement noise variance. In Panel (c) the running mean of source polarization fraction is shown as a function of the mean source spectral index for any source with a spectral index (for the individual sources, not pairs of sources).  \\}
\label{fig:alphaplot}
\end{figure*}

Figure~\ref{fig:sephists} shows the distribution of separations for the two types of pairs. As expected, there are clearly more pairs with small angular separations for physical pairs, with a median angular separation of $1.9{\arcmin}$ for physicals and $14.5{\arcmin}$ for randoms.  Figure~\ref{fig:strfncs1} shows the structure functions (or $(\Delta {\rm RM})^2_{\rm signal}$ vs $\Delta r$) for the two classes, along with all of the data points (top panel). We fit a power-law model to the data in the form of
\begin{equation}
 \Delta {\rm RM(\Delta r)}^{2}_{\rm signal} \, = \, k \, \left ( \frac{\Delta r}{1 \, {\rm arcmin}} \right )^{\gamma},
\label{eq:plaw}
\end{equation}
where $k$ is a normalisation factor with units of rad$^2$ m$^{-4}$. The power-law slopes are found to be $\gamma= 0.6\pm0.1$ for random pairs and $\gamma = 0.8\pm0.2$ for physical pairs, with the amplitudes being $k = 50\pm30$ for the random pairs and $ k = 11\pm15$ for the physical pairs.

\begin{figure*}
\includegraphics[scale=0.36]{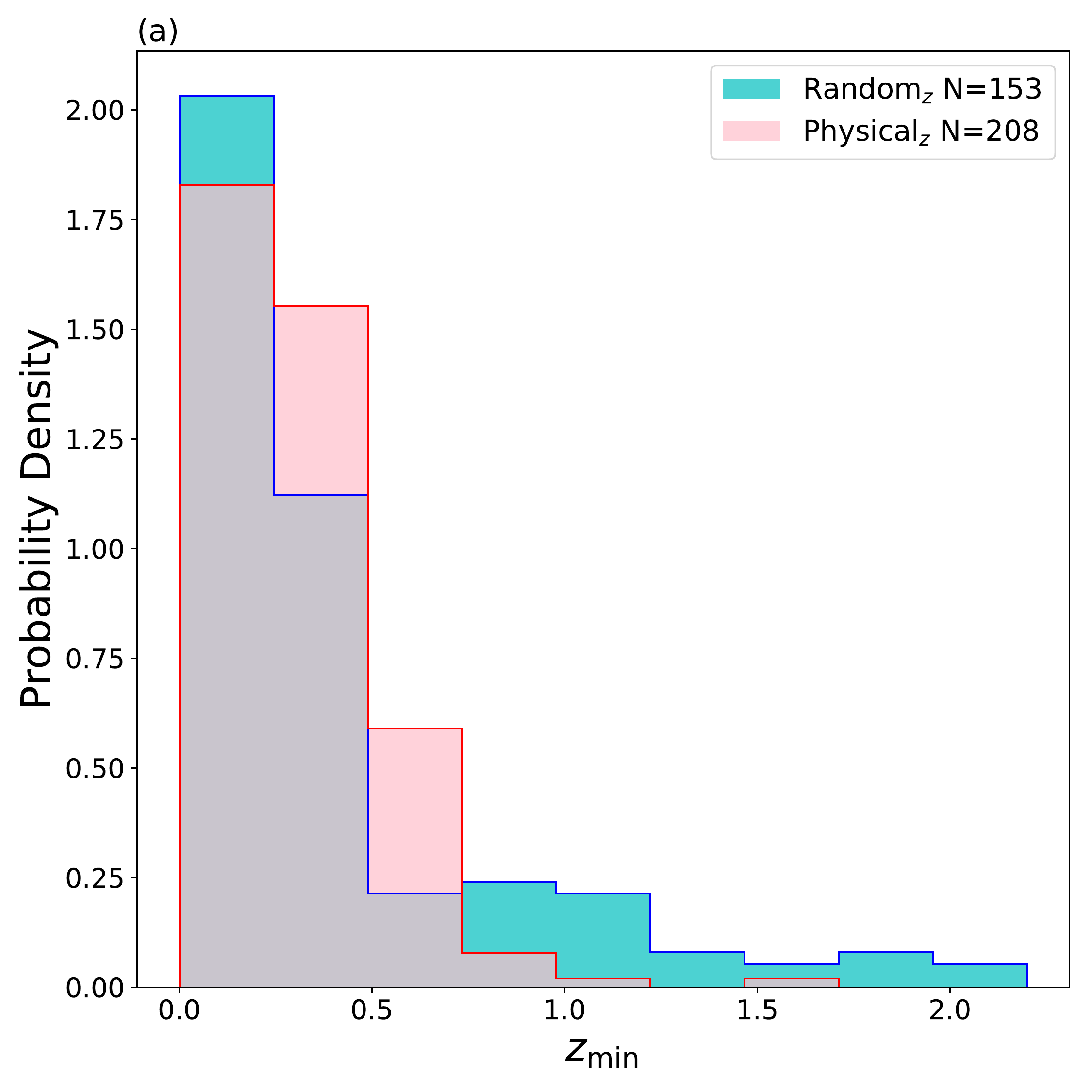}\includegraphics[scale=0.36]{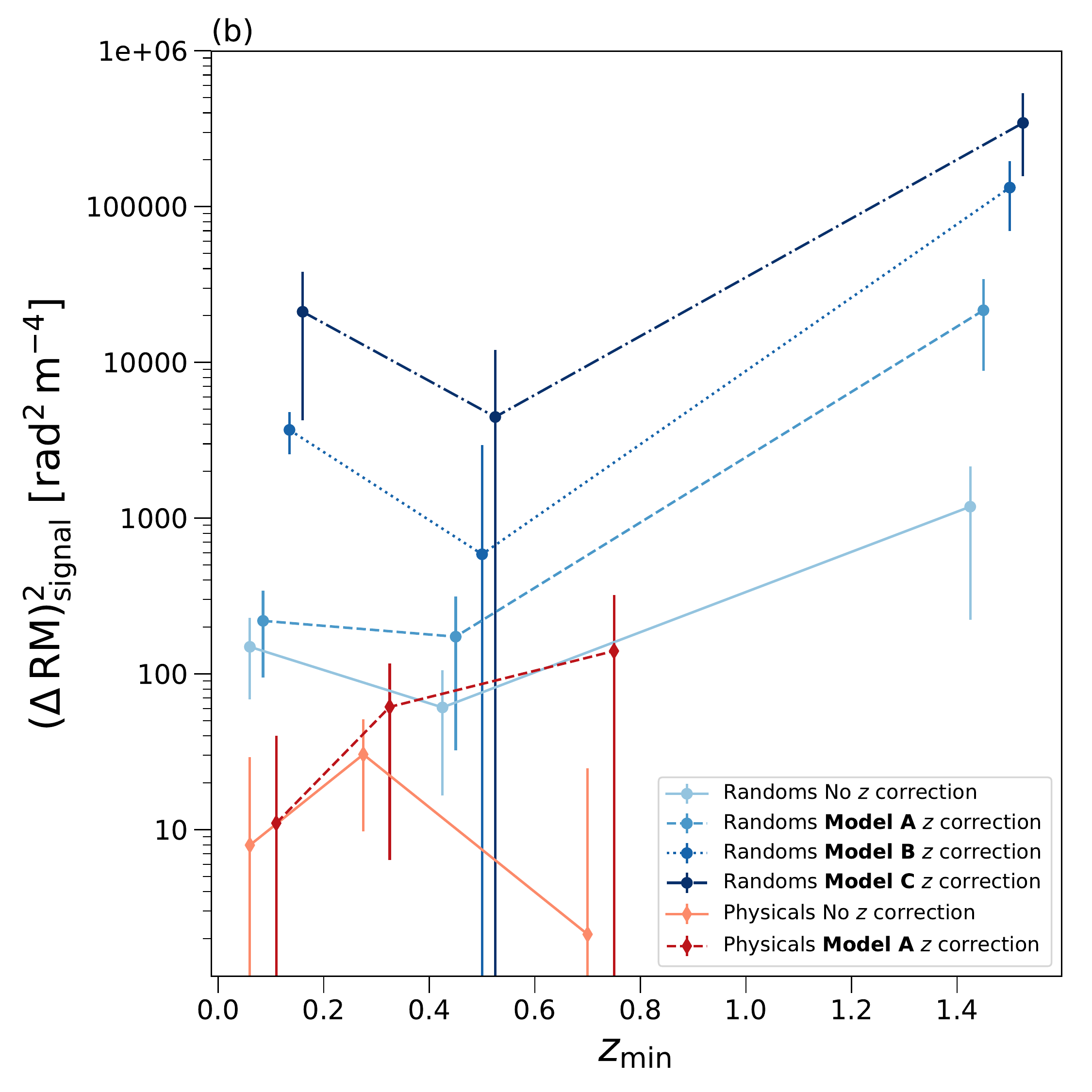}
\includegraphics[scale=0.36]{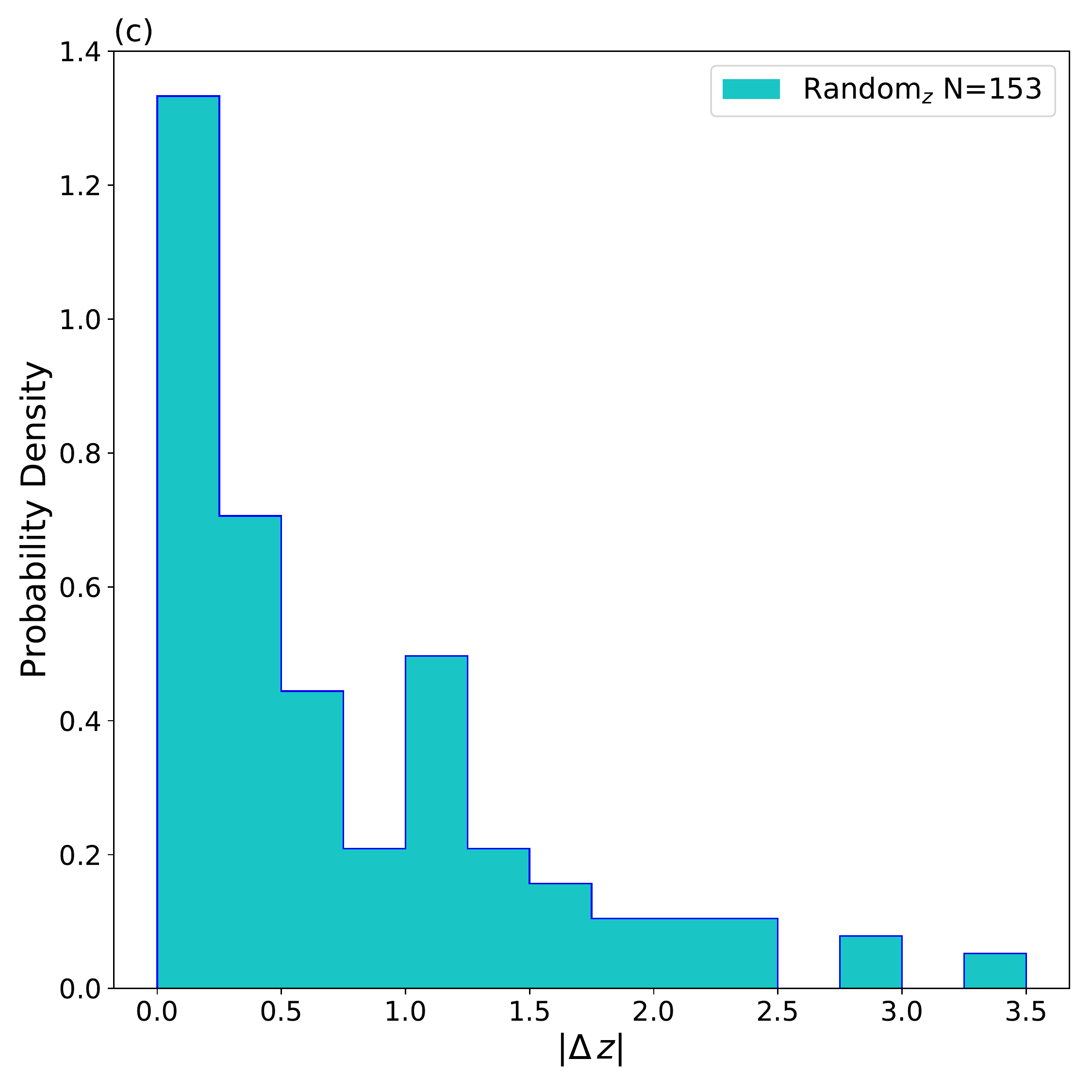}\includegraphics[scale=0.36]{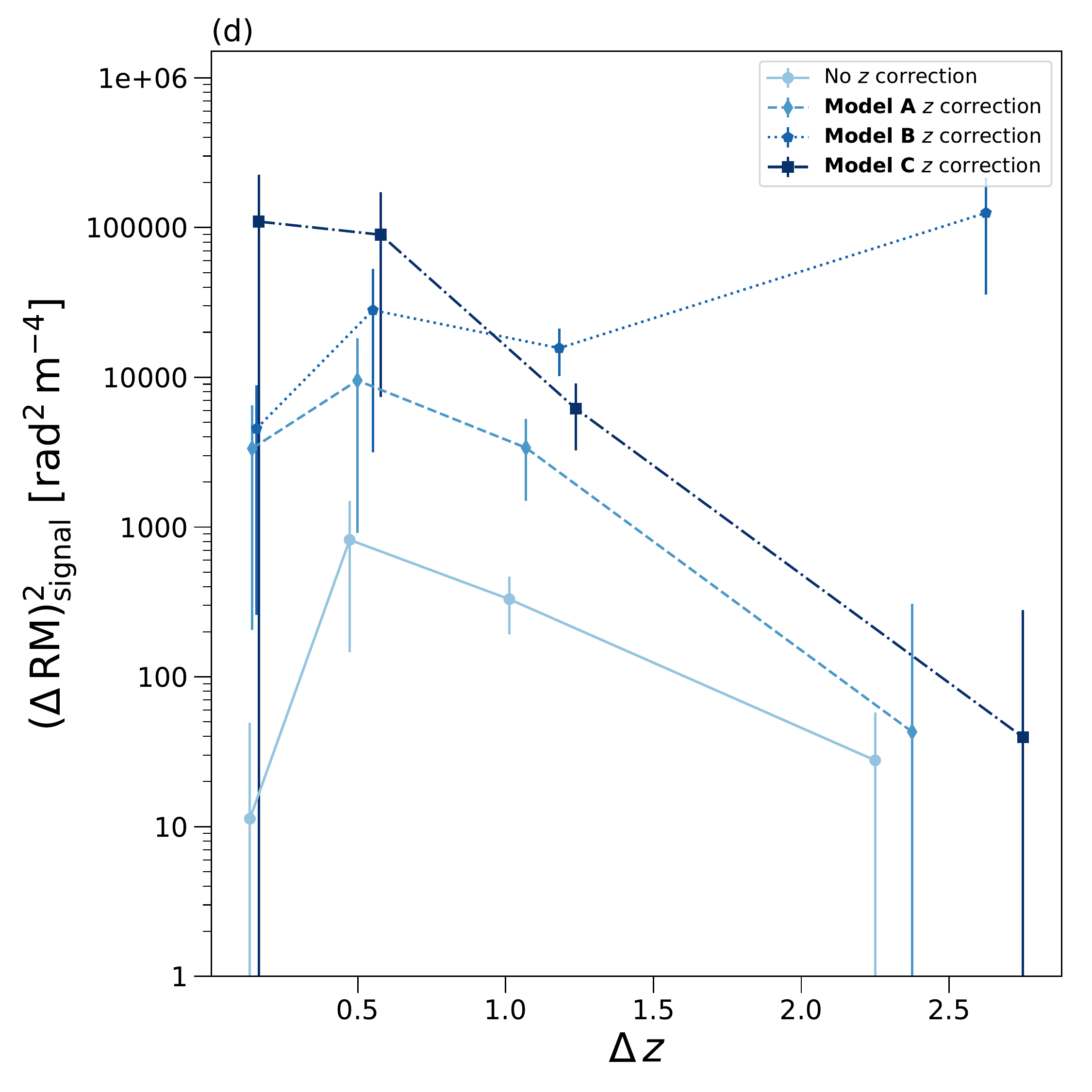}
\caption{Redshift and $\Delta z$ distributions. Panel (a) shows the $z$ (or $z_{\rm min}$ for random pairs) distributions for physical (red) and random (blue) pairs, while panel (c) shows the distribution of $\Delta z$ for random pairs. Panels (b) and (d) show $\langle (\Delta \rm{RM})^2_{\rm signal}\rangle$ vs $z_{\rm min}$ and vs $\Delta z$, respectively, with different possible redshift corrections (models defined in Sec.~\ref{sec:redz}). For panels (b) and (d) the $x$-axis positions have been slightly offset between the different $z$-corrections to avoid overlapping error bars. \\}
\label{fig:zplots2}
\end{figure*}

If we repeat the bootstrap test restricting the pairs to a range of separations that maximizes overlap in $\Delta r$ between the two types, $3{\arcmin}\leq \Delta r \leq 11{\arcmin}$, then the difference in rms is $5.2 \, \rm{rad} \, \rm{m}^{-2}$, which is a $\sim 0.01$ significance level between the rms values of physical and random pairs. This is also shown in Fig.~\ref{fig:boots1}a. For the KS and AD tests, when restricting to pairs in a region of maximized overlap, separations $3{\arcmin}\leq \Delta r \, \leq 11{\arcmin}$, the difference is less significant with $p$-values  around $0.01$. When restricting the separations to a region of more overalap, we do still detect a difference at the 2-$3\sigma$ level. The cumulative probability distributions for restricted $\Delta r$ range can also be seen in Fig~\ref{fig:boots1}b. Fitting the power-law to just this region of greater overlap changes the power-law slopes to $0.8\pm0.2$ and $0.75\pm0.3$, for random and physical pairs respectively. 

To verify that the difference seen is not systematic we randomly shuffled the RMs of the pairs within each group (randoms and physicals) and recomputed $\sqrt{\langle (\Delta \rm{RM})^2_{\rm signal} \rangle}$ 1000 times and looked at the rms as a function of separation. In this case the average structure functions from the 1000 trials were consistent with a flat line (no change with separation) with an rms value of $\sim 43 \, \rm{rad} \, \rm{m}^{-2}$, for both physical and random pairs. We also performed 1000 trials keeping the $(\Delta \rm{RM})^2_{\rm signal}$ of the pairs the same but randomly shuffling the separations (again separately shuffling the random and physical pairs). In this case the average structure functions again show flat lines with rms values of $\sim 5  \, \rm{rad} \, \rm{m}^{-2}$ for physical pairs and $15 \, \rm{rad} \, \rm{m}^{-2}$ for random pairs. This implies that there really is a change in $(\Delta \rm{RM})^2_{\rm signal}$ due to separation and that this change is related to physical aspects of the actual pairs and their RM values.

\subsection{Polarization Fraction}
\label{sec:results_pf}

The RMs may also depend on the source polarization fractions. Figure~\ref{fig:pihists} shows the distributions of the polarized fractions for the sources in the pairs. The sources in physical pairs tend to have higher polarization fractions than those in random pairs, with the physically paired sources having a mean polarized fraction of $11.1\%$ compared to a mean of $7.4\%$ for the randomly paired sources. Looking at the difference in polarization fraction, $\Delta \pi$, for the two types of pairs we get means of $4.5\%$ for random and $4.6\%$ for physical ones. This indicates that while the physical ones may have higher $\pi$s, it does not seem to be the case that one component in the source has a preferentially higher $\pi$ than the other. 

 Figure~\ref{fig:pihists} also shows $\langle(\Delta {\rm RM})^2_{\rm signal}\rangle$ as a function of the minimum polarized fraction of the pair, as well as the minimum polarization fraction as a function of $\langle(\Delta {\rm RM})^2_{\rm signal}\rangle$. The average minimum polization fractions of the pairs are $5.1\%$ and $8.8\%$ for the random and physical pairs, respectively. We can see from these plots a clear difference in the averages of $ \langle (\Delta {\rm RM})^2_{\rm signal} \rangle $ and average $\pi_{\rm min}$s between random and physical pairs. We can also see (panel b) that the values for the physical pairs remain roughly constant as a function of the minimum polarization fraction, whereas for random pairs the average values decrease with increasing minimum polarization fraction. 

 In order to try and account for possible effects from pairs at different separations, $\langle (\Delta {\rm RM})^2_{\rm signal}\rangle$ vs $\pi_{\rm min}$ and $\pi_{\rm min}$ vs $\langle (\Delta {\rm RM})^2_{\rm signal}\rangle$ are also shown in Fig.~\ref{fig:pihists}, with the averages computed separately for those pairs with small and large separations in Fig~\ref{fig:pihists}c,e (i.e. smaller or larger than the median $\Delta r$ for random and physical pairs separately). Regardless of whether the pairs have small or large separations, the difference between physical and random pairs remains and the trends of decreasing $\Delta$RM with increasing $\pi_{\rm min}$ for randoms remain visible. In Fig~\ref{fig:pihists}e, the plot of $\pi_{\rm min}$ as a function of $(\Delta {\rm RM})^2_{\rm signal}$, does show a separation for the physical pairs between those with large vs small $\Delta r$, with the larger separation pair having larger polarization fractions, regardless of $\Delta$RM. 

The relation between $\langle (\Delta {\rm RM})^2_{\rm signal} \rangle$ and $\pi_{\min}$ seems at least as strong, if not stronger, than the relation between $\langle (\Delta {\rm RM})^2_{\rm signal} \rangle$ and $\Delta r$. 

\subsection{Spectral Indices}
\label{sec:results_spec}

Table~\ref{tab:means} lists $N_{\alpha}$, the number of pairs, broken down by physical and random, where both pair components have a spectral index from cross matching with the NVSS-TGSS spectral index catalog by \citet{deGasperin18}. Figure~\ref{fig:alphaplot}a shows the spectral indices of the pairs against one another for both randoms and physicals (for pairs where both components of the pair had a matching source in the spectral index catalog). It is clear that the spectral indices from physical pairs are more tightly correlated, with a correlation coefficient of $\rho=0.73$ for the physical pairs and $\rho=0.08$ for the random pairs. The physical pairs also have steeper spectra on average, with a mean of $-0.78$ compared with a mean of $-0.7$ for the random pair sources. If we restrict our sample to only those pairs where both components have a spectral index then the RM rms values become $15.25\,$rad m$^{-2}$ for random pairs and $0\,$rad m$^{-2}$ for physical pairs (for the physical pairs with spectral indices $\langle (\Delta {\rm RM})^2 \rangle = -8.5$, due to the subtraction of the average noise variance, thus when considering the rms we assume a value of zero). 

From Fig.~\ref{fig:alphaplot}a we can see that the random pairs appear to have a larger number of flat-spectrum sources than the physical pairs, with about $16\,\%$ of the random pairs having at least one source with $\alpha \geq -0.45$, while only about $3\,\%$ of the physical pairs have at least one source with $\alpha\geq -0.45$. If we restrict the sample of random pairs to more closely match the spectral index distribution of the physical pairs, i.e. $\alpha_{\rm max}\la -0.45$, the rms values are $14.9\,$rad m$^{-2}$ for the random pairs and $0\,$rad m$^{-2}$ (or $\langle (\Delta {\rm RM})^2_{\rm signal} \rangle = -18.5$) for the physical pairs.

Panel (b) of Fig.~\ref{fig:alphaplot} shows $\langle(\Delta {\rm RM})^2_{\rm signal}\rangle$ as a function of the average maximum spectral index of the pair. This again shows the difference between the random and physical pairs, with the random pairs having the higher $(\Delta {\rm RM})^2_{\rm signal}$ values, regardless of spectral index. The random pairs do show a trend towards increasing $\langle(\Delta {\rm RM})^2_{\rm signal}\rangle$ with increasing maximum spectral index when $\alpha_{\rm max} \ga -0.2$. For the random pairs, those with the steepest spectra ($\alpha_{\rm max} \la -0.9$) also show a slightly higher variance than those with flatter spectrum sources. 

Panel (c) of Fig.~\ref{fig:alphaplot} shows the running average polarization fraction vs spectral index of a source. For both physical and random pairs $\pi(\alpha)$ remains roughly flat for $\alpha \la -0.7$. For $-0.7 \la \alpha \la -0.5$, $\pi$ increases with increasing $\alpha$ for both physical and random pair sources. For flatter spectrum sources, $\alpha \ga -0.5$ the polarization fraction decreases rapidly with increasing $\alpha$. 

\subsection{Redshifts}
\label{sec:results_red}

\begin{table*}
\caption{Rotation Measure root mean square values for random and physical pairs with redshifts for different redshift corrections. These values have all been corrected for $(\Delta {\rm RM})^2_{\rm noise}$.\\}
\label{tab:meanszz}
\begin{tabular}{ccccc}
\hline
 & No Correction & Model A & Model B & Model C\\
 & $\sqrt{ \langle (\Delta {\rm RM})^2 \rangle }$ & $\sqrt{ \langle (\Delta {\rm RM} (1+z_{\rm min})^2)^2 \rangle }$ &$\sqrt{ \langle [((1+z_1)^2 \, {\rm RM}_1)-((1+z_2)^2 \, \rm{RM}_2)]^2 \rangle }$  &$\sqrt{ \langle [\Delta {\rm RM} / \int_{z_{\rm{min}}}^{z_{\rm{max}}}{(1+z)^{-2}}]^2 \rangle }$  \\
  & [rad m$^{-2}$]  & [rad m$^{-2}$]  & [rad m$^{-2}$] & [rad m$^{-2}$] \\
 
 \hline
 Randoms$_z$&16.5 &55.0 &193.0  & 252.6\\
 Physicals$_z$&5.4  & 4.3 & -- &--\\
\end{tabular}
\end{table*}

If we restrict the pairs to those where both components have redshifts, then in addition to just looking at $\langle (\Delta {\rm RM})^2_{\rm signal} \rangle$ for these pairs, we can also compare the redshift-corrected values $\langle (\Delta {\rm RM})^2_{z, {\rm signal}} \rangle$, or the RM differences corrected by the different possible redshift corrections discussed in Sec.~\ref{sec:redz}. Table~\ref{tab:meanszz} presents the rms values for random and physical pairs with the different redshift corrections. With no corrections, the sample of pairs restricted to those with redshifts yields rms values of $16.5\,$rad m$^{-2}$ and $5.4\,$rad m$^{-2}$ for random and physical pairs respectively, which are similar to the values computed using the full pairs samples (with or without redshifts). However, when the redshift corrections are applied the difference between random and physical pairs changes from $\sim10\,$rad m$^{-2}$  to up to $\sim200\,$rad m$^{-2}$. Comparing, or interpreting, these redshift-corrected rms values requires different assumptions about where the major contribution to $\Delta$RM is coming from, which is discussed further in Sec.~\ref{sec:discussion}. 

Panels (a) and (c) of Fig.~\ref{fig:zplots2} show the distributions of $z_{\rm min}$ and $\Delta z$. Panel (b) shows the binned $\langle (\Delta {\rm RM})^2_{\rm signal} \rangle$ vs $z$ (physical pairs) or $z_{\rm min}$ (random pairs) with no redshift corrections or the possible redshift corrections mentioned from Sec.~\ref{sec:redz}, while panel (d) shows the binned $\langle (\Delta {\rm RM})^2_{\rm signal} \rangle$ vs $\Delta z$ for random pairs with no and possible redshift corrections. In general, trends of decreasing $\langle (\Delta {\rm RM})^2_{\rm signal} \rangle$ with increasing $\Delta z$, and increasing $\langle (\Delta {\rm RM})^2_{\rm signal} \rangle$ with increasing $z$ / $z_{\rm min}$ can be seen.

It is important to note, however, that due to the decreased sample sizes when restricting to those only with redshifts, the uncertainties are much larger. Therefore, any discussion of results from these samples or conclusions drawn are tentative and should be treated with some caution. 

In the following section we look at all of these results in more detail and discuss possible interpretations.

\section{Discussion}
\label{sec:discussion}

In Sec.~\ref{sec:results} it was shown that there is a clear difference in the rms $\Delta$RM$_{\rm signal}$ between random pairs of sources and pairs of physical components, ranging from 5 to $10\,$rad m$^{-2}$, depending on the range of angular separations considered. This difference remains despite different cuts in the data based on spectral index, redshift, leakage fraction, and separation. In addition to there being a difference in the average rms for $\Delta$RM$_{\rm signal}$, from the results presented above and the properties of the samples we can summarize our findings as follows:
\begin{itemize}
\setlength\itemsep{0em}
\item a large fraction of the random pair sources are unresolved at the NVSS resolution of $45{\arcsec}$, while the physical sources are either completely or partially resolved;
\item the distribution of angular separations between physical and random pairs is different, with physical pairs having on average much smaller separations than random pairs (Sec.~\ref{sec:results_sep}, Fig.~\ref{fig:sephists});
\item both groups show a trend of increasing $\langle(\Delta {\rm RM})^2_{\rm signal}\rangle$ with increasing $\Delta r$ (Sec.~\ref{sec:results_sep} Fig.~\ref{fig:strfncs1});
\item the physical pairs have a higher average polarization fraction than the random pairs, with the more largely separated physical pairs having higher polarization fractions than closely separated pairs (Sec.~\ref{sec:results_pf} Fig.~\ref{fig:pihists});
\item the random pairs show a strong trend of decreasing $\langle(\Delta {\rm RM})^2_{\rm signal}\rangle$ with increasing $\pi_{\rm min}$, while the physical pairs show no real trend with $\pi_{\rm min}$ (Sec.~\ref{sec:results_pf} Fig.~\ref{fig:pihists});
\item the physical pairs have steeper spectra on average and have more correlated spectral indices between the components of the pairs (Sec.~\ref{sec:results_spec} Fig.~\ref{fig:alphaplot});
\item the difference in rms $\Delta$RM$_{\rm signal}$ does not significantly change when we cut out pairs containing flat spectrum sources ($\alpha > -0.45$) (Sec.~\ref{sec:results_spec});
\item no significant trend is seen for $\langle(\Delta {\rm RM})^2_{\rm signal}\rangle$ vs $\alpha_{\rm max}$, where $\alpha_{\rm max}$ is the spectral index of the flattest-spectrum component of the pair (Sec.~\ref{sec:results_spec} Fig.~\ref{fig:alphaplot}b);
\item there is a peak in $\pi$ vs $\alpha$ near $\alpha \simeq -0.5$, with polarization fraction decreasing for flatter spectrum sources  (Sec.~\ref{sec:results_spec} Fig.~\ref{fig:alphaplot}c);
\item the random pairs are on average at higher redshifts than the physical pairs (Sec.~\ref{sec:results_red} Fig.~\ref{fig:zplots2});
\item the random pairs show a trend of increasing $\langle(\Delta {\rm RM})^2_{\rm signal}\rangle$ with increasing $z_{\rm min}$, if $z$-corrections are applied (Sec.~\ref{sec:results_red} Fig.~\ref{fig:zplots2});
\item the relation $\langle(\Delta {\rm RM})^2_{\rm signal}\rangle$ vs $\Delta z$ for the random pairs shows some indication of decreasing $\langle(\Delta {\rm RM})^2_{\rm signal}\rangle$ for increasing $\Delta z$, depending on the redshift correction model (Sec.~\ref{sec:results_red} Fig.~\ref{fig:zplots2}).

\end{itemize}
What does the above information tell us about the causes for the difference in $\Delta$RM$_{\rm signal}$ between physical and random pairs?

\begin{figure*}
\includegraphics[scale=0.34]{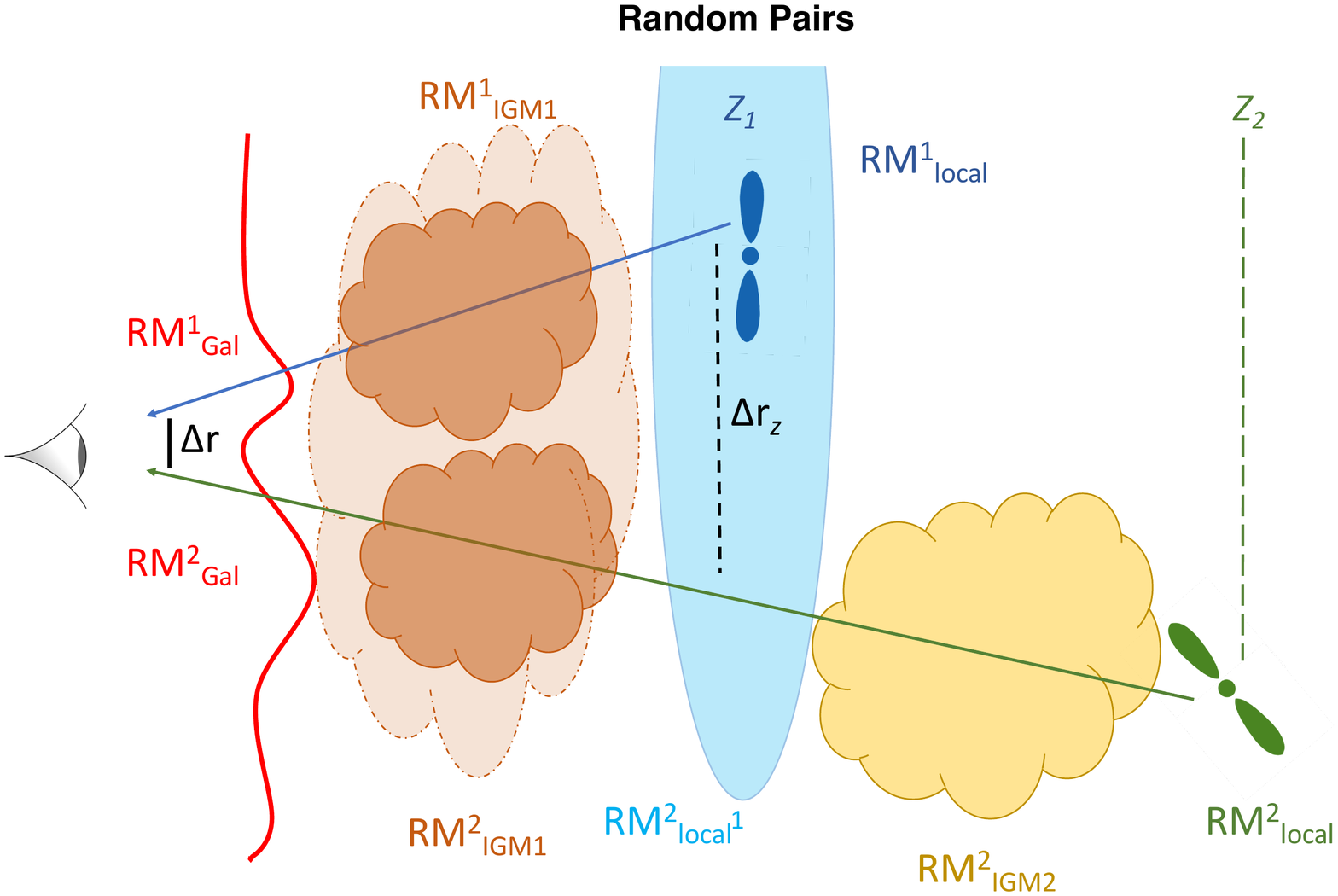}\includegraphics[scale=0.34]{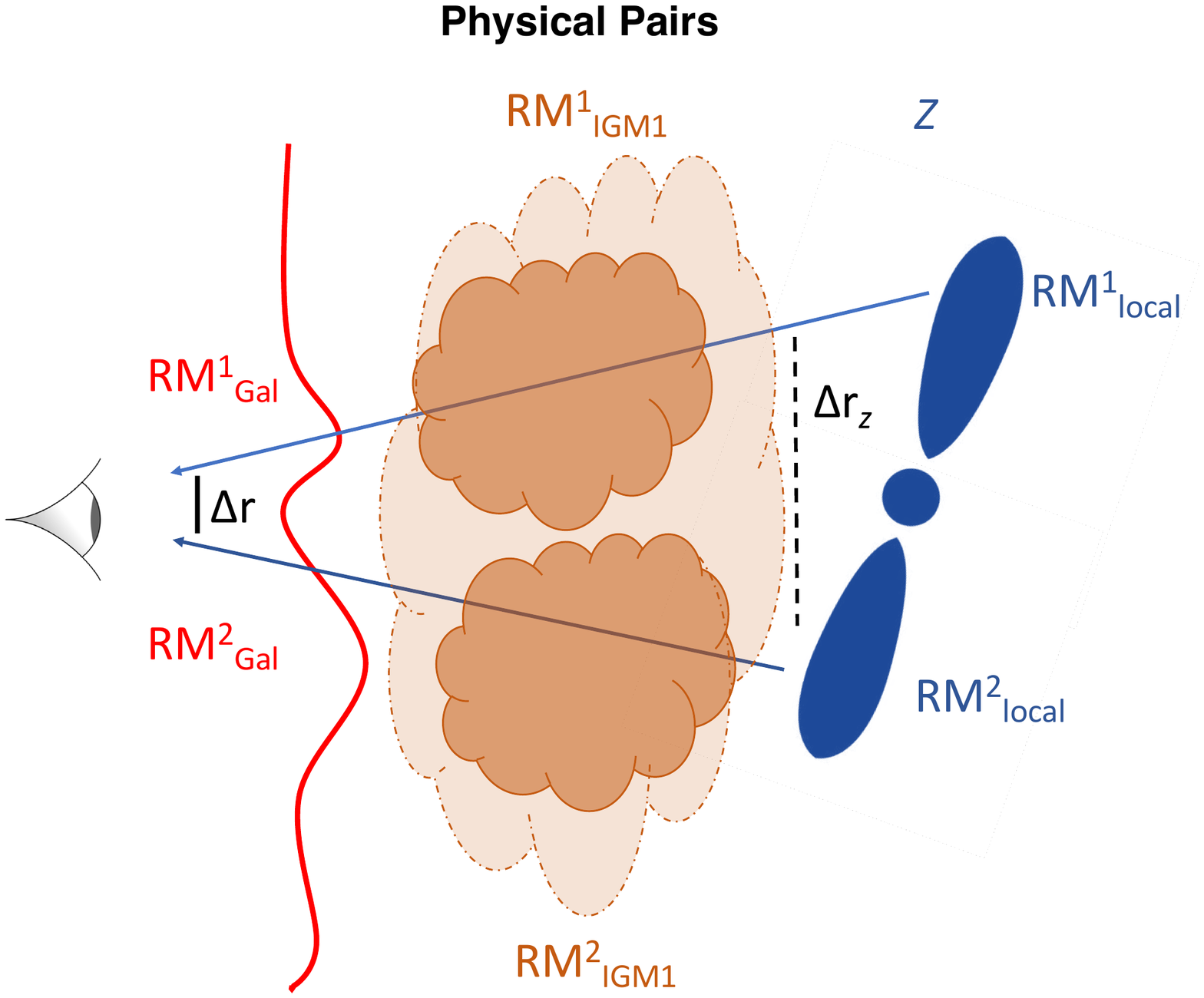}
\caption{Simple visual depictions of the different RM contributors in the random pairs case (left) and physical pairs case (right). The angular separation is labelled as $\Delta r$, while the physical separation (at the same redshift) is labelled as $\Delta r_z$. For the RM$^{1,2}_{IGM1}$, the separate regions enclosed by the larger (dashed-line) region are meant to convey that the IGM along the lines of sight may be different (separate regions) or may be (roughly) the same (larger region enclosed by dashed lines). \\}
\label{fig:cartoons}
\end{figure*}

We can break the RMs down into the different possible contributing regions in a simple model. A simple visual representation of two different cases, one for random pairs and one for physical pairs, and the different contributions is shown in Fig.~\ref{fig:cartoons} (in the random case ``1" refers to the closer source and ``2" is the more distant source). In this figure the different RMs are
\begin{itemize}
\item RM$^{\#}_{\rm Gal}$ : the contribution to the observed RM of source 1 or 2 from our Galaxy along their respective lines of sight;
\item RM$^{\#}_{\rm IGM1}$: the contribution to the observed RM of source 1 or 2 (superscript) from the intervening IGM between us and either the closest source of the pair (for random pairs) or us and the source (for physical pairs) along the respective lines of sight;
\item RM$^{\#}_{\rm local}$: the contribution to the observed RM from the region local to the source (or source components);
\item RM$^{2}_{\rm local^{1}}$: the contribution to the observed RM of source 2 (the more distant source) in a random pair from the region local to source 1 (or the closer source);
\item RM$^2_{\rm IGM2}$: the contribution to the observed RM of source 2 from the intervening IGM from source 1 to source 2 along the line of sight.
\end{itemize}

In this simplistic model, we can rework eq.~(\ref{eq:dsum}) for the difference in RMs to make it slightly more detailed. For random pairs, it becomes
\begin{equation}
\begin{split}
\Delta {\rm RM^{random}_{signal}} \, = \, \Delta {\rm RM}_{\rm Gal} + \Delta {\rm RM}_{\rm IGM1} +\Delta {\rm RM}^{12}_{\rm local}\\ +{\rm RM}^{2}_{\rm local^{1}}+{\rm RM}^{2}_{\rm IGM2}, 
\label{eq:delrmrand}
\end{split}
\end{equation}
and for physical pairs,
\begin{equation}
\begin{split}
\Delta {\rm RM^{physical}_{signal}} \, = \, \Delta {\rm RM}_{\rm Gal} + \Delta {\rm RM}_{\rm IGM1}\\ +\Delta {\rm RM}^{12}_{\rm local} .
\label{eq:delrmphys}
\end{split}
\end{equation}
The different parameters are defined as:
\begin{itemize}
\item $\Delta {\rm RM}_{\rm Gal} = {\rm RM_{Gal}}^{1}-{\rm RM_{Gal}}^{2}$: The difference in RM due to differences in the Galactic foregrounds across the lines of sight;
\item $\Delta {\rm RM}_{\rm IGM1}= {\rm RM_{IGM1}}^{1}-{\rm RM_{IGM1}}^{2}$: The redshift-corrected difference in RM due to the variation in the intervening IGM foreground along the lines of sight between the Galaxy and the source (for physicals) or the closer source (for randoms);
\item $\Delta {\rm RM}^{12}_{\rm local} = {\rm RM_{local}}^{1}-{\rm RM_{local}}^{2}$: The difference in the RM of the sources (or source components) from the differences internal or local to the sources/components; 
\item ${\rm RM}^{2}_{\rm local^{1}}$: The added redshift-corrected RM contribution in random pairs from the background source (source 2) emission passing through or near the closer source or source's environment (source 1);
\item ${\rm RM}^{2}_{\rm IGM2}$: The redshift-corrected RM due to the intervening IGM along the line of sight between the foreground source and the background source for random pairs (the 2 superscript refers to it being only along the line of sight of the more distant source);
\item $\Delta {\rm RM}_{\rm noise}$: The contribution to $\Delta$RM from measurement noise/uncertainty. 
\end{itemize}
It is likely there is some contribution to $\Delta$RM$_{\rm signal}$ from each of these different possible sources. There is no direct way to disentangle each contributing element, but looking at the relationships between $\Delta$RM$_{\rm signal}$ and other variables such as separation, redshift, polarization fraction, etc as above should yield some clues and it may be possible to find some magnetic field upper limits. Below we examine the different possibilities in more detail.

\subsection{Galactic Foreground}
\label{sec:galforeground}

It is known from previously published works that $(\Delta {\rm RM})^2_{\rm signal}$ varies with $\Delta r$ \citep[e.g][]{Leahy87,Mao10,Stil11}, attributed to variations in the Galactic foreground on different angular scales. \citet{Stil11} find that Galactic RM variations range from approximately 10 to $20\,$rad m$^{-2}$  on angular scales of $5{\arcmin}$ to $15{\arcmin}$ (the region of angular separation that was examined by Stil et al., that most closely matches the angular separations examined in this work). For both the random and physical pairs in this work, we see this increasing trend between $(\Delta {\rm RM})^2$ and separation is seen in Fig.~\ref{fig:strfncs1}b, with similar power-law slopes ($0.6\pm 0.1$ for random pairs vs $0.8\pm0.2$ for physical pairs) from eq.~\ref{eq:plaw}.

The Galactic foreground cannot contribute differently to physical pairs compared to random pairs. The positions on the sky of the different populations is uniformly (Fig.~\ref{fig:galcoords}), or randomly, distributed, with neither the physical nor random pairs coming from any preferential or biased region in relation to the Galaxy. 

One could argue that since among the random pairs there are more pairs with larger angular separations, which should have a larger difference in Galactic contribution, that this could cause the discrepancy between the groups. This was checked by comparing $\langle(\Delta {\rm RM})^2_{\rm signal}\rangle$ in bins of $\Delta r$ (so as to compare those in each group with others with the same angular separations), as well as bootstrap and KS tests confining the range in $\Delta r$ to a range where the physical and random distributions had more overlap. Comparing at more similar separations we did see the difference in $\sqrt{\langle(\Delta {\rm RM})^2_{\rm signal}\rangle}$ between random and physical pairs decrease from $\sim 10\,$rad m$^{-2}$ to $5\,$rad m$^{-2}$, but the difference still exists and is significant at the 2 to 3$\sigma$ level. 

Additionally, we can recompute the rms values for the two groups using higher Galactic latitude cuts, or only including pairs/sources further from the Galactic plane where Galactic RM variance should be smaller. If we use a cut of $|b| \geq 50{\degr}$ instead of $|b| \geq 20{\degr}$, the random pairs rms changes from $14.9\,$rad m$^{-2}$ to $13.6\,$rad m$^{-2}$ and the physical pair's rms changes from $4.6\,$rad m$^{-2}$ to $4.03\,$rad m$^{-2}$. For a cut of $|b| \geq 70{\degr}$ we obtain $12.1\,$rad m$^{-2}$ for random pairs and $5.8\,$rad m$^{-2}$ for physical pairs. Thus, the difference in rms values does not disappear when further from the Galactic plane where there would be presumably smaller RM variance from the Galaxy. \citet{Stil11} showed that the RM structure did not vary uniformly with increasing Galactic latitude, and that there are regions of high RM variance that extend to high latitudes. There could be larger concentrations of random pairs from these areas, which would produce an additional excess in the $\Delta$RM compared with the physical pairs, which due to the much smaller numbers of pairs are more sparsely distributed on the sky. The best test for this effect is to increase the sample size of the physical pairs.

While the Galactic RM variance does contribute to the $\Delta$RMs of the two groups, it does not appear that Galactic RM variance is an important factor in the difference between the $\Delta$RMs of the random and physical pairs. Therefore, the difference in the $\Delta$RM rms values can be seen as a detection of an extragalactic signal. In the following sections we examine the possible extragalactic contributions to $\Delta$RM$_{\rm signal}$.

\subsection{Local Source Contribution}
\label{sec:local}

The other possible major contributor to $\Delta$RM$_{\rm signal}$ would be $\Delta {\rm RM}^{12}_{\rm local}$, the difference in RMs from the regions local to (or internal to) the sources. For physical pairs, $\Delta {\rm RM}^{12}_{\rm local}$ could depend on $\Delta r$. Because the two components are part of the same physical object it seems reasonable to think that the two components share similar environments. However, the larger the physical source is (or the larger the separation is between the polarized components of the source) the more likely it is that the environments around the polarized emission are different. 

The same reasoning does not apply to the random pairs, which no matter how close in angular separation they are, are still separated in space. Thus there is no reason for their local environments to be similar. For random pairs, one source is always behind the other in the pair. In this case the light from the background source may pass through the local environment of the foreground source, causing a change in RM (defined above as ${\rm RM}^{2}_{\rm local^{1}}$). This is likely to happen more often and/or have a larger effect when the angular separation between the sources is smaller, i.e. the background emission passes closer to the foreground source. If this is the case, random pairs with small separations should show an increase in $\Delta$RM$_{\rm signal}$ compared to those at larger separations, resulting in a flattening of the structure function. Our sample show an increase in $\Delta$RM$_{\rm signal}$ with increasing $\Delta r$ (Fig.~\ref{fig:strfncs1}). However, this trend is likely due to, or dominated by, variations in the Galactic magnetic field. The contribution from our Galaxy would need to be accounted for before any extragalactic dependence of $\Delta$RM$_{\rm signal}$ on $\Delta r$ could be seen. 

The contribution of  ${\rm RM}^{2}_{\rm local^{1}}$ to $\Delta$RM$_{\rm signal}$ is likely to be very small, especially if the foreground source is unresolved or compact, or if the minimum distance from the line of sight to the foreground source is large ($\ga 1$ or $2\,$Mpc). This is because the foreground source magnetic field and/or electron density usually decrease as a function of distance from the source. The exception would be if the foreground source was situated in a galaxy cluster or high-density environment. 

The fractional polarizations will be influenced by the intrinsic magnetic field disorder in the source, as well as the depolarizing effects of Faraday screens local to the source and in the intervening medium. The physical pairs show no dependence of $\Delta$RM on their fractional polarization (Fig.~\ref{fig:pihists}), which would be expected if the intervening medium were causing both $\Delta$RM and the low fractional polarization. In addition, a scatter of at least $\sim32\,$rad m$^{-2}$ across the face of a component would be required to produce significant depolarization (i.e. $\sim 1/\sqrt{N}$, where $N$ is the number of independent RM patches), while the average $\sqrt{\langle(\Delta {\rm RM})^2_{\rm signal}\rangle}$ is only $\sim 5\,$rad m$^{-2}$ between physical components. It is therefore likely that the fractional polarizations for physical pairs are dominated by magnetic field irregularities.

For the random pairs, there is a strong dependence of $\Delta$RM on the fractional polarization (Fig.~\ref{fig:pihists}), suggesting that they both might be influenced by a Faraday medium. The dependence of the fractional polarization on the spectral index of each component (Fig.~\ref{fig:alphaplot}), suggests that this medium is actually related to the source, since any unrelated intervening screen would not show a spectral index dependence (which could be argued for the physical sources as well, since they also show a relation between $\pi$ and $\alpha$). In addition, flatter spectrum sources are likely to be very compact, and thus probe only very small scales in any intervening medium. The direct connection between spectral index and $\Delta$RM$_{\rm signal}$ is not as apparent, since $\Delta$RM$_{\rm signal}$ depends on the properties of two (unrelated) sources.  However, there is a suggestion of larger values of $\Delta$RM$_{\rm signal}$ when at least one of the components has a flatter spectrum ($\alpha \ga -0.4$). We thus conclude that a substantial component of $\Delta$RM$_{\rm signal}$ comes from a Faraday medium local to the source, although polarization information at other wavelengths and at higher resolution would be important in order to confirm these effects.

It is reasonable to assume that differences in the magnetic fields strengths and electron densities of components of the same physical source would be smaller than differences between random sources that share no physical connections, which is suggested by the $\Delta$RM$_{\rm signal}$ being larger for random pairs. Under the assumption that the source RMs, and correspondingly $\Delta$RM$_{\rm signal}$, are due to a Faraday medium local to the sources, we can take those pairs with redshifts and correct the RMs of each pair component to get the intrinsic RMs and then can convert $\Delta$RM$_{\rm signal}$ to a $\Delta B_{||}$. 

This yields $\sqrt{ \langle \Delta B_{||}^2 \rangle} = 240\, \mu {\rm G} \, (1/n_e) \, (1/L)$ for the random pairs and $140\, \mu {\rm G} \, (1/n_e) \, (1/L)$ for the physical pairs, with $n_e$ the electron density of the source in cm$^{-3}$ and $L$ is the average path length in pc. If we assume $n_e$ and $L$ are the same for all sources (a very simplifying assumption) then the average difference in the local magnetic fields of random pairs of sources is $\sim 1.7$ times larger than the average difference in the magnetic field within a physical source. One might expect the local environment around one source to vary less than the local environments between two completely unrelated sources, which is what we see with our results. Particularly, if the electron densities and or the magnetic fields vary with redshift then this result makes sense as the random pairs will always have one component at a higher redshift, leading to a larger variance in their local environments.


\subsection{The IGM Contribution}
\label{sec:igm}

If the difference in random and physical pairs is not coming from Galactic variations, then it could be due to the intervening medium. It is not known how much of the difference can be attributed to the IGM. However, we can look at some different cases.  

The contribution from the IGM between the closer source and the more distant source in a random pair, ${\rm RM}^{2}_{\rm IGM2}$, should not have any dependence on $\Delta r$. However, it is possible that the contribution from the IGM to $\Delta$RM$_{\rm signal}$ between us and the foreground source (or just the source for physical pairs), $\Delta {\rm RM}_{\rm IGM1}$, would have a positive correlation with separation. The larger the separation between two sources, or source components, the more likely it is that the intervening medium between them is different. The intervening IGM foreground between us and a source (for physical pairs) and between us and the closer of the two sources (random pairs) can vary between the two components of a pair, depending on the positions of the components relative to filaments or clusters. 

This was shown recently for one particular source by \citet{O'Sullivan19}, who attributed the difference in RM between two lobes of an AGN of $2.5\pm 0.1 \,$rad m$^{-2}$ to the IGM, as one lobe was found to be located behind several filaments while the other lobe was not. It is likely that the difference in RM due to a difference in the number of intergalactic filaments along the line of sight is larger for pairs, random or physical, where the separation is larger. Pairs that are very close together have a higher likelihood of having the same or similar foreground. This could explain some of the slope of the structure functions by creating larger RM differences at larger separations. 

Using the very simple assumption that the RM from the IGM contributes uniformly along the line of sight, corresponding to Model C in Sec.~\ref{sec:redz}, with a redshift correction $\Delta {\rm RM}_{\rm signal} / \int_{z_{\rm{max}}}^{z_{\rm{min}}}{(1+z)^{-2} dz}$, and that the IGM is the dominant contribution to the difference in the $\Delta$RMs, then using eq.~(\ref{eq:RM1}) we can get an estimate for the rms line-of-sight magnetic  field strength $\sqrt{<B_{||}^2>}$ for the IGM.

 To first order, $dl/dz$ from eq.~(\ref{eq:RM1}) can be approximated as $c/H_0$. We must also take into account the $N_{\rm rev}$ field reversals along the line of sight. Taking into account possible reversals adds a factor of $N_{\rm rev}^{-1/2}$ to the right hand side of eq.~(\ref{eq:RM1}), with $N_{\rm rev}$ approximated as $N_{\rm rev}\simeq L/l$, where $L$ is the total comoving path length and $l$ is the scale size of the reversals. The difference in the rms $\Delta$RM between physical and random pairs ranges from $\sim 5$-$10\,$rad m$^{-2}$, depending on which criteria are used (e.g. all pairs vs those with separations between $3\arcmin \le \Delta r \le 11\arcmin$). The excess seen for random pairs is roughly a factor of 2 smaller than the $\Delta$RM rms for the random pairs. Thus, if we assume that roughly half of $\Delta {\rm RM}_{\rm signal}$ for each random pair is coming from the IGM that corresponds to attributing the whole of the difference in rms values to the IGM. Putting all this together, for the rms line-of-sight magnetic field from the IGM $\sqrt{ \langle B_{||}^2 \rangle}$, we have
\begin{equation}
\sqrt{ \langle B_{||}^2 \rangle} =\sqrt{\frac{1}{N}  \sum_i \frac{\Delta{\rm RM}^{2}_{{\rm signal}, i}}{4} \left [ \frac{ N_{\rm rev}^{1/2} H_0}{8.12\times 10^{5} \,  n_e \, c \,(1+z)^{-1}|_{ z_{\rm min}  }^{ z_{\rm max}}} \right ]_i^2 },
\label{eq:meanb}
\end{equation}
where the sum is over all random pairs, and $N$ is the number of random pairs. 

The electron density in filaments from simulations varies from $\sim 10^{-6}$ to $10^{-4}\,$cm$^{-3}$ \citep{Ryu08,Vazza15,Akahori10}. Assuming an $ n_e $ for the IGM of $1\times 10^{-5}\,$cm$^{-3}$ and a reversal scale length of $1\,$Mpc we obtain $\sqrt{ \langle B_{||}^2 \rangle}=22 \,$nG, and a total average magnetic field from the IGM of $\langle B_{\rm IGM} \rangle = \sqrt{3} \sqrt{ \langle B_{||}^2 \rangle} =37\,$nG. 

Using this method to estimate the IGM contribution can be viewed as an upper limit for the magnetic field in that it assumes all of the difference in $\Delta$RM between random and physical pairs is coming from the IGM between the foreground and background sources in random pairs, while in reality it is just one contributing factor. However, in another sense it is also a lower limit in that the reversal scale may be much smaller which would increase the magnetic field limit by an order of magnitude or more. 

We can still compare this limit to other limits and estimates. \citet{Vernstrom17} and \citet{Brown17} estimated upper limits on the magnetic field of the IGM using cross-correlation analyses and found limits of $\sim 100$ to $\sim 500 \,$nG. The recent work by \citet{Vacca18} found an average magnetic field in the IGM from simulations of $20-50\,$nG, with previous simulations finding similar values of $\sim 10 - 50\,$nG \citep{Ryu08,Vazza14b,Vazza16}. \citet{O'Sullivan18} obtained estimates of IGM magnetic field strengths of $40\,$nG to $500\,$nG from Faraday depolarization analysis at low frequencies of a large radio source. The sub-$\mu$G limits obtained above are consistent with other estimates. 

If the assumptions from the above paragraphs that the RM from the IGM contributes uniformly along the line of sight and that the IGM is the dominant contribution to the difference in $\Delta$RM$_{\rm signal}$ were correct then applying the redshift correction $[\Delta {\rm RM}_{\rm signal} / \int_{z_{\rm{max}}}^{z_{\rm{min}}}{(1+z)^{-2}}dz]$ would presumably produce a constant average in $\langle(\Delta {\rm RM^{random}_{signal}})^2\rangle$ vs $\Delta z$ or $z_{\rm min}$ (i.e. if the intrinsic IGM RM is a constant then correcting the observed $\Delta$RM for redshift should reveal that constant). However, this is not what is seen in Fig.~\ref{fig:zplots2}. This could mean that either the IGM is not the dominant contributor to $\Delta {\rm RM^{random}_{signal}}$ or that the RM from the IGM has a redshift dependence that has not been taken into account. The shape of the curves in Fig.~\ref{fig:zplots2}b,d, namely the slight increase in $\langle(\Delta {\rm RM})^2_{\rm signal}\rangle$ with $z_{\rm min}$, would argue, in the absence of other confounding variables, for a positive $z$ dependence of either the source RMs, the IGM RM or both. However, other studies \citep{Hammond12,Vernstrom18} have looked at intrinsic source RM as a function of redshift and found no significant dependence.  Given the uncertainty in our sample, due to the small number of pairs with redshifts for both sources, it is difficult to draw firm conclusions. 

The above calculation for the IGM magnetic field limit is a very simplified method for obtaining a limit. For the case of the RM from the IGM, the RM values that would be observed depend on a variety of cosmological parameters, including the inhomogeneous distribution of densities, in clusters, filament and void regions, how the magnetic field strength scales with density, the coherence scale of the field (i.e., over what spatial scale it reverses), and the redshift dependence of all of these. These unknowns make it difficult to calculate an accurate and realistic limit. However, these effects are all embodied in cosmological MHD simulations, and so we compare our results to those, such as summarized recently by \citet{Akahori18}. We find a net value of $\approx 10\,$rad m$^{-2}$ for the $\sqrt{\langle \Delta {\rm RM}^2_{\rm signal} \rangle}$ for random pairs, or an rms contribution of $\approx 7 \,$rad m$^{-2}$ for each component. This is consistent with the asymptotic value of 5-10$\,$rad m$^{-2}$ summarized by \citet{Akahori18} (figure 1 of Akahori) when clusters are excluded from the line of sight, and corresponds to magnetic fields ranging from $0.01\,$nG to $100\,$nG, going from voids into filamentary structures.  When clusters are included, $\sigma_{\rm RM}$ is predicted to be 4-8 times larger, which is not consistent with our data. Since we do not attempt to eliminate clusters along our lines of sight, it appears that the contribution from clusters in the simulations is over-estimated. The simulations also predict that $\sigma_{\rm RM}$ should decrease rapidly below redshifts of approximately 0.5. This effect should be examined when larger samples allow better control for angular separation between pairs, spectral index, etc.

The IGM is expected to have a weak magnetic field and is not expected to be the sole contributor to $\Delta$RM$_{\rm signal}$, which is why being able to account for other confounding variables and RM contributions is necessary. In the following section some caveats from the data used here and areas for improving this method to better isolate the IGM signal are discussed.

\subsection{Uncertainties $\&$ Selection Effects}
\label{sec:selection}

The average uncertainty for source RMs, as measured in this work, is $9.5\,$rad m$^{-2}$ for random pair sources and $6.1\,$rad m$^{-2}$ for physical pair sources. The uncertainties for each $\Delta$RM are then the individual uncertainties added in quadrature, with the average $\langle \sigma_{\Delta \rm{RM}} \rangle = 14.1\,$rad m$^{-2}$ for random pairs and $8.7\,$rad m$^{-2}$ for physical pairs. The difference in the $\Delta$RM$_{\rm signal}$ rms values between physical and random pairs is only $\sim10\,$rad m$^{-2}$, less than the average pair uncertainties. The difference reported for the lobes of a $3.4$-Mpc giant radio galaxy by \citet{O'Sullivan19} (attributed to the IGM) had an uncertainty of $\pm 0.1\,$rad m$^{-2}$. That was using data from the LOFAR telescope, where at low frequencies ($\sim 150\,$MHz) the resolution in Faraday space is approximately $1\,$rad m$^{-2}$. The resolution in Faraday space is determined by the total bandwidth, or $\delta \phi \approx 2 \sqrt{3} / \Delta \lambda^2$. The uncertainty in RM measurements is equal to to $\delta \phi / (2 \,{\rm SNR})$, where SNR is the signal-to-noise ratio of the polarized intensity of the source. In order to isolate different physical contributions to $\Delta$RM, the RMs of the sample need to have smaller uncertainties. This requires either a larger range in $\lambda$ and/or more sensitive data. 

The RMs from NVSS were found from the slope of $\psi$ vs $\lambda^2$ from two adjacent wavelengths. For a single Faraday-thin component this may work well enough. However, if the source has multiple Faraday components or Faraday-thick components this will result in complex, or non-linear behavior in $\psi$ vs $\lambda^2$. If this is the case RM estimates from slope fitting are likely not as accurate. \citet{Farnsworth11} and \citet{O'Sullivan12} showed how RM estimations from narrow-bandwidth observations can yield erroneous results in the presence of multiple interfering Faraday components.

Faraday complexity in the RM spectrum can indicate that RM is measuring the complexity of the source environment rather than probing the intervening foregrounds \citep{Lamee16}. Unfortunately the complexity information is not available from the two-band NVSS polarimetric data. Wide-band data combined with newer techniques such as RM synthesis \citep{Brentjens05} or $QU$-fitting allow for more detailed measurements of the Faraday spectrum (multiple Faraday components, Faraday-thin and -thick components, etc). Additionally, wide-band data would allow for spectral index fitting across the band as well as measures of depolarization. Depolarization and the Faraday complexity are key elements to disentangling $\Delta$RM$_{\rm signal}$ contributions that the current data sample does not provide.  

There is also the issue of completeness of the sample. \citet{Taylor09} do not estimate any completeness levels for the catalog, and there may be sources which should be included but were not. For example, the source discussed in \citet{O'Sullivan19}, J1235+5317, is in the NVSS catalog as a multi-component AGN source, with components separated by $\sim 12{\arcmin}$. The lobes both show detectable polarized emission (of $5$ to $10\%$ fractional polarization) in NVSS and have been detected in polarized emission by others \citep{Vaneck18,O'Sullivan19}, but neither of the source components were included in \citet{Taylor09}. It is unclear how many such sources may be missing in the Taylor catalog. 

Even considering these sources of uncertainties, the difference we find between physical and random pairs is a significant detection. Perhaps an obvious answer to the question of why the physical and random pair $\Delta$RMs are different is that they are very different physical populations that one would expect to have different characteristics. We know that is at least partly the case based on the different separation, redshift, and spectral index distributions between the physical and random pairs. Also the fact that, even at comparable redshifts, the physical pairs are at least partly resolved while the majority of random sources are unresolved indicating different source physical sizes. 

We know that source type can be related to polarization properties. For example, \citet{Pshirkov15} found that high luminosity sources had, on average, higher RMs (from NVSS) than low luminosity sources. Additionally, \citet{Farnes14} showed that compact self-absorbed AGN, which can be identified by their radio morphology and spectra, are known to have different polarization properties. We showed that excluding flat-spectrum sources still results in $\Delta$RM$_{\rm signal}$ rms values similar to those when flat-spectrum sources are included, but there are likely still a range of source types in each sample. 

There are multiple possible contributing sources to $\Delta$RM$_{\rm signal}$ and an even larger number of confounding variables or observables to try and isolate those contributions. To truly isolate the IGM contributions from those local to the source(s), one needs to be comparing physically similar samples broken down over multiple variables. Thus ideally we would want a large sample of source pairs that could then be subdivided by source type, redshift, physical size, luminosity, as well as local environments (i.e. those in or behind clusters vs those in low density environments). Trends in multi-dimensional space could then be examined such as $\langle(\Delta {\rm RM})^2_{\rm signal}\rangle$ as a function of $\Delta r$ and $\pi$ or and $z$. However, with this current sample, such subdivision creates sample sizes too small for statistical comparison.

\section{Conclusions}
\label{sec:conclusions}

Using the NVSS polarimetry catalog by \citet{Taylor09}, we have examined the $\Delta$RMs of pairs of extragalactic radio sources with angular separations $1.5\arcmin \leq \Delta r \leq 20{\arcmin}$ for all sources with Galactic latitude $|b|\geq 20{\degr}$ and at least a $2\%$ polarization fraction. Using radio and multi-wavelength data, a compilation of extended radio galaxies, and a set of criteria we have classified all the pairs as either physically related pairs or randomly associated pairs, resulting in a sample of 317 physical pairs and 5111 random pairs. 

We found that for a significant number of our sample the RM uncertainties reported by \citet{Taylor09} are overestimated. We believe this to be, at least in part, due to the inclusion of real polarized signal in the original noise estimation. Based on a subsample of the original images used by \citet{Taylor09} we use the approximation of $\sigma_{\rm RM} = 150 \, \sigma_{QU}/P$ rad m$^{-2}$, with $\langle \sigma_{QU} \rangle = 0.38 \,$mJy beam$^{-1}$. We find a mean peak polarized intensity of $\langle P \rangle = 20 \,$mJy beam$^{-1}$ for physical pair components and $10\,$mJy beam$^{-1}$ for random pair sources, resulting in average RM uncertainties of $ 9.5\,$rad m$^{-2}$ and $6.1\,$rad m$^{-2}$, for random and physical pair sources respectively.  

The rms for the $\Delta$RM$_{\rm signal}$'s are found to be $14.9\pm0.4\,$rad m$^{-2}$ and $4.6\pm1.1\,$rad m$^{-2}$ for random and physical pairs respectively. Using bootstrap, KS and AD tests, we determine this difference of $\sim 10\,$rad m$^{-2}$ in the $\Delta$RM$_{\rm signal}$ rms of the two populations to be significant at the $\sim3$ to 5 $\sigma$ level, depending on how the significance is measured and if all separations are considered or a more restrictive range in angular separation is used. 

We find both groups show a trend for increasing $\Delta$RM$_{\rm signal}$ with increasing angular separation, attributed to the change in the Galactic magnetic field at different angular scales. However, the difference in $\Delta$RM$_{\rm signal}$ between physical and random pairs remains despite different cuts in Galactic latitude. Thus, the difference in $\Delta$RM$_{\rm signal}$'s can be seen as a detection of an extragalactic signal. 

Under the assumption that the entire difference in $\Delta$RM$_{\rm signal}$ rms can be attributed to the IGM between random sources, we estimate an upper limit on the magnetic field strength of the IGM to be $\sim 0.037 \, \mu$G, assuming $n_e=10^{-5}\,$cm$^{-3}$, and assuming no redshift dependence of $n_e$ or $B$. This upper limit is consistent with other estimates and limits from previous studies and simulations.

Based on the trends seen in $\langle(\Delta {\rm RM})^2_{\rm signal}\rangle$, $\pi$, and spectral index $\alpha$, it is likely that the local source environments contribute significantly to the $\Delta$RM$_{\rm signal}$ rms for each of the two groups of pairs. Attributing the RM difference to regions local to the sources results in estimates for $\langle \Delta B_{\||}\rangle$ roughly 1.7 times larger for random than physical pairs, assuming a constant $n_e$ and path length, i.e. the local environments around individual sources vary less than the local environments of two unrelated sources. 

Currently the sample size is too small to properly break down and compare the sources in groups matched by physical characteristics such as source type, physical size, luminosity, environment, etc. To disentangle contributions from the local source environments and the IGM, more sensitive wide-band polarimetric data are required, ideally with higher resolution for the random pairs. The Polarization Sky Survey of the Universe's Magnetism \citep[POSSUM,][]{Gaensler10}, with the Australia SKA Pathfinder (ASKAP) is one upcoming polarization survey that should help. The survey will cover the whole sky south of declination $30{\degr}$ and is expected to yield RMs for $\sim10^6$ sources. The VLASS survey, which is already underway, will cover the entire sky north of declination $-40{\degr}$ and is predicted to measure at least $200,000$ RMs and fractional polarizations. 

This increase in sample size is a key element to breaking the parameter degeneracies in determining the contributions to $\Delta$RM, as large sample sizes decrease uncertainties and allow for the samples to be subdivided by different criteria (e.g. source type, size, luminosity, redshift, etc) and $\Delta$RM to be looked at as a function of multiple variables at once. Additionally, the larger bandwidth of these surveys will decrease RM uncertainties, yielding more accurate measurements. Using the approximation of the RM uncertainty $\propto 1 / (\Delta \lambda^2 SNR)$, where SNR is the signal-to-noise ratio, and $\Delta \lambda^2=0.017$ for VLASS and $0.028$ for full POSSUM, we can expect RM uncertainties of $\sim 7\,$rad m$^{-2}$ for sources with $\sim$mJy polarized flux for VLASS and $\sim 1\,$rad m$^{-2}$ for $\sim$mJy sources for POSSUM. This is an improvement over the $\sim 10\,$rad m$^{-2}$ average uncertainty for NVSS RMs and the increased resolution of POSSUM and VLASS will also enable us to get RMs measurements across spatially resolved lobes for many sources, which will add a lot more information about what is going on local to the sources. Also, new work on machine learning algorithms for classification of physical pairs will be instrumental in making use of these new data. 

While more and better data are required to truly understand the results shown here, this work is the first to find an unambiguous extragalactic signal in RMs and show that separating of sources into physical and random pairs can be a useful and important method for accounting for Galactic foregrounds and statistically analysing extragalactic magnetic fields.

\section{Acknowledgments}
The Dunlap Institute is funded through an endowment established by the David Dunlap family and the University of Toronto. T.V. and B.M.G. acknowledge the support of the Natural Sciences and Engineering Research Council of Canada (NSERC) through grant RGPIN-2015-05948, and of the Canada Research Chairs program. H.A. has benefited from grant DAIP 66/2018 of Universidad de Guanajuato, Mexico. Partial support for L.R. comes from the U.S. National Science Foundation grant AST-1714205 to the University of Minnesota. We acknowledge the work of volunteers (see \url{http://rgzauthors.galaxyzoo.org } for a full list) on the Radio Galaxy Zoo project and their contributions to finding and classifying extended radio galaxies and their counterparts. We also thank Shane O'Sullivan, Jackie Ma, and Jeroen Stil for their shared information and suggestions, as well as the anonymous referee for helpful feedback. 

\appendix 

\counterwithin{figure}{section}

\section{RM Uncertainties}
\label{sec:rmuncert}

It was reported by \citet{Stil11} that the RM uncertainties in the \citet{Taylor09} catalog may not be correct and explicit details on how these uncertainties were obtained were not included. Given that the uncertainties on the RMs are important for our analysis we wanted to investigate this issue. We found that for a significant fraction of our sample $\Delta {\rm RM}^2_{\rm obs} \ll \Delta {\rm RM}^2_{\rm noise}$ (where $\Delta {\rm RM}^2_{\rm obs} = ({\rm RM_1} -{\rm RM_2})^2$ without any corrections applied), indicating that the reported RM uncertainties are significantly overestimated for at least some fraction of sources, particularly for the physical pairs (unlike Stil et al. who found the uncertainties to be underestimated). To check this for our sample of sources, we obtained a subset of maps for 100 pairs (200 sources) of stokes $I$, $Q$, and $U$ at the two frequency bands (1365 MHz and 1435 MHz) used in the Taylor paper, courtesy of J. Stil. These 100 pairs consisted of 75 physical pairs and 25 random pairs, all with separations less than $10{\arcmin}$.

\begin{figure}
\includegraphics[scale=0.37]{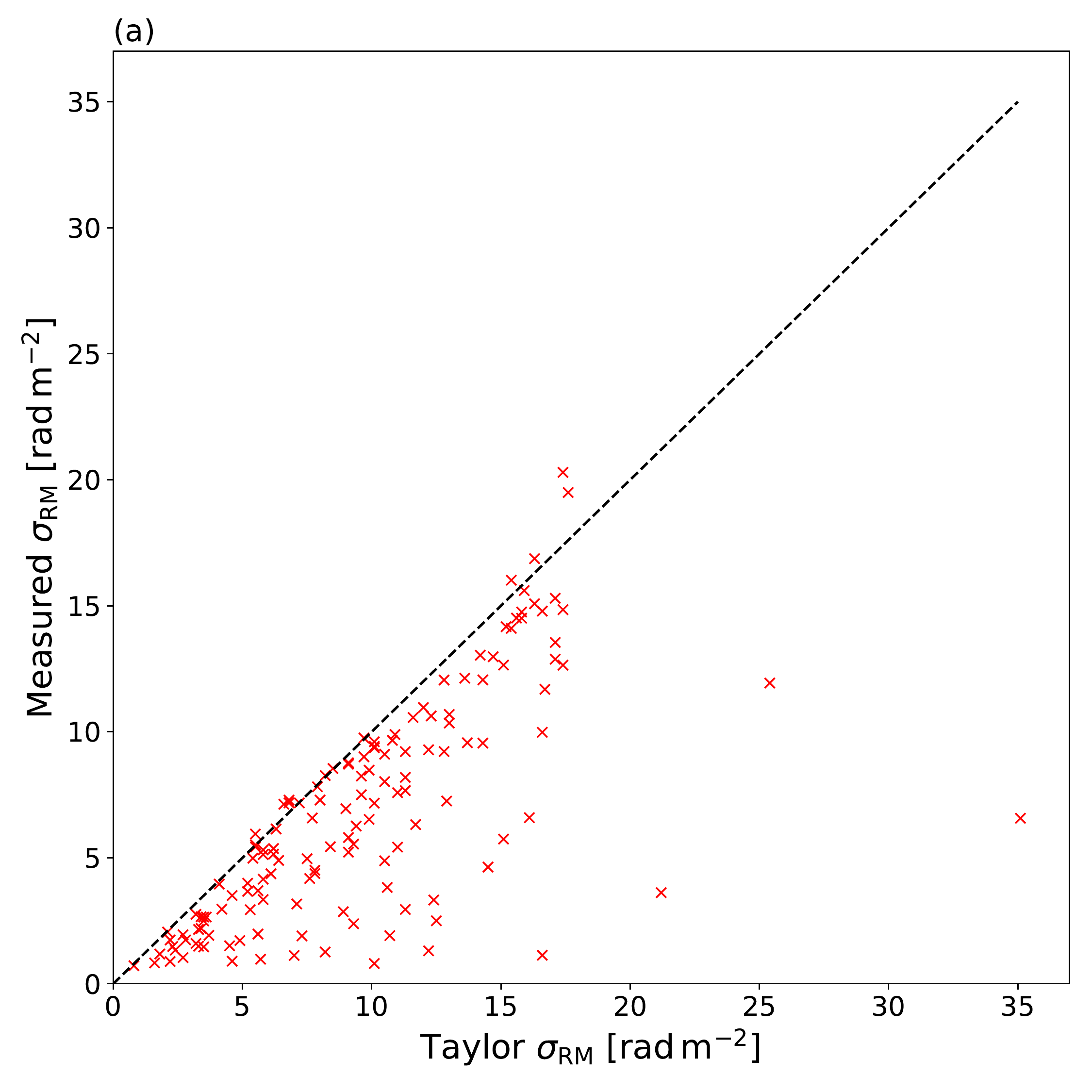}
\includegraphics[scale=0.37]{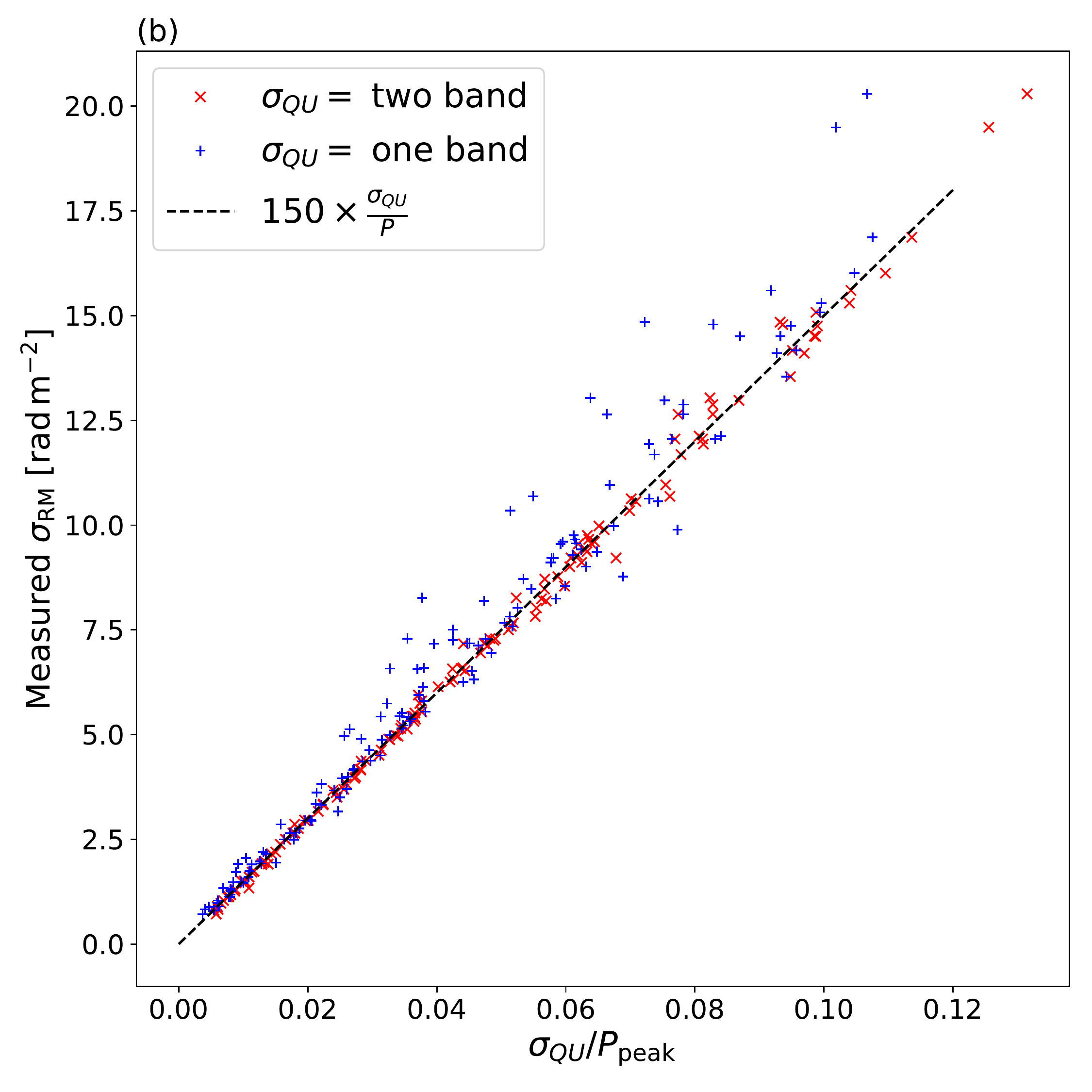}
\caption{Panel (a): RM uncertainties as measured by us compared to the reported RM uncertainties in the \citet{Taylor09} catalog for 100 pairs of sources. The black dashed line shows a one-to-one relation. Panel (b): RM uncertainties as measured by us for 100 pairs of sources vs the average $Q$ and $U$ image noise, $\sigma_{QU}$, (as measured by us) over the peak polarized intensity, $P$. The blue pluses are from measuring, and averaging, the $Q$ and $U$ noise in each at each of the two frequencies. The red crosses have $\sigma_{QU}$ as the average from the the single frequency, originally processed, NVSS data. The dashed black line shows the best fit line to the data of $150 \times \sigma_{QU}/P$. \\}
\label{fig:noise1}
\end{figure}

The RM from the two frequencies is determined by
\begin{equation}
{\rm RM} = C \, \frac{\Psi_{\lambda 1} -\Psi_{\lambda 2}}{\lambda_1^2-\lambda_2^2}.
\label{eq:a1}
\end{equation}
Here $C$ is a correction factor that accounts for the effect of the finite width of the bands on the effective center wavelength in $\lambda^2$ space that Taylor et al. report as $C=0.96$. In this case $\Psi_{\lambda i}$ is the polarization angle at $\lambda_i$ and is equal to
\begin{equation}
\Psi_{\lambda i} = \frac{1}{2} \arctan{\frac{U_{\lambda i}}{Q_{\lambda i}}},
\label{eq:A2}
\end{equation}
where $U_{\lambda i}$ and $Q_{\lambda i}$ are the measured intensities in the Stokes $U$ and $Q$ images at $\lambda_i$. To determine the uncertainty in RM, $\sigma_{\rm RM}$, one may use propagation of uncertainty from equation~(\ref{eq:A2}) such that
\begin{equation}
\sigma_{\Psi_{\lambda i}} = \frac{1}{2}\, \frac{Q_{\lambda i}U_{\lambda i}}{Q_{\lambda i}^2+U_{\lambda i}^2} \, \,\sqrt{ \left ( \frac{\sigma_{U_{\lambda i} }}{Q_{\lambda i} } \right )^2+ \left ( \frac{\sigma_{Q_{\lambda i} }}{U_{\lambda i} } \right )^2} .
\label{eq:A4}
\end{equation}
with $\sigma_{U_{\lambda i} }$ and $\sigma_{Q_{\lambda i}} $ being the measured rms noise in the $Q_{\lambda i}$ and $U_{\lambda i}$ images. Another way of writing eq.~\ref{eq:A4} if $U_{\lambda i}$ and $Q_{\lambda i}$ have approximately the same noise is 
\begin{equation}
\sigma_{\Psi_{\lambda i}} =  \frac{\sigma_{QU_{\lambda i}}}{2 P_{\lambda i}},
\label{eq:a44}
\end{equation}
where $\sigma_{QU_{\lambda i}}=\sigma_{Q_{\lambda i}} = \sigma_{U_{\lambda i}}$ \citep[see Appendix A of ][]{Brentjens05}. Finally, the uncertainty in the RM is
\begin{equation}
\sigma_{\rm RM} = C \, \frac{\sqrt{\sigma_{\Psi_{\lambda 1}}^2 +\sigma_{\Psi_{\lambda 2}}^2 }}{\lambda_1^2-\lambda_2^2}.
\label{eq:a5}
\end{equation}
Using these equations and the 100 maps, we measured the RM and RM uncertainties. We were able to recover the RMs reported by Taylor for the subsample of sources, just not the RM uncertainties.

The reported Taylor uncertainties for these 200 sources (100 pairs) compared to our measured uncertainties are shown in Fig.~\ref{fig:noise1}a. This panel clearly shows that the uncertainties for these sources are indeed overestimated in \citet{Taylor09}. This issue is not discussed in \citet{Taylor09} or any subsequent works (other than in \citet{Stil11}, where the uncertainties are thought to be underestimated). Therefore, we are not sure of the reason for this overestimation. We believe a likely explanation is that the measured Stokes $Q$ and $U$ image noise is significantly overestimated in the catalog of \citet{Taylor09}. \citet{Taylor09} states that the noise level $\sigma$ for $Q$ and $U$ was determined by calculating the rms variation about the mean in an annulus around each source in the mosaic images. It is not stated how large the annulus around each source was. Therefore, it is possible that for sources with a second component nearby, the noise-estimation regions for one component often included real polarized emission from the other (nearby) component in the pair, resulting in overestimated image noise. This would explain why the overestimation is seen more drastically in the sample of physical pair sources, as the physical pair sources have smaller separations. 

\begin{figure*}
\includegraphics[scale=0.34]{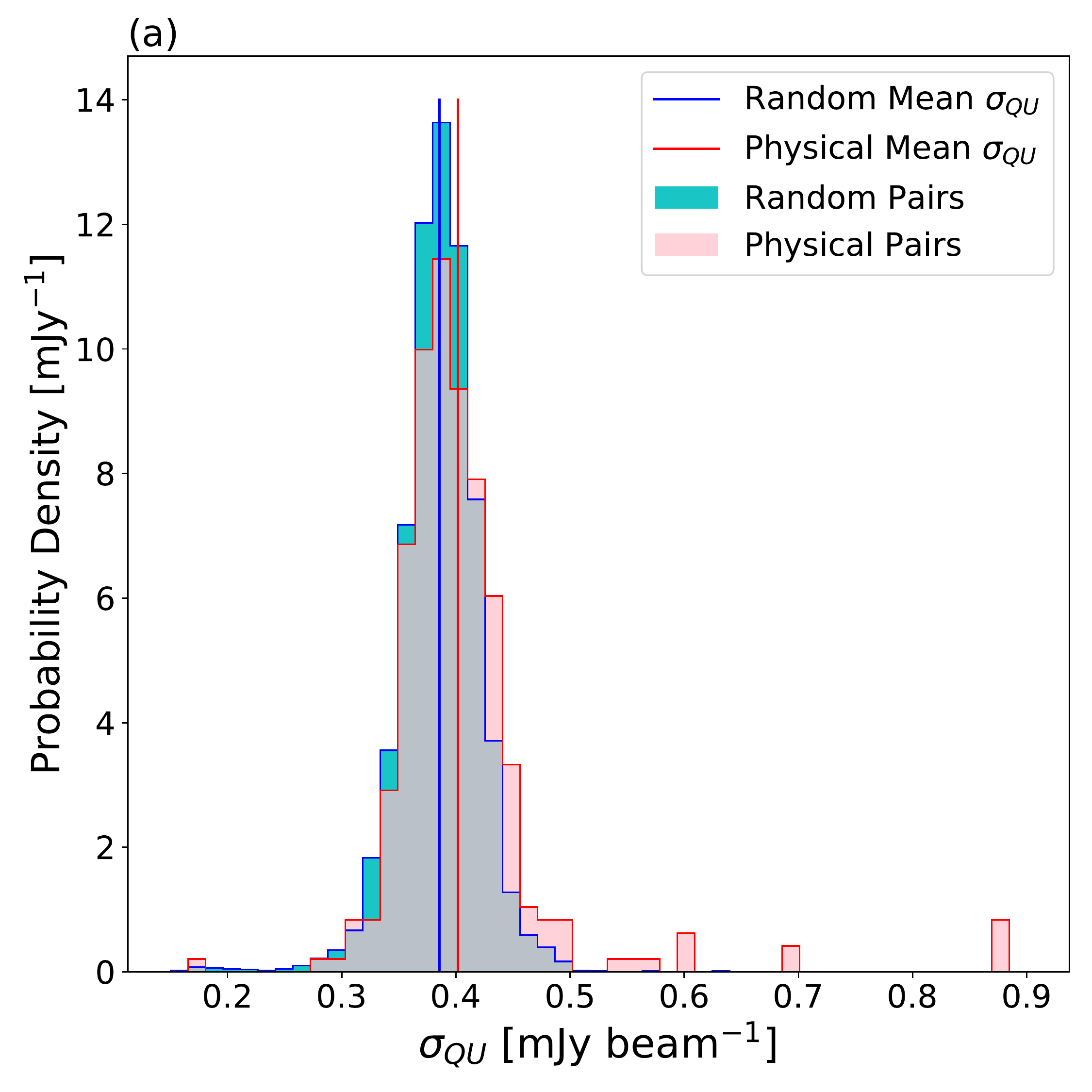}\includegraphics[scale=0.34]{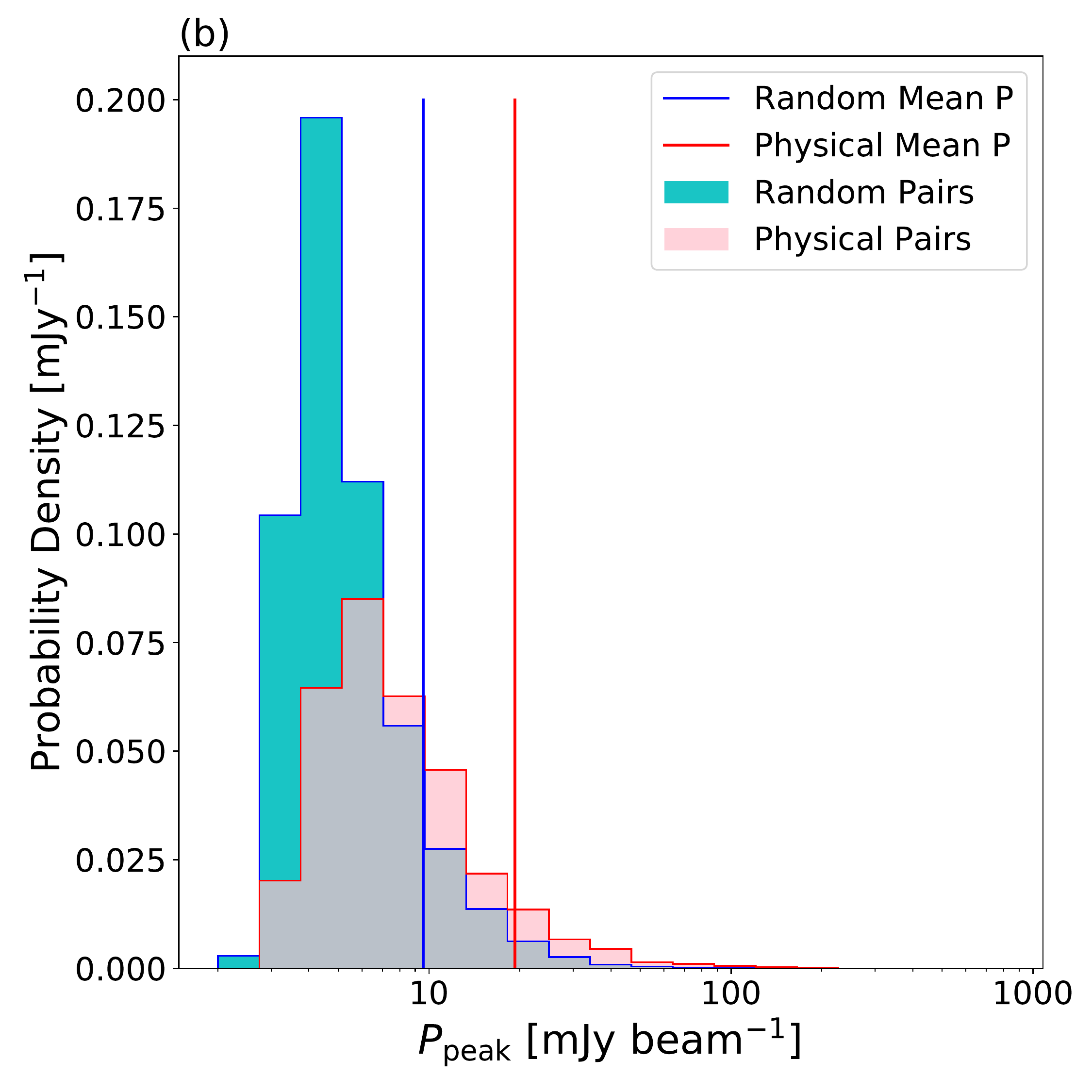}
\includegraphics[scale=0.34]{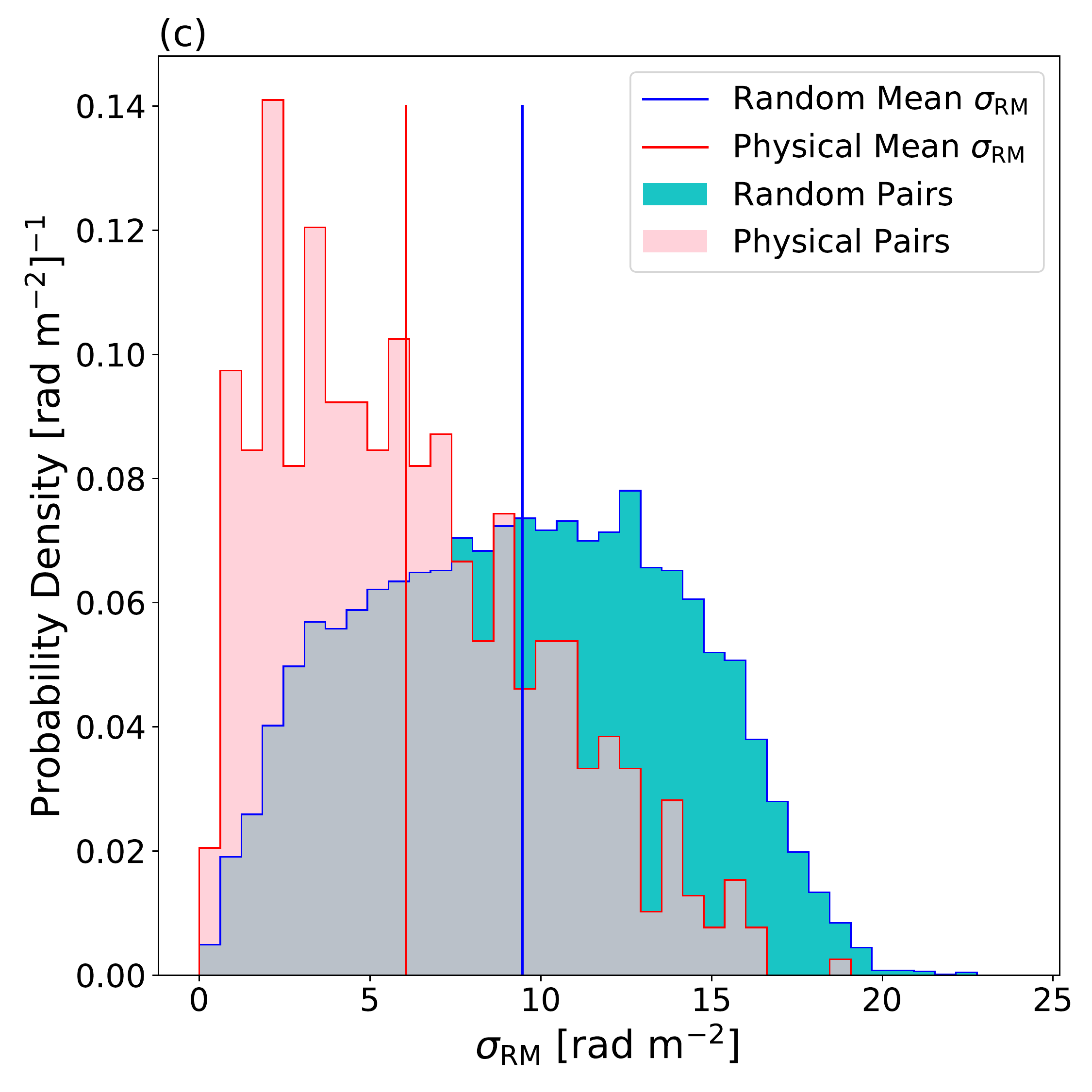}\includegraphics[scale=0.34]{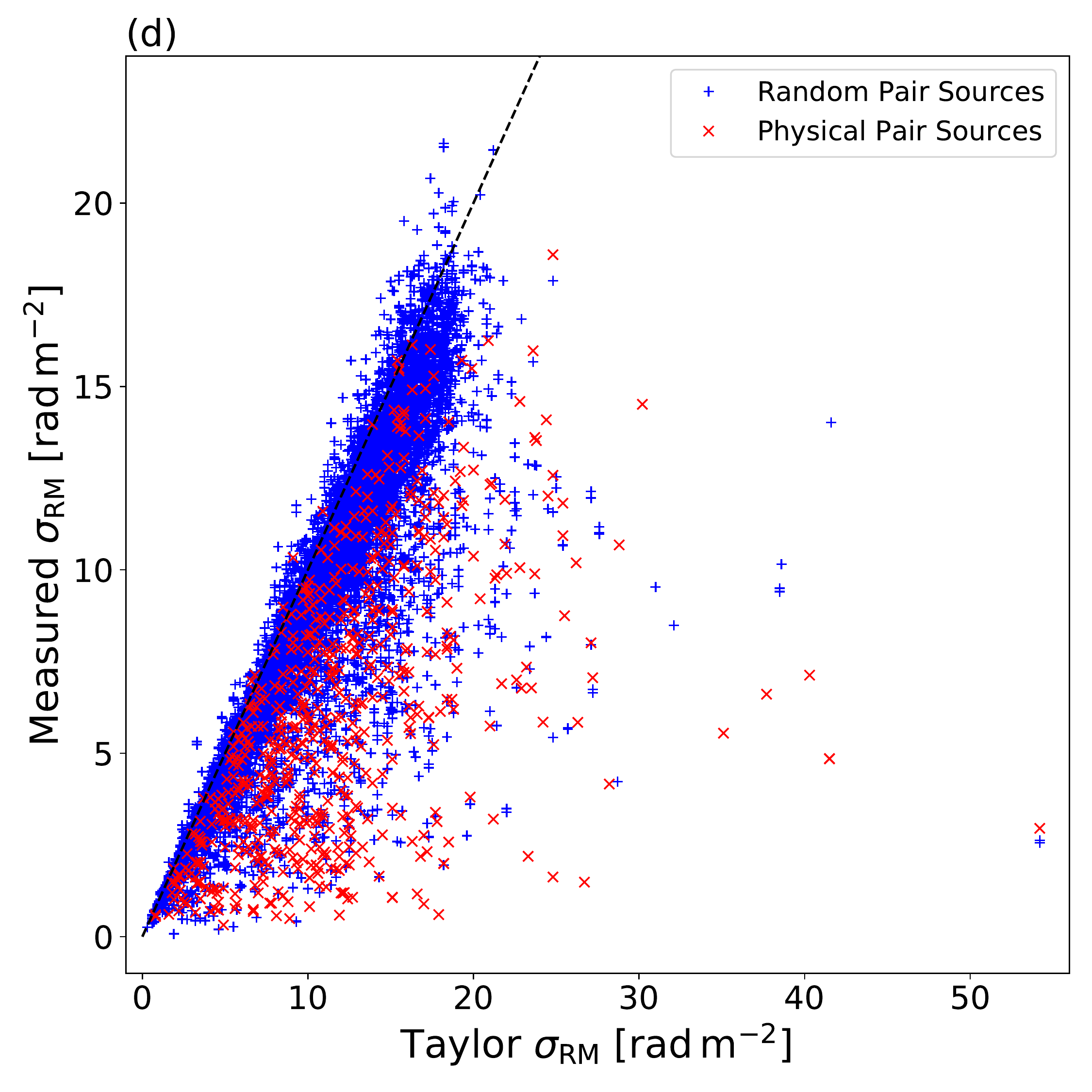}
\caption{Panel (a): Distributions of $\sigma_{QU}$, the average of the $Q$ and $U$ image noise for each source in our sample measured from the band-averaged $Q$ and $U$ NVSS images multiplied by $\sqrt{2}$ (to approximate the noise in the multi-frequency images). Panel (b): Distributions of the peak polarized intensities averaged over the two frequencies for all sources in our sample, as reported in the \citet{Taylor09} catalog. Panel (c): Distributions of $\sigma_{\rm RM}$ derived using eq.~(\ref{eq:A6}), the measured $\sigma_{QU}$ values, and the peak polarized intensities from \citet{Taylor09}. Panel (d): The reported uncertainties from the Taylor et al. catalog compared to those measured using eq.~(\ref{eq:A6}). The black dashed line shows a one-to-one relation. In all panels red is for physical pair sources and blue is for random pair sources and in panels a-c the vertical solid lines show the means the of the distributions. \\}
\label{fig:histss}
\end{figure*}

Without access to all of the two-frequency data used in \citet{Taylor09} to remeasure the RM uncertainties for our entire sample, it is necessary to estimate or approximate the RM noise for each source. The RM uncertainty should go as $K \times \sigma_{QU}/P$, where $P$ is the peak polarized intensity, $\sigma_{QU}$ is the average Stokes $Q$ and $U$ image noise and $K$ is a some factor. Using the subset of 100 pairs we find $K \simeq 150 \pm 5\,$ rad m$^{-2}$. This empirically derived factor matches what we expect from equations~(\ref{eq:a44}) and (\ref{eq:a5}), by simplifying $\sqrt{\sigma_{\Psi_{\lambda 1}}^2 +\sigma_{\Psi_{\lambda 2}}^2 } =$$ {\sqrt{2}\sigma_{QU}}\over{2\, P}$ assuming ${\sigma_{QU_{\lambda 1}}}/{P_{\lambda 1}} ={\sigma_{QU_{\lambda 2}}}/{P_{\lambda 2}} ={\sigma_{QU}}/{P}$, then 
\begin{equation}
\sigma_{\rm RM} = \frac{\sqrt{2} C}{2 \Delta \lambda^2} \frac{\sigma_{QU}}{P}, 
\end{equation}
and here ${(\sqrt{2} C)}/{(2 \Delta \lambda^2}) \simeq 147$. Figure~\ref{fig:noise1}b shows the measured RM uncertainties vs $\sigma_{QU}/P$ for the subsample of sources, along with the best fit model. For the subsample of data where the images were available at both frequencies, $\sigma_{QU}$ is taken as the average over the measured $\sigma$ of $Q_{\lambda 1}$,$Q_{\lambda 2}$, $U_{\lambda 1}$, and $U_{\lambda 2}$.

The frequency-averaged peak polarized intensities for all the sources are reported in the \citet{Taylor09} catalog. However, we did not have all of the two-frequency images to measure $\sigma_{{QU}_{\lambda i}}$ or $P_{\lambda i}$. The original processing of NVSS \citep{Condon98} did include Stokes $Q$ and $U$ and images are available from a postage stamp server. \footnote{\url{https://www.cv.nrao.edu/nvss/postage.shtml}} We assumed that the noise and in $Q$ and $U$ at the combined single frequency is approximately equal to the average noise in the two frequency bands divided by $\sqrt{2}$. We obtained postage stamp $Q$ and $U$ images at least $20{\arcmin}$ in size, covering all of our sources and measured the image noise in multiple regions, avoiding any polarized emission, in both $Q$ and $U$, and took the average values. \citet{Taylor09} reports the band averaged peak polarized intensities. Figure~\ref{fig:noise1}b shows the single-band $\sigma_{QU}/P$ as blue pluses, where in this case $\sigma_{QU} = \sigma_{QU_{{\rm center}}} \times \sqrt{2}$. There is more scatter when the single-band data is used to measure $\sigma_{QU}$, but still results in a best fit slope of 150. The additional scatter can likely be attributed to possible differences in the processing, or reprocessing, or the NVSS data such that the single band noise may not exactly equal $\sqrt{2}$ the multi-band noise. 

For this work we computed the RM uncertainties, $\sigma_{\rm RM}$, of all the RM sources used as 
 \begin{equation}
 \sigma_{\rm RM} = 150 \times \frac{\sqrt{2} \,\,\sigma_{QU_{{\rm center}}}}{P_{\rm center}}.
 \label{eq:A6}
 \end{equation}
 Figure~\ref{fig:histss} shows distributions of the measured $\sigma_{QU}$ values, peak polarized intensities $P$, and the derived $\sigma_{\rm RM}$'s, with all of these quantities being given in Table~\ref{tab:pairs1}. These uncertainties were then used to compute the noise corrections to $\Delta {\rm RM}^2$ as detailed in Sec.~\ref{sec:rmdifs}.

\bibliographystyle{aasjournal}


\label{lastpage}
\end{document}